\documentclass[aps,prl,twocolumn,superscriptaddress]{revtex4-2}
\usepackage{amsmath,amssymb,bm,mathrsfs,mathtools,graphicx}

\usepackage{mathtools, physics2, derivative, fixdif}
\newtheorem{theorem}{Theorem}
\usephysicsmodule{ab}
\usephysicsmodule{ab.braket}
\newcommand{\davg}[1]{%
  \left\langle\!\left\langle #1
  \right\rangle\!\right\rangle
}

\usepackage[colorlinks=true,citecolor=blue,linkcolor=blue,urlcolor=blue]{hyperref}
\newcommand{\sectionprl}[1]{\par\noindent\textit{#1.---}}
\newcommand{\E}{\mathbb E}
\newcommand{\Var}{\operatorname{Var}}

\newcommand{\dd}{\mathrm d}

\newcommand{\originalfigure}[1]{%
  \IfFileExists{#1}{\includegraphics[width=.99\linewidth]{#1}}{%
  \fbox{\parbox[c][30mm][c]{.91\linewidth}{\centering
  Figure file: \texttt{\detokenize{#1}}.\\
  Available in the accompanying source archive.}}}}

\begin{document}
\title{Thermodynamic signatures of the period-doubling route to chemical chaos}
\author{Kyota Tamano}
\email{tamano.kyota.45w@st.kyoto-u.ac.jp}
\affiliation{Department of Physics, Kyoto University, Kyoto 606-8502, Japan}
\author{Keiji Saito}
\email{keiji.saitoh@scphys.kyoto-u.ac.jp}
\affiliation{Department of Physics, Kyoto University, Kyoto 606-8502, Japan}
\date{\today}

\begin{abstract}
Chemical reaction networks can evolve from stable oscillations to chaos through successive period doublings. We investigate general properties of entropy-production fluctuations near the onset of chaos and within the chaotic regime in reversible, well-mixed reaction networks. Near an isolated supercritical period-doubling bifurcation, we find a one-sided divergence of the diffusion constant with exponent one. For chaotic attractors satisfying the stated conditions, the entropy variance per unit volume grows exponentially when the macroscopic limit is taken first. Its exponential growth rate is $2\lambda_{\max}$, where $\lambda_{\max}$ is the largest Lyapunov exponent, implying a divergent diffusion constant. We also discuss the volume dependence of the fluctuations and the underlying physical mechanisms.
\end{abstract}
\maketitle

\sectionprl{Introduction}Deterministic chaos is one of the most striking manifestations of nonlinear classical dynamics. Although the governing equations are fully deterministic, exponential sensitivity to initial conditions imposes an intrinsic limit on long-time predictability and gives rise to irregular, aperiodic motion. Chaos plays a central role in dynamical-systems descriptions of the transition to turbulence and is also prominent in a broad range of driven dissipative systems~\cite{ruelle1971nature,Lorenz1963DeterministicNonperiodicFlow}. In chemical kinetics, chaotic dynamics has been observed in the Belousov--Zhabotinsky and peroxidase--oxidase reactions along routes that include period doublings~\cite{zhang1993deterministic,geest1992period,hauser1997routes}.

The thermodynamic consequences of the onset of chaos are, however, far less well understood. 
Stochastic thermodynamics has endowed entropy production with a trajectory-level meaning 
and has uncovered universal constraints far from equilibrium, 
most notably fluctuation theorems, thermodynamic speed limits and thermodynamic uncertainty relations~\cite{Seifert_2025, Seifert_2012, seifert2005entropy, CrooksPhysRevE.60.2721CrooksRelation, EvansPhysRevLett.71.2401SecondLawViolation, GallavottiPhysRevLett.74.2694, Lebowitz1999GCsymmetry, shiraishi2018speed, dechant2022minimum, lee2022speed, van2023thermodynamic, funo2019speed, BaratoPhysRevLett.114.158101TUR, GingrichPhysRevLett.116.120601TUR}. 
It has also provided a systematic thermodynamic formulation for open chemical reaction networks driven by chemostatted species~\cite{RaoPhysRevX.6.041064ChemicalThermo,schmiedl2007stochastic, 865m-cxxg}. 
Recent studies have also revealed that entropy-production
fluctuations can exhibit singular behavior near
nonequilibrium phase transitions~\cite{NguyenPhysRevE.102.022101Schlogl, Remlein10.1063/5.0203659, Nguyen10.1063/1.5032104ChemicalOscillator, FiorePhysRevE.104.064123DiscontinuousPhaseTransition, OberreiterPhysRevLett.126.020603TimeCristal, chudak2025synchronizationthermodynamicallyconsistentstochastic, KewmingPhysRevA.106.033707Kerr, ShimPhysRevE.93.012113ActiveMatterBased, PtaszynskiPhysRevE.111.034125CurieWeiss,tamano2026universal}.
Although the mean entropy production in chemically chaotic networks has previously been investigated~\cite{Gaspard10.1063/5.0025350ChaosEntropyProduction}, the singular behavior of entropy-production fluctuations near the onset of chaos remains largely unexplored. In particular, it is not known in general how the critical mode of a local bifurcation is encoded in thermodynamic fluctuations, or how the subsequent amplification of those fluctuations is related to Lyapunov instability.

\begin{figure}[t]
\begin{center}
\includegraphics[width=0.77\linewidth]{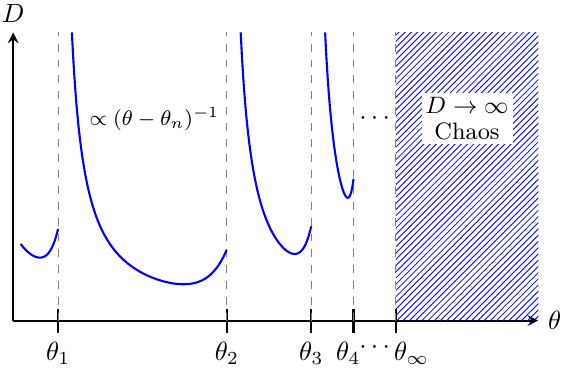}
\end{center}
  \vspace*{-0.3cm}
\caption{\label{fig1}Schematic picture of the diffusion constant $D$ defined in Eq. (\ref{diffusion_constant}) along a period-doubling cascade. At each bifurcation, a one-sided inverse-distance singularity emerges. The point $\theta_\infty$ is the transition point to chaos, beyond which chaos emerges, accompanied by $D\to\infty$.}
\end{figure}

Here, we address these questions for reversible, well-mixed chemical reaction networks in a volume $\Omega$, as defined below in Eq.~(\ref{che_meq}). As a control parameter $\theta$ is varied, a stable chemical oscillation undergoes successive period-doubling bifurcations at $\theta_1,\theta_2,\ldots$, accumulating at $\theta_\infty$ along a period-doubling route to chaos~\cite{feigenbaum1978quantitative}. The mean entropy-production rate remains finite at each isolated doubling, whereas its fluctuations can be singular.
We first study entropy fluctuations near an isolated period-doubling bifurcation. Under the nondegeneracy and Floquet regularity conditions, the diffusion constant per unit volume (see Eq.~(\ref{diffusion_constant})) shows a one-sided divergence with exponent $1$, as shown schematically in Fig.~\ref{fig1}. The analysis combines the system-size expansion with the center-manifold normal form. We next consider chaos for $\theta > \theta_{\infty}$, where unstable stretching in phase space produces a divergent diffusion constant. This phenomenon is explained through volume-dependent diffusion as well as the time dependence of the fluctuations, as summarized in Table~\ref{table1}. The  results provide a first step toward understanding {\it chaos} with thermodynamic fluctuations.

\begin{table}[t]
  \vspace*{-0.4cm}
\begin{ruledtabular}
\begin{tabular}{lccc}
  Attractor & $D$     & $\mathfrak{D}_{\Omega}$   & $\upsilon_{{\bm x}_0} (\tau ) $    \\
\hline
  Fixed point & finite      & $O(\Omega)$ & $O(\tau)$    \\
  Limit cycle & finite       & $O(\Omega)$ & $O(\tau)$    \\
Chaos       & $\infty$  & $O(\Omega^2)$  & $e^{2 \lambda_{\rm max} \tau + o(\tau)}$
\end{tabular}
\end{ruledtabular}
  \vspace*{-0.1cm}
\caption{\label{table1}Scalings for invariant attractors. See Eqs.~\eqref{diffusion_constant}-\eqref{time_depe_constant} for the definitions.}
\end{table}

\sectionprl{Setup}
We consider a spatially homogeneous chemical reaction network with reversible reaction channels:
\begin{align}
    \sum_{\ell' \in \mathcal{S}_c}
    \bar{\nabla}_{+\rho}^{\ell'} A_{\ell'}
    +
    \sum_{\ell \in \mathcal{S}}
    \nabla_{+\rho}^{\ell} X_{\ell}
   &
    \overset{k_{+\rho}}{\underset{k_{-\rho}}{\rightleftharpoons}}
    \sum_{\ell \in \mathcal{S}}
    \nabla_{-\rho}^{\ell} X_{\ell}
    +
    \sum_{\ell' \in \mathcal{S}_c}
    \bar{\nabla}_{-\rho}^{\ell'} A_{\ell'} . \label{che_meq}
\end{align}
Here, $X_{\ell}$ denotes the $\ell$th dynamical chemical species, whereas
$A_{\ell'}$ denotes a chemostatted species whose concentration is maintained
at a prescribed constant value.
We denote the corresponding sets of species by $\mathcal{S}$ and
$\mathcal{S}_c$, respectively, and the set of reaction channels by
$\mathcal{R}$, with $\rho \in \mathcal{R}$.
The nonnegative integers $\nabla_{\pm\rho}^{\ell}$ and $\bar{\nabla}_{\pm\rho}^{\ell'} \in \mathbb{Z}_{\geq 0}$ are the stoichiometric coefficients of the forward and backward reactions, and $k_{\pm\rho}$ are the corresponding rate constants. We set $k_{\mathrm B}=1$ and the bath temperature to unity. Concentrations and rate constants are expressed in fixed reference units, and local detailed balance is assumed. If conservation laws are present, all dynamics and stability conditions are restricted to a stoichiometric compatibility class. In this paper, we assume that the chemical reaction network has no conservation laws. 

We first introduce the dynamics of the chemical reaction network. Let $\Omega$ denote the volume of the reaction vessel. In the macroscopic limit $\Omega \to \infty$, the dynamics is governed by the law of mass action~\cite{GrootBA03510754NonequilibriumThermo, Epstein10.1093/oso/9780195096705.001.0001NonlinearChemicalDynamics}:
\begin{align}
\begin{split}
\dot{x}_{\ell} = F_{\ell} ({\bm x}(t)),& ~~~ F_{\ell} ({\bm x}(t)) = \sum_{\rho \in {\cal R}} \nabla_{\rho}^{\ell} (J_{+\rho} - J_{-\rho}) \, , \\
J_{\pm \rho} ({\bm x}(t))&:=k_{\pm \rho} \prod_{\ell' \in \mathcal{S}_c} a_{\ell'}^{ \bar{\nabla}_{\pm\rho}^{\ell'}} \prod_{\ell \in \mathcal{S}} [x_{\ell}(t)]^{\nabla_{\pm\rho}^{\ell}} \, , 
\end{split}
\label{deterministic eq}
\end{align}
Here, $x_{\ell}(t)$, the $\ell$th component of $\bm{x}(t)$, denotes the concentration of species $X_{\ell}$ at time $t$, and $a_{\ell'}$ denotes the fixed concentration of the chemostatted species $A_{\ell'}$. We define $\nabla_{\rho}^{\ell} := \nabla_{-\rho}^{\ell}  - \nabla_{+\rho}^{\ell}$.

For finite $\Omega$, molecular discreteness gives rise to intrinsic reaction noise, and the dynamics is instead described by the chemical master equation
\begin{multline}
    \frac{\partial}{\partial t}P(\bm{n},t)
    =
    \sum_{\rho\in\mathcal{R}}\sum_{s=\pm}
    \Bigl[
    W^{\rho}_{\bm{n},\,\bm{n}+s\nabla_{\rho}}\,
    P(\bm{n}+s\nabla_{\rho},t)
    \\
    -
    W^{\rho}_{\bm{n}+s\nabla_{\rho},\,\bm{n}}\,
    P(\bm{n},t)
    \Bigr] .
\end{multline}
The components $n_{\ell}$ of the vector $\bm{n}$ represent the numbers of
molecules of species $X_{\ell}$, whereas $\nabla_{\rho}$ denotes the
stoichiometric change vector with components $\nabla_{\rho}^{\ell}$.
The explicit expressions for the transition rates ${\bm W}$ are given in the footnote
\footnote{The transition rates are written as
\begin{align*}
 W^{\rho}_{\bm{n},\,\bm{n}\mp\nabla_{\rho}}  &= \Omega k_{\pm\rho}
    \prod_{\ell'\in\mathcal{S}_c}
    a_{\ell'}^{\bar{\nabla}_{\pm\rho}^{\ell'}}
    \prod_{\ell\in\mathcal{S}}
    \frac{(n_{\ell}\mp\nabla_{\rho}^{\ell})!
    }{
        (n_{\ell}-\nabla_{\mp\rho}^{\ell})!
    }
     \Omega^{-\nabla_{\pm\rho}^{\ell}} 
\end{align*}
}.
We choose a control parameter $\theta$ along which period-doubling bifurcations occur. One choice is to vary the chemostat concentrations as $a_{\ell}(\theta)=a_{\ell}^{(0)}+\theta\hat{a}_{\ell}$, where $\hat{\bm a}$ specifies the direction of variation from the reference concentrations $\bm a^{(0)}$. Another choice, adopted in the numerical model below, is to vary a reaction rate constant. When its forward-to-backward ratio changes, the entropy increment must also be varied consistently; this changes a thermodynamic affinity rather than only a kinetic timescale.

Entropy fluctuations are formulated in terms of trajectories of the stochastic process. A trajectory $\Gamma$ over the time interval $[0,\tau]$ is specified by
$ \Gamma:
    \bm{n}_0
    \xrightarrow{(\rho_1,s_1,t_1)}
    \bm{n}_1
    \cdots
    \xrightarrow{(\rho_i,s_i,t_i)}
    \bm{n}_i
    \cdots
    \xrightarrow{(\rho_r,s_r,t_r)}
    \bm{n}_r$, 
where $t_i$ is the time of the $i$th reaction event, $\rho_i$ specifies the reaction channel, and $s_i=\pm1$ specifies the reaction direction. Thus, $\bm{n}_i=\bm{n}_{i-1}+s_i\nabla_{\rho_i}$, and $r$ denotes the total number of reaction events along the trajectory. The entropy transferred to the environment along $\Gamma$ is
\begin{align}
    \Sigma_e(\Gamma)
    &:=
    \sum_{\rho\in\mathcal{R}}
    \eta_{\rho} Z_{\rho}(\Gamma),
    \qquad
    Z_{\rho}(\Gamma)
    :=
    \sum_{i=1}^{r}
    s_i\,\delta_{\rho_i,\rho},
\label{eq:entropy}
\end{align}
where $Z_{\rho}(\Gamma)$ is the extent of reaction
$\rho$ along the trajectory.
The entropy transferred to the environment by a single forward reaction $(+\rho)$ is denoted by $\eta_{\rho}$ and is given by
$ \eta_{\rho}  =
    \ln [ ({k_{+\rho}}/{k_{-\rho}})
    \prod_{\ell'\in\mathcal{S}_c}
    a_{\ell'}^{( \bar{\nabla}_{+\rho}^{\ell'} - \bar{\nabla}_{-\rho}^{\ell'}) } ]$.
Accordingly, $\Sigma_e$ is the medium entropy with the intrinsic molecular-state entropy treated separately~\cite{schmiedl2007stochastic}. It differs from the sum of logarithmic propensity ratios by a temporal boundary term~\cite{tamano2026universal}. Such a term does not change a finite stationary long-time variance rate when its second moment is finite, but it need not be negligible. In this paper, we consider Eq.~\eqref{eq:entropy} throughout. We write $J_\rho=J_{+\rho}-J_{-\rho}$, ${\mathcal A}_\rho=J_{+\rho}+J_{-\rho}$, and $j_{\mathrm d}(\bm x)=\sum_\rho\eta_\rho J_\rho(\bm x)$ for the deterministic entropy rate per volume. The deterministic flow starting at $\bm x_0$ is denoted by $\phi_t\bm x_0$.

\sectionprl{Definitions of entropy production and its fluctuations}
We assume that the finite-volume process is nonexplosive and has a stationary distribution $\pi_\Omega$ with the moments used below. Starting from a fixed molecular state $\bm n_0$, we first average over trajectories. The average entropy-production rate is defined as $ \sigma = \lim_{\tau\to \infty} {1\over \tau} \lim_{\Omega\to\infty}{1\over\Omega }\E_{\pi_\Omega}    \E_{\Gamma} [\Sigma_e (\Gamma | {\bm n}_0 )  ] $, where $(\Gamma | {\bm n}_0)$ denotes a trajectory starting from $\bm n_0$. The averages $\E_\Gamma$ and $\E_{\pi_\Omega}$ are over trajectories conditional on that state and over stationary initial states, respectively. We define the conditional variance by
\begin{align}
  V_{{\bm n}_0}(\Omega,\tau)  &:=   \Var_{\Gamma} \Sigma_e (\Gamma | {\bm n}_0 ) \, , 
\end{align}
where $\Var_{\Gamma}$ is the variance over trajectories at fixed $\bm n_0$. We then define three quantities:
\begin{align}
  D &:= \lim_{\tau\to \infty} {1\over \tau} \lim_{\Omega\to\infty}{1\over \Omega} \E_{\pi_\Omega} [ V_{{\bm n}_0}(\Omega,\tau) ]      \, , \label{diffusion_constant}  \\
  \mathfrak{D}_{\Omega}    &:= \lim_{\tau\to\infty}{1\over \tau}  \E_{\pi_\Omega} [ V_{{\bm n}_0}(\Omega,\tau) ]
                          \, , \label{volume_depe_constant} \\
  \upsilon_{{\bm x}_0} (\tau )  &:=  \lim_{\Omega\to\infty}{1\over \Omega} V_{{\bm n}_0}(\Omega,\tau)   \, , \label{time_depe_constant} 
\end{align}
For a limit cycle, the outer time limit defining $D$ is taken along $\tau=mT_\theta\, (m\in I)$, with $T_\theta$ the actual period at the parameter $\theta$. The quantity $\upsilon_{{\bm x}_0} (\tau)$ describes the time dependence, and $D=\lim_{\tau\to\infty} \E_{\mu}[\upsilon_{{\bm x}_0} (\tau) ]/\tau$ with an invariant measure $\mu$ in the macroscopic limit. These quantities are defined in the infinite-volume limit. In contrast, $\mathfrak{D}_\Omega$ is the long-time variance rate at fixed volume. Numerical calculations provide estimates of the long-time variance rate $\mathfrak{D}_\Omega$ at different volumes.

\sectionprl{Critical behavior along the period-doubling route}
We consider an isolated supercritical period-doubling bifurcation at $\theta_c$. Center-manifold theory reduces the Poincar\'e return map to~\cite{Kuznetsov2023Elements}
\begin{equation}
 P(\xi)=-(1+\varepsilon)\xi+c\xi^3+\cdots,
 \quad \varepsilon=c_\theta(\theta-\theta_c),\quad c,c_\theta>0.
 \label{eq:normalform}
\end{equation}
For $\varepsilon<0$, the fixed point $\xi=0$ represents the stable parent cycle. For $\varepsilon>0$, the points $\xi_\pm=\pm\sqrt{\varepsilon/c}+O(\varepsilon)$ form a stable two-cycle. Correspondingly, the new orbit differs from the parent by a displacement of order $\sqrt\varepsilon$, which changes sign after one parent period. Averaging entropy production and return time over the doubled orbit cancels their leading odd-amplitude terms. The average entropy production rate $\sigma$ is therefore finite and continuous at $\theta_c$, although its derivative can change.

To analize the entropy-production fluctuations, we use the linear noise picture given by the system-size expansion $\bm n/\Omega\sim \bm x(t)+{\bm y}/\sqrt\Omega$ with a Langevin dynamics
\begin{align}
  d\boldsymbol{y}
 &=[\partial_{\boldsymbol{x}}\boldsymbol{F}(\boldsymbol{x}(t) )    ] \boldsymbol{y}\,dt
+\sum_\rho\boldsymbol{\nabla}_\rho(A_\rho(\boldsymbol{x} (t) ))^{1/2}\,dw_\rho \, , \label{ylangevin}
\end{align}
where $d w_{\rho}$ stands for an independent Wiener process. This picture is rigorously guaranteed by Kurtz's central limit theorem in the macroscopic limit~\cite{tamano2026universal,Kurtz10.1063/1.1678692,Kurtz_1971,KURTZ1978223}. The entropy change is then govened by $ d\Sigma_{\rm e} \simeq\Omega j_{\rm d}(\boldsymbol{x} )\,dt +\sqrt{\Omega}\,[  {\bm g}^{T}\boldsymbol{y}\,dt +\sum_\rho\eta_\rho({A_\rho(\boldsymbol{x})  })^{1/2}\,dw_\rho ] $ where ${\bm g}=\partial_{\bm x} j_d ({\bm x})$, which leads to the formula derived in Ref.~\cite{tamano2026universal}:
\begin{align}
\begin{split}
  \upsilon_{{\bm x}_0} (\tau ) 
  & \! =\!\int_0^\tau \dd s \sum_\rho {\mathcal A}_\rho(\bm x(s)) \nu_\rho(\tau,s)^2 ,  \\
 \nu_\rho(\tau,s)&=\eta_\rho+\!\!\sum_{\ell, \ell' \in {\cal S}}\int_s^\tau \!\!\dd t   \partial_{x_{\ell}}  j_{\mathrm d}(\bm x(t)) \Phi_{\ell,\ell'}(t,s) \nabla^{\ell'}_\rho \,  
\end{split}
\label{lcformula}
\end{align}
The microscopic initial state is fixed with $\bm n_0/\Omega\to\bm x_0$ and vanishing initial fluctuation in the central-limit scaling. The reference path is $\bm x(t)=\phi_t\bm x_0$, and ${\bm \Phi}(t,s)$ solves $\partial_t{\bm \Phi}={\partial_{\bm x}}{\bm F}(\bm x(t))\Phi$ with ${\bm \Phi} (s,s)={\mathbb I}$, where ${\mathbb I}$ is the identity matrix.

\begin{figure}[ht]
    \centering
    \includegraphics[width = 0.9\linewidth]{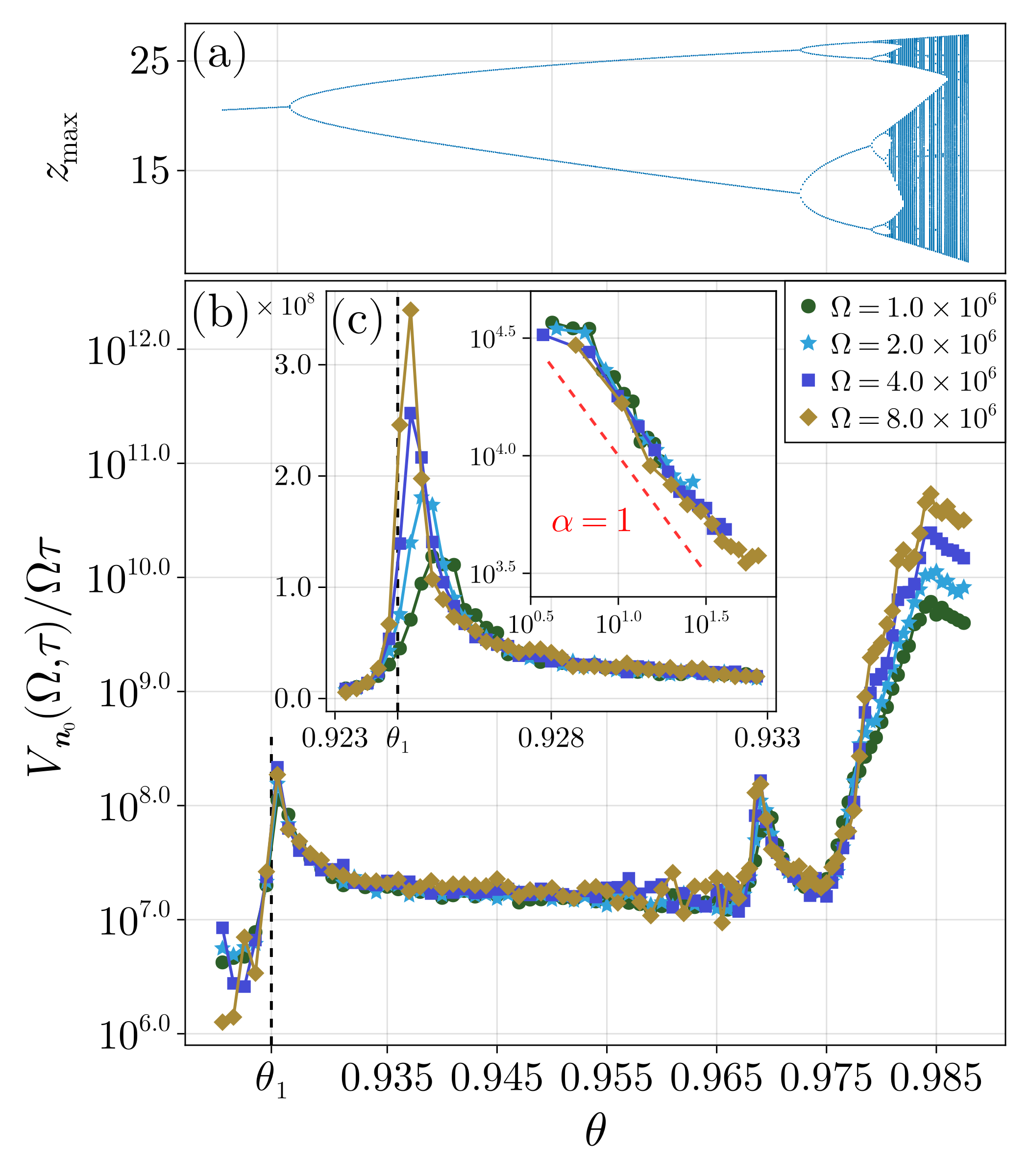}
    \caption{
        Period-doubling sequence and entropy-production fluctuations
        in the Willamowski--R\"ossler model~\eqref{WRmodel}.
        (a) Successive local maxima of $z$, the concentration of chemical species $Z$, in the deterministic dynamics, illustrating the bifurcation structure.
        (b) $V_{\bm{n}_0}/(\Omega\tau)$ for the indicated system volumes, plotted on a logarithmic scale.
        (c) Entropy-production fluctuations near a period-doubling point, shown on a linear scale, 
        and finite-size scaling in the corresponding period-two regime.
        The finite-size scaling analysis assumes $V_{\bm{n}_0}(\theta)/(\Omega\tau)=\Omega^y f(\Omega^x(\theta-\theta_1)), f(x) = x^{-\alpha}$, with $\theta_1=0.92445$, $x=0.62$, and $y=0.59$.
        The ratio $\alpha = y / x \approx 0.952$ is consistent with the analytical prediction $\alpha = 1$. 
        The red dashed lines indicate a slope of unity.
        The control parameter and observation time are $\theta=k_{+2}$ and $\tau=400000$, respectively.
        See EM and SM~\cite{supple} for further details.
    }
    \label{fig:period}
\end{figure}

To obtain the critical behavior of the fluctuations, we use Floquet theory for the monodromy matrix and the properties of its Floquet multipliers implied by the normal form in Eq.~\eqref{eq:normalform}. The derivation is outlined in the End Matter (EM), with further details given in the Supplemental Material (SM) \cite{supple}. Here, we state the resulting critical behavior:
\begin{equation}
 D =
 \begin{cases}
 O(1), &\varepsilon\uparrow0,\\
 c_+ \varepsilon^{-1}, &\varepsilon\downarrow0,
 \end{cases}
 \label{eq:critical}
\end{equation}
where $c_+$ is a positive constant. This indicates a one-sided divergence with exponent $1$. This behavior resembles that at a Hopf bifurcation~\cite{tamano2026universal}, but its antiperiodicity mechanism is different. This local result applies to successive isolated doublings along the period-doubling cascade satisfying these conditions, giving the schematic picture in Fig.~\ref{fig1}.
We numerically investigate this behavior using the Willamowski--R\"{o}ssler (WR) model~\cite{WillamowskiRoessler1980}, whose reactions and parameters are given in Eq.~(\ref{WRmodel}) in the EM. Fig.~\ref{fig:period} shows finite-time estimates of the entropy-variance rate per unit volume for several volumes. Clear peaks are found near $\theta_1$ and $\theta_2$, while later peaks are smeared at the present resolution. At finite volume, the linear-noise regime requires $\Omega\varepsilon^2\gg1$ and observation times longer than the critical relaxation time. The inset magnifies the peaks around $\theta_1$, and the $\Omega$ dependence is investigated in the log-log plot. The overall structure in the inset supports the theory.  

\begin{figure}[t]
    \centering
    \includegraphics[width=0.9\linewidth]{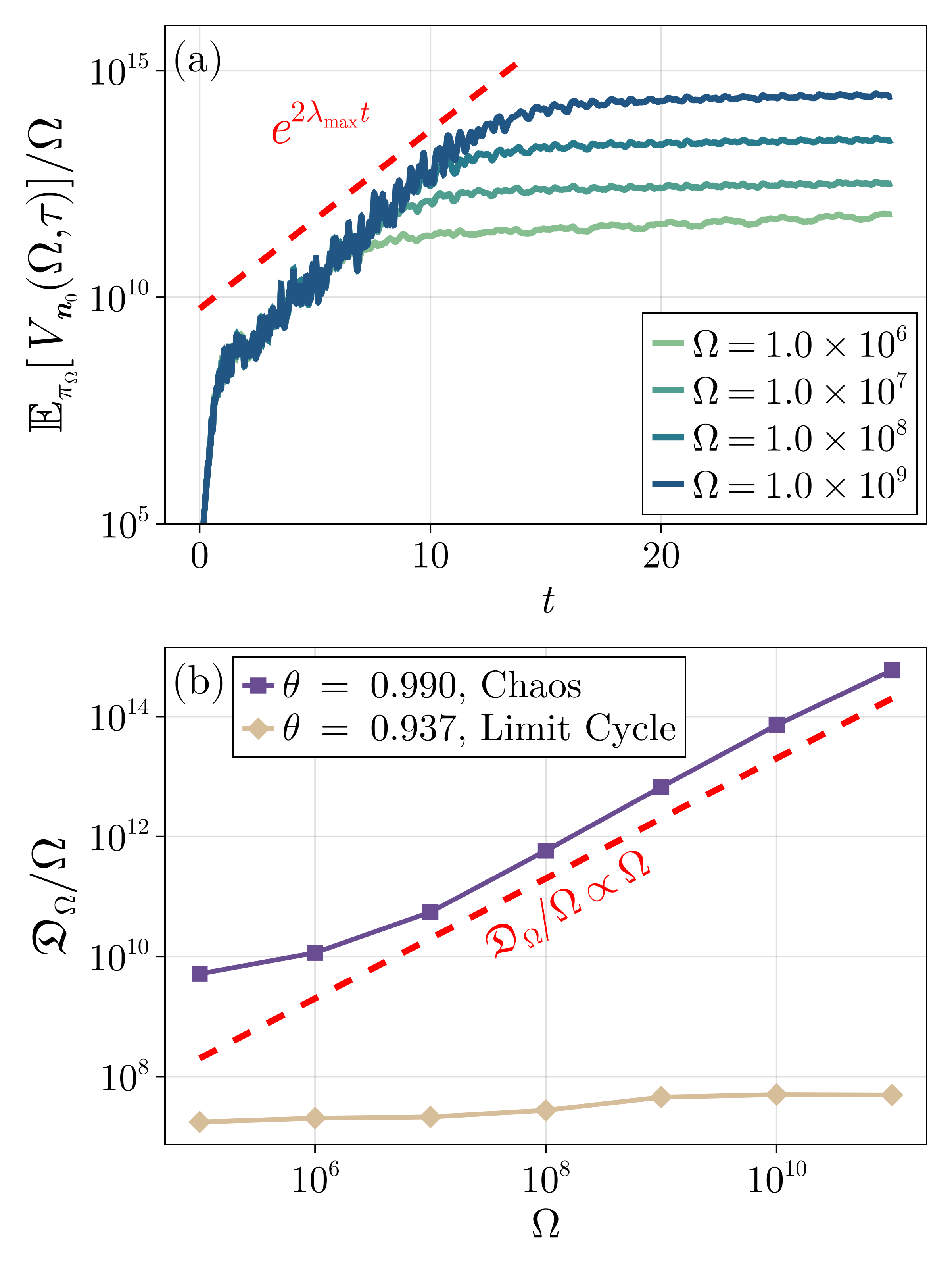}
    \caption{
        Entropy-production fluctuations in the chaotic regime
        ($\theta=0.990$, $\lambda_{\mathrm{max}}=0.453$)
        of the Willamowski--R\"ossler model~\eqref{WRmodel}.
        (a) $\mathbb{E}_{\pi_\Omega}[V_{\bm n_0}(\Omega,\tau)]/\Omega$
        for several system volumes, estimated by averaging over initial states
        sampled from the deterministic invariant measure $\mu$.
        The vertical axis is logarithmic, and the red dashed line
        shows $e^{2\lambda_{\mathrm{max}}\tau}$ as a reference growth law.
        (b) $\mathfrak{D}_{\Omega}/\Omega$ as a function of system volume
        $\Omega$ in the chaotic and limit-cycle ($\theta=0.937$) regimes.
        Both axes are logarithmic, and the red dashed line
        indicates the scaling $\mathfrak{D}_{\Omega}/\Omega\propto\Omega$.
        For (b), the observation time is $\tau=400000\gg\tau_E$.
        See EM and SM~\cite{supple} for further details.
    }
 \label{fig:volume}
\end{figure}

\sectionprl{Chaotic phase}
We now consider a parameter $\theta > \theta_{\infty}$ for which the deterministic attractor is chaotic. As $\Omega\to\infty$, we assume convergence of the stationary concentration process to the deterministic process with Sinai--Ruelle--Bowen (SRB) measure $\mu_{\rm SRB}$ \cite{Young2002}, including convergence of the averaged conditional second moments in Eq.~\eqref{lcformula}~\cite{Kifer1986RandomPerturbations,Young1986StochasticStability}. Numerically, we have confirmed that the average entropy production rate itself has no singular behavior as shown in the SM \cite{supple}. However, the entropy fluctuation generically shows anomalous behavior as stated in the following theorem:
\begin{theorem}\label{theo1}
Let $\mu=\mu_{\rm SRB}$ be an ergodic invariant measure supported on a compact invariant set, with positive concentrations $\mu$-almost everywhere and largest Lyapunov exponent $\lambda_{\rm max}>0$.
Let $E(\bm x)$ be the Oseledets subspace associated with $\lambda_{\rm max}$.
Assume that the entropy rate responds to perturbations in this subspace, $\bm g(\bm x)^T|_{E(\bm x)}\neq0$, on a set of positive $\mu$-measure.
Then
\begin{equation}
\frac{1}{\tau}\ln\upsilon_{\bm x_0}(\tau)
\xrightarrow[\tau\to\infty]{\mu\text{-prob.}}2\lambda_{\rm max}.
\label{timedepe_upsilon}
\end{equation}
Under the stationary-limit assumption above, this implies $D=\infty$.
\end{theorem}
The probability is over stationary initial states: the fraction whose logarithmic growth rate differs from $2\lambda_{\rm max}$ by more than any fixed tolerance tends to zero. An outline of the proof is given in the EM and detailed proof is provided in the SM~\cite{supple}.
The theorem concerns the entropy transferred to the environment in Eq.~\eqref{eq:entropy}; it determines a typical growth rate and an exponential lower bound on the initial-state average.
Figure~\ref{fig:volume}(a) shows the averaged variance $\E_{\pi_\Omega}[V_{\bm n_0}(\Omega,\tau)]/\Omega$, with $2\lambda_{\rm max}$ as a reference rate.
The interval of initial exponential amplification broadens as the volume increases.

In Fig.~\ref{fig:volume}(a), we observe linear time dependence after the exponential growth for the WR model. The slope of this linear time dependence corresponds to the value $\mathfrak{D}_{\Omega}/\Omega$. Let us now consider the volume dependence of these values. Fig.~\ref{fig:volume}(b) shows $\mathfrak{D}_{\Omega}/\Omega$ as a function of volume $\Omega$. In this figure, the results for the limit-cycle case are also shown. As can be clearly seen, the chaotic phase shows $\mathfrak{D}_{\Omega} = O(\Omega^2)$, while the limit cycle has $O(\Omega)$.

We consider this anomalous volume dependence in the chaotic regime using a {\it physical argument}. At finite volume, the initial exponential growth crosses over when the noise-amplified concentration displacement becomes macroscopic. A physical estimate uses the Ehrenfest time $\tau_{\rm E}$: the initial displacement is of order $\Omega^{-1/2}$, and typical unstable stretching makes it of order unity at
\begin{align}
  \tau_{\rm E} &\simeq {1\over 2 \lambda_{\rm max}} \ln \Omega \, .
\end{align}
The linear-noise picture is expected to describe the initial amplification before this crossover. If the integrated entropy responds to the same instability, its variance reaches an order-$\Omega^2$ scale near $\tau_{\rm E}$. Subsequent stationary entropy-production fluctuations of order $\Omega$, with a correlation time of order unity, then suggest $\mathfrak{D}_\Omega=O(\Omega^2)$. This is a {\it physical} estimate of the volume exponent; the Ehrenfest time alone does not determine the long-time coefficient. Nevertheless, this argument provides a consistent account of the observed volume dependence. 

  We give a more systematic, though still physical, argument for the volume dependence of the diffusion constant using correlation functions. We consider both the chaotic and nonchaotic phases in a unified way. Let us consider the decomposition of the long-time variance rate: $D_{\rm tot}:= \mathfrak{D}_\Omega + \mathfrak{D}_\Omega^{\rm init}$, where $\mathfrak{D}_\Omega^{\rm init}:= \lim_{\tau\to\infty} \Var_{\pi_\Omega}\E_\Gamma[\Sigma_e\mid\bm n_0]/\tau$. The quantity $D_{\rm tot}$ is the long-time growth rate of the unconditional variance over both initial states sampled from the stationary distribution and the trajectories starting from them, measured relative to the overall mean. This is just the law of total variance \cite{blitzstein2019introduction}. We focus on dynamics for which the dependence of the conditional mean entropy-production rate on the initial state decays sufficiently rapidly. More precisely, at each finite volume, the deviation of the conditional mean entropy-production rate from its stationary value has an $L^2(\pi_\Omega)$ norm that is integrable in time (see Eq.~(\ref{l2pinorm}) in the SM). This ensures that $\mathfrak{D}_\Omega^{\rm init}=0$, as shown in the EM~\cite{supple}. We therefore discuss the diffusion constant $D_{\rm tot} = \mathfrak{D}_\Omega$ as follows. Introduce the point-process entropy current $\hat{\cal J\,}_{\!\!\Sigma_e}(t)=\sum_{i=1}^r s_i\eta_{\rho_i}\delta(t-t_i)$, so that $\Sigma_e(\Gamma)=\int_0^\tau\hat{\cal J\,}_{\!\!\Sigma_e}(t)\dd t$. Its stationary covariance contains a contact term,
\begin{align}
  \begin{split}
 C_\Omega(s)&\!=\!\langle\delta\hat{\cal J\,}_{\!\!\Sigma_e}\!(s)
              \delta\hat{\cal J\,}_{\!\!\Sigma_e}              (0)\rangle_{\Gamma, \pi_\Omega}
    \! =\!\Omega \,A \delta(s)\!+\!C_\Omega^{\rm reg}(s),\\
 A&=\bigl< \sum_{\rho,\epsilon=\pm1}
    \eta_\rho^2w_{\rho,\epsilon}^\Omega \bigr>_{\pi_\Omega}/\Omega \, ,
  \end{split}\label{eq:contact}  
\end{align}
where $w_{\rho,\epsilon}^\Omega(\bm n):=W^\rho_{\bm n+\epsilon\bm\nabla_\rho,\bm n}$ and $\delta\hat{{\cal J}\,}_{\!\!\Sigma_e}\!=\hat{{\cal J}\,}_{\!\!\Sigma_e}\!-\langle\hat{{\cal J}\,}_{\!\!\Sigma_e}\!\rangle_{\pi_\Omega}$. The constant $A$ is $O(1)$. The regular part includes correlations between distinct reaction events. When integrable, it gives the formula
\begin{align}
  \mathfrak{D}_{\Omega}  &=A \,\Omega + 2 \lim_{\tau\to\infty}\int_{0}^\tau \left(1 - {s \over \tau}\right) C_\Omega^{\rm reg}(s)\dd s.
 \label{eq:Kubo}
\end{align}
Note that the Kurtz's central limit theory guarantees the linear noise picture for a given time in the infinite volume limit \cite{Kurtz10.1063/1.1678692,Kurtz_1971,KURTZ1978223}. In the fixed-point and limit-cycle cases, 
however, we assume that the linear-noise picture is still available due to the stability of the unique fixed point and limit-cycle \cite{bressloff2020phase,giacomin2018small}. 
Then, the physical argument gives the correlation functions in the forms $\Omega c e^{- b s}$ and $\Omega^2 c' e^{-\kappa s/\Omega } \cos (\omega_0 s )$, respectively. See the EM for an outline of the derivation. 
This immediately yields the volume dependence $O(\Omega)$ in $\mathfrak{D}_{\Omega}$ for both cases.

In the chaotic phase, we connect the correlation $C_\Omega^{\rm reg}(s)$ to detreministic correlation. We take the invariant measure in the deterministic limit to be the SRB measure and note the relation $\Omega^{-2}C_\Omega^{\rm reg}(s) \to C_{\mathrm d}(s)$ as $\Omega \to \infty$, where $C_{\mathrm d}(s)$ is the deterministic correlation function given by $C_{\mathrm d}(s)=\langle \delta j_{\mathrm d}(\phi_s\bm x_0)\, \delta j_{\mathrm d}(\bm x_0) \rangle_{\mu_{\rm SRB}}$. Here, $\delta j_{\mathrm d}(\phi_s\bm x_0):= j_{\mathrm d}(\phi_s\bm x_0) - \langle j_{\mathrm d} \rangle_{\mu_{\rm SRB}}$. Let us define the diffusion constant in the deterministic limit:
\begin{align}
  \mathfrak{D}_{\mathrm d}  &=2 \lim_{\tau\to\infty}\int_{0}^\tau \left(1 - {s \over \tau} \right) C_{\mathrm d}(s) \dd s \, .
 \label{eq:detKubo}
\end{align}
We assume that the limits $\tau\to \infty$ and $\Omega\to\infty$ can be interchanged to relate $\mathfrak{D}_{\Omega}$ to $\mathfrak{D}_{\mathrm d}$. Under this assumption, the following asymptotic behavior is obtained:
\begin{align}
  \mathfrak{D}_{\Omega}  &=\Omega^2  \mathfrak{D}_{\mathrm d}     + o(\Omega^2)\, .
 \label{eq:quadratic}
\end{align}
The deterministic diffusion constant $\mathfrak{D}_{\mathrm d}$ is genrally finite when the average is taken over the SRB measure $\mu\to\mu_{\rm SRB}$, which explains the anomalous volume dependence $\mathfrak{D}_{\Omega}  \propto \Omega^2$. 
For the WR model, our numerical simulations yield $\mathfrak{D}_{\Omega}/(\mathfrak{D}_{\mathrm{d}}\Omega^2) \simeq 0.65$, $0.75$, $0.81$, and $0.66$ for $\Omega=10^8, 10^9, 10^{10}$, and $10^{11}$, respectively.
These results lend numerical support to the argument leading to Eq.~\eqref{eq:quadratic}.
Further details of this comparison, along with a check of the finiteness of $\mathfrak{D}_{\mathrm{d}}$, are provided in the SM~\cite{supple}.

\sectionprl{Concluding remarks}
We study thermodynamic fluctuations underlying chemical reactions near the onset of chaos and within the chaotic regime. Our findings are summarized in Fig.~\ref{fig1} and Table~\ref{table1}. Entropy-production fluctuations reveal both local bifurcation singularities and chaotic amplification. Under the stated conditions, period doubling produces a one-sided divergence with exponent one, while the logarithmic growth rate of the entropy variance on a chaotic attractor converges in probability to $2\lambda_{\rm max}$ over stationary initial states when the macroscopic limit is taken first. 
Taking the long-time limit first instead yields ${\mathfrak D}_{\Omega}=\Omega^2 \mathfrak{D}_d+o(\Omega^2)$ under the stationary correlation conditions. These results connect thermodynamic fluctuations to dynamical instability while distinguishing the roles of the entropy observable and the order of limits.
\begin{acknowledgments}
K.T. acknowledges support from JST SPRING, Grant Number JPMJSP2110.
This work was supported by JSPS KAKENHI Grant Nos.~JP26H02015 and JP26H00388.
\end{acknowledgments}

\bibliography{cite}

\clearpage
\section*{End Matter}
\setcounter{equation}{0}
\renewcommand{\theequation}{E\arabic{equation}}
\renewcommand{\theHequation}{E\arabic{equation}}

\sectionprl{Outline of the derivation of the critical behavior}
We briefly derive the critical behavior near a period-doubling bifurcation.
With the limits taken in the order $\Omega \to \infty$ followed by $\tau \to \infty$, the entropy-production fluctuations can be evaluated using Eq.~\eqref{lcformula}.

Let us consider the period-one regime $\theta < \theta_c$ and denote the monodromy matrix of the period-one orbit $\bm{x}_{\mathrm{P1}}$ by $M_\theta$.
We define the left and right eigenvectors by $ M_\theta \bm{v}_{i,\theta} = \mu_i \bm{v}_{i,\theta}$, $\bm{w}_{i,\theta}^\dagger M_\theta = \mu_i \bm{w}_{i,\theta}^\dagger$ and 
$\bm{w}_{i,\theta}^\dagger \bm{v}_{j,\theta} = \delta_{ij}$, where $\mu_{1}, \mu_{2}, \ldots, \mu_{n}$ denote the eigenvalues of $M_\theta$.
By Floquet theory, we can decompose the fundamental matrix $\Phi^{\mathrm{P1}}_\theta(t, s) := \mathcal{T} \exp\ab(\int_{s}^{t} \d u \mathbb{S}_\theta(\bm{x}_{\mathrm{P1}}(u ,\theta)))$ as
$\Phi^{\mathrm{P1}}_\theta(t, s) = \sum_{i} \Phi^{\mathrm{P1}}_{i, \theta}(t, s)$
where each mode is given by $\Phi_{i, \theta}^{\mathrm{P1}}(t, s) := e^{\lambda_i (t - s)} \bm{v}_{i, \theta}(t) \bm{w}^\dagger_{i, \theta}(s)$. 
We define $\bm{v}_{i, \theta}(t) := e^{- \lambda_i t} \Phi^{\mathrm{P1}}_\theta(t, 0) \bm{v}_{i, \theta}, \bm{w}^\dagger_{i, \theta}(t) := e^{\lambda_i t} \bm{w}^\dagger_{i, \theta} \Phi^{\mathrm{P1}}_\theta(0, t)$, and $ \lambda_{i} := \frac{1}{T} \mathrm{Log} \mu_{i}$.  
We take $i = 1$ to label the critical mode, $i = 2, \dots, n - 1$ the noncritical modes, and $i = n$ the phase mode.
Focusing on the critical-mode contribution, the dominant contribution to the entropy-production variance is given by
\begin{equation}
    D_{1}
    = 
    \sum_{m, m' = -\infty}^{\infty} 
    \frac{
        \mathcal{F}_{2m+1} \mathcal{D}_{m' - m } \mathcal{F}_{2m'+1}^*
    }{
        \ab(\lambda_{\mathrm{P1}} + i \omega (m + \frac{1}{2}))
        \ab(\lambda_{\mathrm{P1}} - i \omega (m' + \frac{1}{2}))
    },
    \label{EM:ep-fluctuation-period-one}
\end{equation}
where $\lambda_{\mathrm{P1}} := \Re \lambda_{1}$, $\omega := 2\pi / T$, and the coefficients are defined by
\begin{equation}
    \begin{aligned}
        \mathcal{F}_{2m + 1} 
        & := \frac{1}{2T} \int_{0}^{2T} \d u  \sum_{\ell} \pdv{}!{x_\ell} 
        j_{\mathrm{d}}(\bm{x}_{\mathrm{P1}}(u)) \ab[\bm{v}_{1, \theta}(u)]_{\ell} e^{i\pi u / T} \\
        & \times e^{- i\omega (m + \frac{1}{2}) u }, \\
        \mathcal{D}_{n} 
        & := \frac{1}{T} \int_{0}^{T} \d s \sum_{\rho} \sum_{\ell, \ell'} 
        \ab[\bm{w}^\dagger_{1, \theta}(s)]_{\ell} 
        \nabla_{\rho}^{\ell}
        \mathcal{A}_{\rho}(\bm{x}_{\mathrm{P1}}(s)) \\
        & \times \nabla_{\rho}^{\ell'}
        \ab[\bm{w}^\dagger_{1, \theta}(s)]^*_{\ell'} e^{- i n \omega s}. 
    \end{aligned}
\end{equation}
Although the period-one limit cycle loses linear stability at $\theta = \theta_c$, so that $\lambda_{\mathrm{P1}} \to 0$ as $\theta \to \theta_c$, Eq.~\eqref{EM:ep-fluctuation-period-one} remains finite because $m,m'$ are integers, and hence neither $m + \frac{1}{2}$ nor $m' + \frac{1}{2}$ can vanish.
Therefore, the entropy-production fluctuations remain finite as $\theta \to \theta_c$, implying $\alpha_{-} = 0$.

Next, we consider the period-two regime $\theta > \theta_{c}$. 
We denote the monodromy matrix of the period-two orbit $\bm{x}_{\mathrm{P2}}$ by $N_\theta$. 
Let us define the eigenvalues and eigenvectors by $N_\theta \bm{u}_{i, \theta} = \nu_{i} \bm{u}_{i, \theta}$,  
$\bm{s}^{\dagger}_{i, \theta} N_\theta = \nu_i \bm{s}^{\dagger}_{i, \theta}$ and $\bm{s}_{i, \theta}^{\dagger} \bm{u}_{j, \theta} = \delta_{ij}$.
The critical-mode contribution to the fundamental matrix is given by $\Phi_{1, \theta}^{\mathrm{P2}}(t,s) = e^{\lambda_{\mathrm{P2}}(t-s)} \tilde{\bm{u}}_\theta(t) \tilde{\bm{s}}_\theta^\dagger(s)$. 
Here, we define $\tilde{\bm{u}}_\theta(t) := e^{- \lambda_{\mathrm{P2}} t} \Phi^{\mathrm{P2}}_\theta(t, 0) \bm{u}_{1, \theta}$, 
$\tilde{\bm{s}}_\theta^\dagger(s) := e^{\lambda_{\mathrm{P2}} s} \bm{s}^\dagger_{1, \theta} \Phi^{\mathrm{P2}}_\theta(0, s)$ and $\lambda_{\mathrm{P2}} := \frac{1}{2T} \mathrm{Log}\, \nu_1 \approx -\frac{2}{T} (\theta - \theta_c)$. 
The critical-mode contribution to the entropy-production fluctuations is given by
\begin{equation}
    D_{1} = \sum_{k, k' = -\infty}^{\infty} \frac{\mathcal{G}_{k} \mathcal{H}_{k' - k} \mathcal{G}_{k'}^*}{(\lambda_{\mathrm{P2}} + i \omega k / 2 ) (\lambda_{\mathrm{P2}} - i \omega k' / 2 )},
    \label{EM:ep-fluctuation-period-two}
\end{equation}
where 
\begin{equation}
    \begin{aligned}
        \mathcal{G}_{k} & := \frac{1}{2T} \int_{0}^{2T} \d u \sum_{\ell} \pdv{}!{x_\ell} j_{\mathrm{d}}(\bm{x}_{\mathrm{P2}}(u)) \ab[\tilde{\bm{u}}_\theta(u)]_{\ell} e^{- i k \omega u / 2},  \\        
        \mathcal{H}_{n} & := \frac{1}{2T} \int_{0}^{2T} \d s \sum_{\rho} \sum_{\ell, \ell'} \ab[\tilde{\bm{s}}^\dagger_\theta(s)]_{\ell} \nabla_{\rho}^{\ell} \mathcal{A}_{\rho}(\bm{x}_{\mathrm{P2}}(s)) \\
        & \times \nabla_{\rho}^{\ell'} \ab[\tilde{\bm{s}}^\dagger_\theta(s)]^*_{\ell'} e^{- i n \omega s / 2}. 
    \end{aligned}
\end{equation}
The $k=k'=0$ term in Eq.~\eqref{EM:ep-fluctuation-period-two} can give rise to a divergence.
An asymptotic expansion of $\bm{x}_{\mathrm{P2}}$ near $\theta=\theta_c$ shows that the numerator of the dominant contribution scales as
$\mathcal{G}_{0}\mathcal{H}_{0}\mathcal{G}_{0}^* = O(\theta-\theta_c)$.
Since $\lambda_{\mathrm{P2}}=O(\theta-\theta_c)$, the dominant term scales as $ \ab[\mathrm{Var}\,\sigma]_1 = O\ab((\theta-\theta_c)^{-1}). $
Therefore, we obtain $\alpha_+=1$. Further details are provided in the SM \cite{supple}.

\sectionprl{The Willamowski--R\"ossler model}
The Willamowski--R\"ossler model \cite{WillamowskiRoessler1980} exhibits chaotic behavior through a period-doubling cascade.
Its reaction scheme is given by
\begin{equation}
\begin{aligned}
    A_{1} + X & \xrightleftharpoons[k_{-1}]{k_{+1}} 2 X, & 
    X + Y &\xrightleftharpoons[k_{-2}]{k_{+2}} 2Y, & 
    A_5 + Y &\xrightleftharpoons[k_{-3}]{k_{+3}} A_{2}, \\
    X + Z &\xrightleftharpoons[k_{-4}]{k_{+4}} A_{3}, & 
    A_{4} + Z &\xrightleftharpoons[k_{-5}]{k_{+5}} 2Z
\end{aligned}
\label{WRmodel}
\end{equation}
where the concentrations of $X$, $Y$, and $Z$ vary in time, whereas $A_i$, $i=1,\dots,5$, are chemostatted species.
The deterministic dynamics are governed by the following rate equations:
\begin{equation}
    \begin{aligned}
        \dot{x} & = k_{+1} a_{1} x - k_{-1} x^2 - k_{+2} x y + k_{-2} y^2 - k_{+4} x z + k_{-4} a_{3} \\
        \dot{y} & = k_{+2} x y - k_{-2} y^2 - k_{+3} a_{5} y + k_{-3} a_{2}  \\
        \dot{z} & = - k_{+4} x z + k_{-4} a_{3} + k_{+5} a_{4} z - k_{-5} z^2 \\
    \end{aligned}
\end{equation}
The model parameters used in the simulations are $a_i=1$ for $i=1,\dots,5$, $k_{+1}=30.0$, $k_{-1}=0.25$, $k_{-2}=10^{-4}$, $k_{+3}=10.0$, $k_{-3}=10^{-3}$, $k_{+4}=1.0$, $k_{-4}=10^{-3}$, $k_{+5}=16.5$, and $k_{-5}=0.5$.
With these parameters fixed, the model undergoes a period-doubling cascade as the control parameter $\theta=k_{+2}$ is increased.

We numerically estimate the entropy-production variance
$V_{\bm{n}_0}(\Omega,\tau)$
from $N_{\mathrm e}$ independent realizations of the chemical Langevin equations (CLEs), integrated using the Euler--Maruyama method [see Eq.~\eqref{current-CLE} in the Supplemental Material (SM)~\cite{supple}], with the initial state held fixed within each ensemble.
For Figs.~\ref{fig:period} and \ref{fig:volume}(b), we use $N_{\mathrm e}=512$ independent trajectories with a single fixed initial state and an observation time of $\tau=400000$.
For Fig.~\ref{fig:volume}(a), we evaluate the time dependence of $V_{\bm{n}_0}(\Omega,\tau)$ over $\tau\in[0,30]$ using $N_{\mathrm e}=100$ independent trajectories for each of $100$ initial states. We then average the estimated variances over these initial states, sampled from the deterministic invariant measure $\mu$ as an approximation to the stationary distribution.
Further details of the numerical procedures and additional results for the Gaspard model, which undergoes a transition to chaos via a Shilnikov homoclinic bifurcation, are provided in the SM~\cite{supple}.

\vspace*{0.2cm}
\sectionprl{Outline of the proof of Theorem~\ref{theo1}}
The detailed proof is provided in the SM~\cite{supple}. Write $\bm g(t)=\partial_{\bm x}j_d(\bm x(t))$ and $E(t)=E(\bm x(t))$.
Lyapunov regularity holds in both time directions for $\mu$-almost every initial state.
Define $R_\tau(s)=\int_s^\tau\bm g(t)^T\bm\Phi(t,s)\,dt$ and the response on $E(t)$,
\begin{equation}
\bm q(t):=\int_0^\infty\!du\,
\bm g(t-u)^T\left.\bm\Phi(t-u,t)\right|_{E(t)}.
\label{eq:theo1_q_def}
\end{equation}
This row vector acts on the fastest unstable perturbations; backward contraction at rate $\lambda_{\rm max}$ makes the integral finite.
The composition law gives
\begin{equation}
\left.R_\tau(s)\right|_{E(s)}
=\bm q(\tau)\left.\bm\Phi(\tau,s)\right|_{E(s)}-\bm q(s).
\label{eq:theo1_R_q}
\end{equation}

Mass action makes $\bm F$ and $j_d$ polynomial in the concentrations.
Hence the row vector
$\bm H(t):=\bm q(t)\bm\Phi(t,0)|_{E(0)}=\bm q(0)+R_t(0)|_{E(0)}$
is analytic in time, with derivative $\bm g(t)^T\bm\Phi(t,0)|_{E(0)}$.
Ergodicity and the coupling assumption ensure that $\bm H$ is not identically zero along a typical trajectory.
Its zeros are therefore isolated. Invariance of $\mu$, followed by integration over time, gives $\mu\{\|\bm q(0)\|=0\}=0$.
Consequently, for every $\epsilon>0$,
\begin{equation}
\mu\{\|\bm q(\tau)\|<e^{-\epsilon\tau}\}
=\mu\{\|\bm q(0)\|<e^{-\epsilon\tau}\}\longrightarrow0.
\label{eq:theo1_q_probability}
\end{equation}

Let $m_\tau$ be the smallest singular value of $\bm\Phi(\tau,0)|_{E(0)}$, the least stretching within $E(0)$.
Regularity gives $m_\tau=e^{\lambda_{\rm max}\tau+o(\tau)}$ and, for $0<\epsilon<\lambda_{\rm max}$,
$\|\bm\Phi(t,s)\|\le K_\epsilon e^{\epsilon s}e^{(\lambda_{\rm max}+\epsilon/2)(t-s)}$ for $t\ge s\ge0$.
Substitution in Eq.~\eqref{lcformula} yields the upper bound below.
For the lower bound, retain a fixed initial interval $[0,T]$.
Positive reaction activities and the absence of conservation laws give
$\bm B(s):=\sum_\rho A_\rho(\bm x(s))\bm\nabla_\rho\bm\nabla_\rho^T\ge bI$ there, with $b>0$.
Using $R_\tau(s)\bm\Phi(s,0)=R_\tau(0)-R_s(0)$ and $(a-b)^2\ge a^2/2-b^2$ in Eq.~\eqref{lcformula}, together with Eq.~\eqref{eq:theo1_R_q}, gives
\begin{equation}
\begin{split}
c\,m_\tau^2\|\bm q(\tau)\|^2-C
&\le\upsilon_{\bm x_0}(\tau)\\
&\le C_0\tau+C_\epsilon e^{(2\lambda_{\rm max}+\epsilon)\tau}.
\end{split}
\label{eq:theo1_variance_bounds}
\end{equation}
All constants are finite, with $c>0$, and may depend on $\bm x_0$ but not on $\tau$.
Equation~\eqref{eq:theo1_q_probability} prevents exponential suppression of the lower bound in probability.
Since $\lambda_{\rm max}>0$, the finite subtraction is immaterial in probability; the two bounds and $\epsilon\downarrow0$ prove Eq.~\eqref{timedepe_upsilon}.
Finally, convergence in probability implies
$\E_\mu[\upsilon_{\bm x_0}(\tau)]\ge\tfrac12 e^{(2\lambda_{\rm max}-\epsilon)\tau}$
for every $0<\epsilon<2\lambda_{\rm max}$ and all sufficiently large $\tau$, proving $D=\infty$.
\vspace*{0.2cm}

\sectionprl{On $\mathfrak{D}_\Omega^{\rm init}$ in the total variance}
At fixed volume $\Omega$, let ${\mathbb L}$ be the Markov generator governing the probability vector $\bm p(t)$ through $\partial_t\bm p={\mathbb L}\bm p$, with stationary distribution $\pi_\Omega$. Its adjoint generator ${\mathbb L}^\dagger$ generates conditional expectations of state observables. Let $j_{\bm n}$ be the local mean entropy-production rate in state ${\bm n}$, and set $\bm f=\bm j-j_{\rm ss}\bm 1$, where $j_{\rm ss}=\sum_{{\bm n}} \pi_{\Omega,{{\bm n}}} j_{{\bm n}}$. We define the $L^2 (\pi_{\Omega})$ norm as $\| {\bm a} \|_{L^2(\pi_\Omega)}^2 = \sum_{{\bm n}} a_{{\bm n}}^2 \pi_{\Omega,{{\bm n}}}$ for an arbitrary real vector ${\bm a}$. 

We consider dynamics in which this dependence decays sufficiently rapidly. We measure its magnitude through $\bigl\|e^{{\mathbb L}^\dagger s}\bm f\bigr\|_{L^2(\pi_\Omega)}^2$. Exponential relaxation in this norm is sufficient for the following condition. We assume 
\begin{equation}
\int_0^\infty ds\,
\bigl\|e^{{\mathbb L}^\dagger s}\bm f\bigr\|_{L^2(\pi_\Omega)}
<\infty. \label{l2pinorm}
\end{equation}
This is an assumption on relaxation under the stochastic dynamics at fixed volume.
Under this condition, $\bm h:=-\int_0^\infty ds\, e^{{\mathbb L}^\dagger s}\bm f$ exists in $L^2(\pi_\Omega)$ and satisfies ${\mathbb L}^\dagger\bm h=\bm f$. Thus, 
$\E_\Gamma[\Sigma_e\!\mid\!\bm n_0]-j_{\rm ss}\tau =\!\int_0^{\tau} dt [e^{{\mathbb L}^\dagger\tau} {\bm f}]_{{\bm n}_0} \!=\!\bigl[ e^{{\mathbb L}^\dagger\tau}\bm h\!-\!\bm h\bigr]_{{\bm n}_0}$, where the rightmost expression has zero stationary mean. Then, we have 
\begin{align}
\Var_{\pi_\Omega}\E_\Gamma[\Sigma_e\!\mid\! \bm n_0]
&=\bigl\|e^{{\mathbb L}^\dagger\tau}\bm h-\bm h
  \bigr\|_{L^2(\pi_\Omega)}^2 \nonumber \\
  &\le 4\|\bm h\|_{L^2(\pi_\Omega)}^2 \, , 
\end{align}
where we use $\|e^{{\mathbb L}^\dagger\tau}\bm h\|_{L^2(\pi_\Omega)} \le\|\bm h\|_{L^2(\pi_\Omega)}$. This leads to $\mathfrak{D}_\Omega^{\rm init}=0$.

\vspace*{0.2cm}
\sectionprl{Typical correlation functions for nonchaotic cases}
Near a stable fixed point \(\boldsymbol{x}_*\), write \(\boldsymbol{n}/\Omega=\boldsymbol{x}_*+\Omega^{-1/2}\boldsymbol{y}\). The linear-noise approximation (see also (S3) in the SM~\cite{supple}) reads
\begin{align}
  d\boldsymbol{y}
  &=[\partial_{\boldsymbol{x}}\boldsymbol{F}(\boldsymbol{x}_*)    ] \boldsymbol{y}\,dt
+\sum_\rho\boldsymbol{\nabla}_\rho(A_\rho(\boldsymbol{x}_*))^{1/2}\,dw_\rho \, , \label{ylangevin} \\
d\Sigma_{\rm e}
&\simeq\Omega j_{\rm d}(\boldsymbol{x}_*)\,dt
+\sqrt{\Omega}\,[
  g_*^{T}\boldsymbol{y}\,dt
  +\sum_\rho\eta_\rho({A_\rho(\boldsymbol{x}_*)  })^{1/2}\,dw_\rho ] \nonumber
\end{align}
where ${\bm g}_*:=\partial_{\boldsymbol{x}}j_{\rm d}(\boldsymbol{x}_*)$ and \(w_\rho\) are independent Wiener processes. With the stationary covariance \(Q_*=\langle\boldsymbol{y}\boldsymbol{y}^{T}\rangle\), the regular current correlation is given by
\begin{align}
  C_\Omega^{\rm reg}(s)&\simeq \Omega\,\boldsymbol{g}_*^{T}e^{[\partial_{\boldsymbol{x}}\boldsymbol{F}(\boldsymbol{x}_*)] s} [ Q_* \boldsymbol{g}_* +\sum_\rho\eta_\rho A_\rho(\boldsymbol{x}_*)\boldsymbol{\nabla}_\rho ] \, .
\end{align}
A real relaxation mode with eigenvalue \(-b<0\) gives a contribution \(\Omega c e^{-bs}\), with \(b\) and \(c\) independent of \(\Omega\).

For a stable limit cycle, the phase-averaged diffusion approximation is given by~\cite{bressloff2020phase,giacomin2018small}
\begin{align}
  d\phi=\omega_0dt+\sqrt{2\kappa/\Omega}\,dw  \, ,
\end{align}
where \(\omega_0\ne0\) and \(\kappa>0\) are \(O(1)\). Retaining the first harmonic of the deterministic entropy rate, write \(j_{\rm d}(\boldsymbol{x}_*(\phi))=\bar j+a\cos\phi\). The stationary phase is uniform, and its Gaussian increments give $\left\langle\cos\phi(s)\cos\phi(0)\right\rangle =(1/2) e^{-\kappa s/\Omega}\cos(\omega_0s)$. Consequently, this leads to
\begin{align}
  C_\Omega^{\rm reg}(s) & \sim \Omega^2c'\,e^{-\kappa s/\Omega}\cos(\omega_0s) \, ,
\end{align}
where $c'=a^2/2$.

\clearpage
\onecolumngrid

\setcounter{secnumdepth}{2} 
\setcounter{section}{0}
\renewcommand{\thesection}{S.\Roman{section}}

\setcounter{equation}{0}
\renewcommand{\theequation}{S\arabic{equation}}
\renewcommand{\theHequation}{S\arabic{equation}}

\begin{center}
{\large \bf Supplementary Material for \protect \\ ``Thermodynamic Signatures of the Period-Doubling Route to Chemical Chaos''}\\
\vspace*{0.3cm}
Kyota Tamano$^{1}$ and Keiji Saito$^{1}$ \\
\vspace*{0.1cm}
 $^{1}${\small {\it Department of Physics, Kyoto University, Kyoto 606-8502, Japan}}
\end{center}

\section{System Size Expansion and Entropy-Production Fluctuation Formula} \label{fluctuation-formula}
Here, we derive the formula for the entropy-production fluctuation $D$. 
The derivation essentially follows that in Ref.~\cite{tamano2026universal}.
To derive the formula, we utilize the system-size expansion \cite{Gardiner2010StochasticMethods}:
\begin{equation}
    \begin{aligned}
        n_{\ell}(\Gamma) / \Omega = x_{\ell}(t) + y_{\ell}(t) / \sqrt{\Omega}, && Z_{\rho}(\Gamma) / \Omega = z_{\rho}(t) + w_{\rho}(t) / \sqrt{\Omega},
    \end{aligned}
\end{equation}
where $x_{\ell}(t)$ and $z_{\rho}(t)$ are the deterministic components, whose time evolution obeys the deterministic equations
\begin{equation}
    \begin{aligned}
        \dot{x}_{\ell} = \sum_{\rho \in \mathcal{R}} \nabla_{\rho}^{\ell} J_{\rho}, && \dot{z}_{\rho} = J_{\rho}.
    \end{aligned}
\end{equation}
$y_{\ell}(t) / \sqrt{\Omega}$ and $ w_{\rho}(t) / \sqrt{\Omega}$ are the stochastic fluctuations
and their time evolution obeys the Langevin equations \cite{Horowitz10.1063/1.4927395, KURTZ1978223}:
\begin{equation}
    \begin{aligned}
        \dot{y}_{\ell} &= \sum_{\ell'} \pdv{F_{\ell}(\bm{x}(t))}{x_{\ell'}} y_{\ell'} + \sum_{\rho} \nabla_{\rho}^{\ell} \sqrt{\mathcal{A}_{\rho}(\bm{x}(t))} \, \xi_{\rho}(t), \\
        \dot{w}_{\rho} &= \sum_{\ell'} \pdv{J_{\ell}(\bm{x}(t))}{x_{\ell'}} y_{\ell'} + \sqrt{\mathcal{A}_{\rho}(\bm{x}(t))} \, \xi_{\rho}(t), \\
    \end{aligned}
    \label{langevin-eq-fluc-var}
\end{equation}
where $\xi_{\rho}$ are independent Gaussian white noises satisfying $\davg{\xi_{\rho}} = 0, \davg{\xi_{\rho}(t) \xi_{\rho'}(t')} = \delta_{\rho\rho'} \delta(t - t')$.
$\davg{\cdots}$ denotes the average over noise realizations. The system-size expansion has been rigorously justified for a class of density-dependent Markov jump processes in Refs.~\cite{Kurtz10.1063/1.1678692, Kurtz_1971, KURTZ1978223}, 
provided that the time $\tau$ is fixed and the macroscopic limit $\Omega \to \infty$ is taken first.
The formal solution of \eqref{langevin-eq-fluc-var} is obtained for the initial conditions $y_{\ell}(0) = 0$ and $w_{\rho}(0) = 0$: 
\begin{equation}
    \begin{aligned}
        y_{\ell}(t) & = \sum_{\rho} \sum_{\ell'} \int_{0}^{t} \d s \Phi_{\ell, \ell'}(t, s) \nabla^{\ell'}_{\rho} \sqrt{\mathcal{A}_{\rho}(\bm{x}(s))} \, \xi_{\rho}(s), \\
        w_{\rho}(t) & = \sum_{\rho'} \int_{0}^{t} \d s G_{\rho, \rho'}(t, s) \xi_{\rho'}(s), 
    \end{aligned}
    \label{formal-solution}
\end{equation}
where $G_{\rho, \rho'}(t, s)$ is the Green's function
\begin{equation}
    G_{\rho, \rho'}(t, s) = \ab(\delta_{\rho, \rho'} + \int_{s}^{t} \d u \sum_{\ell, \ell'} \pdv{J_{\rho}(\bm{x}(u))}{x_{\ell}} \Phi_{\ell,\ell'}(u, s) \nabla_{\rho'}^{\ell'} ) \sqrt{\mathcal{A}_{\rho'}(\bm{x}(s))}.
\end{equation}

Using the fluctuation variable $w_{\rho}$, we obtain
\begin{equation}
    \begin{aligned}
        v_{\bm{x}_0}(\tau)
        = \lim_{\Omega \to \infty}\frac{1}{\Omega} \ab<\ab(\sum_{\rho} \eta_{\rho} Z_{\rho}(\Gamma) - \ab<\sum_{\rho} \eta_{\rho} Z_{\rho}(\Gamma)> )^2>
        = \sum_{\rho, \rho'} \eta_{\rho} \davg{w_{\rho} w_{\rho'}} \eta_{\rho'}. 
    \end{aligned}
    \label{fluc-ep-noise-representation}
\end{equation}
where we have replaced the path-probability average by the noise average. Substituting the formal solution into the right-hand side of \eqref{fluc-ep-noise-representation}, we obtain 
\begin{equation}
    \begin{aligned}
        D & = \lim_{\tau \to \infty} \frac{1}{\tau} \mathbb{E}_{\mu} 
        \ab[\int_{0}^{\tau} \d s \sum_{\rho} \nu^2_{\rho}(\bm{x}(s)) \mathcal{A}_{\rho}(\bm{x}(s))],  \\
        \nu_{\rho}(\bm{x}(s)) & = \eta_{\rho} + \sum_{\ell, \ell'} \int_{s}^{\tau} \d u \pdv{}!{x_\ell} j_{d}(\bm{x}(u)) \Phi_{\ell, \ell'}(u, s) \nabla_{\rho}^{\ell'}.
    \end{aligned}
    \label{fluctuation-formula-ep}
\end{equation}

\section{Critical Behavior near Period-Doubling Bifurcation}
In this section, we investigate the behavior of entropy-production fluctuations and the parametric response 
to the system's control parameter $\theta$ near a period-doubling bifurcation point. 
The goal of this section is to determine the critical exponents characterizing the divergences of entropy-production fluctuations and the parametric response of the entropy production. 
We define the critical exponents of entropy-production fluctuations by
\begin{equation}
    \begin{aligned}
        D \propto |\theta - \theta_c|^{ - \alpha_{-}} & \text{ for } \theta < \theta_c \\
        D \propto |\theta - \theta_c|^{ - \alpha_{+}} & \text{ for } \theta_c < \theta
    \end{aligned}
    \label{ep-fluctuation-critical-exponent}
\end{equation}
and the parametric response of the entropy production by
\begin{equation}
    \begin{aligned}
        \pdv{}!{\theta} \sigma \propto  |\theta - \theta_c|^{ - \beta_{-}} & \text{ for } \theta < \theta_c \\
        \pdv{}!{\theta} \sigma \propto |\theta - \theta_c|^{ - \beta_{+}} & \text{ for } \theta_c < \theta
    \end{aligned}
\end{equation}
where $\theta_c$ denotes the period-doubling bifurcation point. 

\subsection{Period-doubling bifurcation and reduced dynamics: A review} \label{pd-review}
We briefly review the period-doubling bifurcation and introduce the setup.
Emphasizing the dependence of the system on the control parameter $\theta$, we consider
\begin{equation}
    \begin{aligned}
        \dot{\bm{x}} &= \bm{F}_{\theta}(\bm{x}),
        &
        \bm{x} &\in \mathbb{R}^{|\mathcal{S}|},
        &
        \theta &\in \mathbb{R}.
    \end{aligned}
    \label{original-dynamics-system}
\end{equation}
We assume that a supercritical period-doubling bifurcation occurs at $\theta=\theta_c$. 
As $\theta$ passes through $\theta_c$, the stable solution of the system undergoes a transition from a period-one limit cycle to a period-two limit cycle. 
We denote the critical orbit at $\theta = \theta_c$ by $\bm{x}_c(t)$ and its minimal period by $T$. 
Depending on the orientation of the control parameter, the period-one regime may lie on either side of the bifurcation point $\theta=\theta_c$. 
For definiteness, we choose $\theta<\theta_c$ to correspond to the period-one regime and $\theta>\theta_c$ to correspond to the period-two regime.
We denote the period-one limit-cycle solution by $\bm{x}_{\mathrm{P1}}(t,\theta)$, 
whose period is $T_{\mathrm{P1}}(\theta) = T + O(|\theta-\theta_c|)$.
Similarly, we denote the corresponding period-two limit-cycle solution by $\bm{x}_{\mathrm{P2}}(t,\theta)$, whose period is $ T_{\mathrm{P2}}(\theta) = 2T + O(|\theta-\theta_c|)$.
Since we consider a supercritical period-doubling bifurcation, the critical orbit satisfies $\bm{x}_c(t) = \bm{x}_{\mathrm{P1}}(t,\theta_c) = \bm{x}_{\mathrm{P2}}(t,\theta_c)$.

\begin{figure}[ht]
    \centering
    \includegraphics[scale = 1.0]{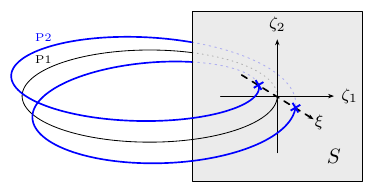}
    \caption{Schematic diagram of a Poincar\'e section.}
    \label{poincare-section}
\end{figure}

We introduce a Poincar\'e section $S$ transverse to the period-one orbit
$\bm{x}_{\mathrm{P1}}(t, \theta)$, as illustrated in Fig.~\ref{poincare-section}.
Let $\bm{\zeta}$ denote local coordinates on $S$, and choose the origin such
that the intersection of $\bm{x}_{\mathrm{P1}}(t,\theta)$ with $S$
corresponds to $\bm{\zeta}=0$.
The successive intersections of an orbit with $S$ define the Poincar\'e
return map
\begin{equation}
    \bm{\zeta}_{i+1}
    =
    P_{\theta}(\bm{\zeta}_i),
    \label{poincare-map}
\end{equation}
which provides a reduced discrete-time description of the dynamics on $S$.
Here, $\bm{\zeta}_i$ denotes the $i$th intersection of the orbit with $S$.
A period-one limit cycle of the original continuous-time system corresponds
to a fixed point of the Poincar\'e map.
Since the origin is chosen to correspond to
$\bm{x}_{\mathrm{P1}}(t,\theta)$, we have
\begin{equation}
    P_{\theta}(0)=0.
\end{equation}
Bifurcations of limit cycles in the original system can be
analyzed through the corresponding bifurcations of invariant orbits of the
Poincar\'e map.

The stability of a limit cycle is equivalent to the stability of the fixed point of the Poincar\'e map. 
Therefore, the stability of the limit cycle can be evaluated by analyzing the Jacobian of the Poincar\'e map 
\begin{equation}
    \mathbb{J}_{P_\theta}(\bm{\zeta}_*) := \left.\pdv{P_\theta}{\bm{\zeta}} \right|_{\bm{\zeta} = \bm{\zeta}_*}
\end{equation}
where $\bm{\zeta}_*$ is the corresponding fixed point. 
The eigenvalues $\mu_1,\mu_2, \dots \mu_{n-1} \in \mathbb{C}$ of $\mathbb{J}_{P_\theta}(\bm{\zeta}_*)$ 
determine the linear stability. The eigenvalues are independent of the choice of Poincar\'e section and of the local coordinates on it \cite{Kuznetsov2023Elements}.
If all the eigenvalues satisfy $|\mu_i| < 1$, then the limit cycle is stable. 
On the other hand, the limit cycle loses its linear stability and the system typically undergoes a bifurcation when any eigenvalue $\mu$ crosses the unit circle $|\mu| = 1$. 

In the case of a period-doubling bifurcation, one eigenvalue of $\mathbb{J}_{P_\theta}(\bm{\zeta} = 0)$ crosses $\mu = -1$ along the real axis. 
We denote this critical eigenvalue by $\mu_{1}(\theta)$ and assume the nondegeneracy of the linear term:
\begin{equation}
    \begin{aligned}
        \mu_1(\theta) = - (1 + a (\theta - \theta_c)) + O( (\theta - \theta_c)^2 ), && a > 0.  
    \end{aligned}
\end{equation}
and noncriticality $|\mu_{i}| < 1$ for the other eigenvalues $i = 2, \dots, n-1$ near the critical point $\theta_c$. 
Hereafter, we set $a = 1$ by rescaling the control parameter $\theta$. 
According to the \textit{center manifold theorem} \cite{Kuznetsov2023Elements}, 
the $(n-1)$-dimensional dynamics \eqref{poincare-map} can be reduced to
one-dimensional effective dynamics on the center manifold. 
After an appropriate coordinate transformation and rescaling, the reduced dynamics can be written in the normal form
\begin{equation}
    \begin{aligned}
        \xi \mapsto Q_\theta(\xi) = \mu_{1}(\theta) \xi + \xi^3, && \xi \in \mathbb{R}
    \end{aligned}
    \label{normal-form-flip}    
\end{equation}
where $\xi$ denotes a one-dimensional local coordinate on the center manifold. 
In discrete dynamical systems, a bifurcation described by the normal form in Eq.~\eqref{normal-form-flip} is known as a flip bifurcation. 
A period-doubling bifurcation of a limit cycle corresponds precisely to such a flip bifurcation of the Poincar\'e map.

We describe the behavior of the reduced dynamics \eqref{normal-form-flip}. 
The system \eqref{normal-form-flip} has a fixed point 
\begin{equation}
    \xi_* = 0
\end{equation} 
corresponding to the period-one limit cycle $\bm{x}_{\mathrm{P1}}(t, \theta)$ throughout the entire parameter regime.
The Jacobian at the fixed point $\xi_* = 0$ is given by
\begin{equation}
    \mu(\theta) := \left. \pdv{Q_\theta(\xi)}{\xi} \right|_{\xi = \xi_*} = \mu_1(\theta) = - (1 + \theta - \theta_c). 
\end{equation}
Therefore, the fixed point $\xi_* = 0$ is stable for $\theta < \theta_c$ and unstable for $\theta > \theta_c$. 

In the regime $\theta > \theta_c$, two additional fixed points of the second iterate $Q_\theta^2 := Q_\theta \circ Q_\theta$
\begin{equation}
    \xi_{\pm} \approx \pm \sqrt{\theta - \theta_c},
\end{equation}
emerge. The fixed points $\xi_{\pm}$ satisfy $\xi_{\mp}=Q_\theta(\xi_{\pm})$ and thus form a period-two orbit of the reduced map $Q_\theta$. 
The corresponding period-two orbit of the Poincar\'{e} map $P_\theta$ represents the period-two limit cycle $\bm{x}_{\mathrm{P2}}(t,\theta)$ of the original system.
Figure \ref{poincare-section} schematically illustrates the relationship among the period-two orbit of the original system, the Poincar\'{e} section $S$, and the center manifold within $S$.

The Jacobian of $Q^2_\theta$ at the fixed points $\xi_{\pm}$ is given by
\begin{equation}
    \begin{aligned}
        \kappa (\theta)
        :=
        \left.
        \odv{Q_{\theta}^2}{\xi}
        \right|_{\xi=\xi_{\pm}}
        =
        1 - 4\ab(\theta-\theta_c)
        + O((\theta-\theta_c)^2).
    \end{aligned}
\end{equation}
Since $|\kappa(\theta)|<1$ for $\theta>\theta_c$ sufficiently close to $\theta_c$, the period-two orbit $\xi_{+}, \xi_{-}$ is stable within the center manifold. 
Let us denote the period-two orbit of the full Poincar\'e map $P_\theta$ by $\bm{\zeta}_{\pm}$
and the eigenvalues of the Jacobian of the second iterate $\mathbb{J}_{P^2_\theta}(\bm{\zeta}_{\pm}) := \left.\pdv{}!{\bm{\zeta}} P_\theta^2(\bm{\zeta})\right|_{\bm{\zeta} = \bm{\zeta}_{\pm}}$ by $\kappa_1, \kappa_2, \dots \kappa_{n-1}$.  
We can label the eigenvalue associated with the critical mode as $\kappa_1 = \kappa(\theta)$. The other eigenvalues $\kappa_{2}, \dots, \kappa_{n-1}$ remain noncritical, i.e., $|\kappa_{i}| < 1$ for $i = 2, \dots, n-1$ for $\theta$ sufficiently close to $\theta_c$
by continuity of $\pdv{}!{\bm{\zeta}} P_\theta^2(\bm{\zeta})$ and of its spectrum.

\subsection{Critical Floquet mode and asymptotic expansion}
Here, we introduce the theoretical framework for the subsequent analysis of the critical behavior of entropy production in Secs.~\ref{ep-fluctuation} and \ref{ep-response}. 
In Sec.~\ref{pd-review}, we introduced the reduced dynamics near the period-doubling bifurcation and described its critical behavior. 
Based on this reduced description, we characterize the critical Floquet mode of the original continuous-time system 
and derive asymptotic expansions of the limit-cycle solutions around the critical orbit.

\subsubsection{Floquet-mode decomposition}
We consider the behavior of $\Phi_\theta(t,s)$, where the subscript $\theta$ emphasizes the parameter dependence, associated with the limit-cycle
solutions $\bm{x}_{\mathrm{P1}}(t,\theta)$ and $\bm{x}_{\mathrm{P2}}(t,\theta)$.

We first focus on the regime $\theta<\theta_c$.
In this regime, $\bm{x}_{\mathrm{P1}}(t,\theta)$ is stable. We define the fundamental matrix along the solution $\bm{x}_{\mathrm{P1}}(t, \theta)$
\begin{equation}
    \Phi_\theta^{\mathrm{P1}}(t, s) := \mathcal{T} \exp\ab(\int_{s}^{t} \d u \mathbb{S}_\theta(\bm{x}_{\mathrm{P1}}(u ,\theta))).
\end{equation}
We define the monodromy matrix by
\begin{equation}
    M_\theta := \Phi^{\mathrm{P1}}_\theta(T,0).
\end{equation}
We assume that $M_\theta$ is diagonalizable.
The monodromy matrix $M_\theta$ has a trivial eigenvalue $1$ with an
eigenvector proportional to $\dot{\bm{x}}(0)$.
The remaining nontrivial eigenvalues coincide with the eigenvalues of the
Jacobian of the Poincar\'e map $\mu_1,\dots,\mu_{n-1}$ \cite{Kuznetsov2023Elements}.
We define the corresponding right and left eigenvectors by
\begin{equation}
    \begin{aligned}
        M_\theta \bm{v}_{i,\theta}
        &= \mu_i \bm{v}_{i,\theta},
        &&
        \bm{w}_{i,\theta}^\dagger M_\theta
        = \mu_i \bm{w}_{i,\theta}^\dagger,
        &&
        \bm{w}_{i,\theta}^\dagger \bm{v}_{j,\theta}
        = \delta_{ij}.
    \end{aligned}
\end{equation}
Here, we set $\mu_n=1$ and $\bm{v}_{n,\theta}\propto\dot{\bm{x}}_{P1}(0, \theta)$. We define the time-dependent vectors
\begin{equation}
    \begin{aligned}
        \bm{v}_{i, \theta}(t) := e^{- \lambda_i t} \Phi^{\mathrm{P1}}_\theta(t, 0) \bm{v}_{i, \theta}, && \bm{w}^\dagger_{i, \theta}(t) := e^{\lambda_i t} \bm{w}^\dagger_{i, \theta} \Phi^{\mathrm{P1}}_\theta(0, t), && \lambda_{i} := \frac{1}{T} \mathrm{Log} \mu_{i}. 
    \end{aligned}
\end{equation}
Here, $\mathrm{Log}$ denotes the principal logarithm, whose imaginary part is chosen to lie in $(-\pi,\pi]$.
$\bm{v}_{i, \theta}(t)$ and $\bm{w}^\dagger_{i, \theta}(t)$ are $T$-periodic in time and satisfy $\bm{w}^\dagger_{i, \theta}(t)\bm{v}_{j, \theta}(t) = \delta_{ij}$. 
Using the vectors $\bm{v}_{i, \theta}(t)$ and $\bm{w}^\dagger_{i, \theta}(t)$, we can decompose the fundamental matrix into Floquet modes $\Phi^{\mathrm{P1}}_{i, \theta}(t, s)$ associated with the corresponding eigenvalues $\mu_{i}$: 
\begin{equation}
    \begin{aligned}
        \Phi^{\mathrm{P1}}_\theta(t, s) = \sum_{i} \Phi^{\mathrm{P1}}_{i, \theta}(t, s), && \Phi_{i, \theta}^{\mathrm{P1}}(t, s) := e^{\lambda_i (t - s)} \bm{v}_{i, \theta}(t) \bm{w}^\dagger_{i, \theta}(s). 
    \end{aligned}
\end{equation}
Since we assume that the eigenvalues have the form $\mu_1(\theta) = - (1 + \theta - \theta_c)$ and $|\mu_i| < 1$ for $i = 2, \dots, n - 1$, we obtain  
\begin{equation}
    \begin{aligned}
        \lambda_{1} \approx \frac{1}{T} (\theta - \theta_c) + i \frac{\pi}{T}, && \Re \lambda_{i} < 0 \text{ for } i = 2, \dots n-1. 
    \end{aligned}
    \label{period-one-floquet-exponents}
\end{equation}
We now focus on the critical mode corresponding to the eigenvalue $\mu_1$. 
We define new vectors that absorb the phase factor associated with the imaginary part of $\lambda_1$: 
\begin{equation}
    \begin{aligned}
    \tilde{\bm{v}}_\theta(t) := e^{i\pi t / T} \bm{v}_{1, \theta}(t), && \tilde{\bm{w}}^\dagger_\theta(s) = e^{ - i \pi s / T} \bm{w}_{1, \theta}^\dagger(s). 
    \end{aligned}
\end{equation}
The vectors $\tilde{\bm{v}}_\theta(t)$ and $\tilde{\bm{w}}^\dagger_\theta(t)$ are $T$-antiperiodic, i.e., 
$\tilde{\bm{v}}_\theta(t+T)=-\tilde{\bm{v}}_\theta(t)$ and $\tilde{\bm{w}}^\dagger_\theta(t+T)=-\tilde{\bm{w}}^\dagger_\theta(t)$.
Using these vectors and redefining the real part of $\lambda_1$ as $\lambda_{\mathrm{P1}} := (\theta - \theta_c) / T$, 
we can rewrite the critical mode in the form
\begin{equation}
    \Phi^{\mathrm{P1}}_{1, \theta}(t, s) =  e^{\lambda_{\mathrm{P1}}(t - s)} \tilde{\bm{v}}_\theta(t)\tilde{\bm{w}}^\dagger_\theta(s).
\end{equation}

We next consider the regime $\theta > \theta_c$. In this regime, the period-two limit cycle $\bm{x}_{\mathrm{P2}}(t, \theta)$ is stable. 
We take $\Phi^{\mathrm{P2}}_\theta(t, s)$ to denote the fundamental matrix evaluated along the period-two limit cycle $\bm{x}_{\mathrm{P2}}(t, \theta)$. 
We define the monodromy matrix 
\begin{equation}
    N_\theta = \Phi^{\mathrm{P2}}_\theta(2T, 0). 
\end{equation}
We define the left and right eigenvectors for the eigenvalues $\kappa_1, \kappa_2, \dots \kappa_n$
\begin{equation}
    \begin{aligned}
        N_\theta \bm{u}_{i, \theta} = \kappa_{i} \bm{u}_{i, \theta}, && \bm{s}^{\dagger}_{i, \theta} N_\theta = \kappa_i \bm{s}^{\dagger}_{i, \theta}, && \bm{s}_{i, \theta}^{\dagger} \bm{u}_{j, \theta} = \delta_{ij}. 
    \end{aligned}
\end{equation}
where we set $\kappa_n = 1, \bm{u}_{n, \theta} \propto \dot{\bm{x}}_{\mathrm{P2}}(0, \theta)$ 
and choose the eigenvectors to satisfy $\bm{u}_{i, \theta_c} = \bm{v}_{i, \theta_c}, \bm{s}^\dagger_{i, \theta_c} = \bm{w}^\dagger_{i, \theta_c}$.  
By an argument analogous to the decomposition of $\Phi^{\mathrm{P1}}(t,s)$ in the regime $\theta<\theta_c$, the critical mode in the regime $\theta>\theta_c$ is given by 
\begin{equation} 
    \Phi_{1, \theta}^{\mathrm{P2}}(t,s) = e^{\lambda_{\mathrm{P2}}(t-s)} \tilde{\bm{u}}_\theta(t) \tilde{\bm{s}}_\theta^\dagger(s). 
\end{equation}
Here, we define
\begin{equation}
    \begin{aligned}
        \tilde{\bm{u}}_\theta(t) := e^{- \lambda_{\mathrm{P2}} t} \Phi^{\mathrm{P2}}_\theta(t, 0) \bm{u}_{1, \theta}, && \tilde{\bm{s}}_\theta^\dagger(s) := e^{\lambda_{\mathrm{P2}} s} \bm{s}^\dagger_{1, \theta} \Phi^{\mathrm{P2}}_\theta(0, s), &&  \lambda_{\mathrm{P2}} := \frac{1}{2T} \mathrm{Log}\, \kappa_1 \approx - \frac{2}{T} (\theta - \theta_c). 
    \end{aligned}
\end{equation}
$\tilde{\bm{u}}_\theta(t)$ and $\tilde{\bm{s}}_\theta^\dagger(s)$ are $2T$-periodic and $\tilde{\bm{u}}_{\theta_c}(t)$ and $\tilde{\bm{s}}_{\theta_c}^\dagger(s)$ are $T$-antiperiodic.

\subsubsection{Expansion around the critical orbit}

We now derive asymptotic expansions of the stable limit-cycle solutions in each regime near the critical point $\theta_c$.
We first consider the regime $\theta<\theta_c$.
The period-one limit-cycle branch depends smoothly on the control parameter $\theta$ through the bifurcation point $\theta=\theta_c$.
Therefore, expanding the period-one limit-cycle solution around $\theta_c$, we obtain
\begin{equation}
    \bm{x}_{\mathrm{P1}}(t,\theta)
    =
    \bm{x}_c(t)
    +
    (\theta-\theta_c)
    \left.
    \pdv{\bm{x}_{\mathrm{P1}}(t,\theta)}{\theta}
    \right|_{\theta=\theta_c}
    +
    O(|\theta-\theta_c|^2).
    \label{asymptotic-period-one-solution}
\end{equation}
The period-one solution given by Eq.~\eqref{asymptotic-period-one-solution} is stable for $\theta<\theta_c$.
For $\theta>\theta_c$, the same branch of period-one solutions continues to exist but becomes unstable.

We next consider the regime $\theta>\theta_c$.
As discussed in Sec.~\ref{pd-review}, near the bifurcation point, the
period-two orbit emerges on the center manifold with
$\xi_\pm
=
\pm \sqrt{\theta-\theta_c}
+
O(\theta-\theta_c)$.
The center manifold is tangent to the critical direction associated with
the eigenvalue $\mu_1=-1$ at the bifurcation point.
To identify this critical direction along the continuous-time orbit, we consider the monodromy matrix with its base point shifted from $\bm{x}_{\mathrm{P1}}(0,\theta)$ to $\bm{x}_{\mathrm{P1}}(t,\theta)$:
\begin{equation}
    M_\theta(t)
    :=
    \Phi^{\mathrm{P1}}_\theta(t,0)
    M_\theta
    \Phi^{\mathrm{P1}}_\theta(0,t)
    =
    \Phi^{\mathrm{P1}}_\theta(t+T,t).
\end{equation}
Since $M_\theta(t)$ is similar to $M_\theta$, it has the same eigenvalues.
From the definition of the time-dependent vector $\bm{v}_{i,\theta}(t)$, we have
\begin{equation}
    M_\theta(t)\bm{v}_{i,\theta}(t)
    =
    \mu_i \bm{v}_{i,\theta}(t).
\end{equation}
Thus, $\bm{v}_{i,\theta}(t)$ represents the eigendirection associated with $\mu_i$ at the point $\bm{x}_{\mathrm{P1}}(t,\theta)$ along the orbit.
Since $\tilde{\bm{v}}_\theta(t)$ differs from $\bm{v}_{1,\theta}(t)$ only by a scalar phase factor, it represents the same eigendirection.
In particular, at $\theta=\theta_c$, $\tilde{\bm{v}}_{\theta_c}(t)$ represents the critical direction associated with $\mu_1=-1$ along the critical orbit.

It then follows from the center-manifold reduction that the displacement of the period-two limit cycle from the period-one limit cycle is, to leading order, along this critical direction:
\begin{equation}
    \bm{x}_{\mathrm{P2}}(t,\theta)
    -
    \bm{x}_{\mathrm{P1}}(t,\theta)
    =
    c\sqrt{\theta-\theta_c}\,
    \tilde{\bm{v}}_{\theta_c}(t)
    +
    O(\theta-\theta_c),
\end{equation}
where $c\neq0$ depends on the normalization of the center-manifold coordinate and the critical mode.
Hereafter, we choose the normalization of the critical mode such that $c=1$.
The relation
$\tilde{\bm{v}}_{\theta_c}(t+T)
=
-\tilde{\bm{v}}_{\theta_c}(t)$
is consistent with the period-doubled nature of the solution.
Since
$\bm{x}_{\mathrm{P1}}(t,\theta)-\bm{x}_c(t)
=O(\theta-\theta_c)$,
we finally obtain
\begin{equation}
    \bm{x}_{\mathrm{P2}}(t,\theta)
    =
    \bm{x}_c(t)
    + 
    \sqrt{\theta-\theta_c}\,
    \tilde{\bm{v}}_{\theta_c}(t)
    +
    O(\theta-\theta_c).
    \label{double-period-solution}
\end{equation}
This geometric argument can be made rigorous by applying the periodic center-manifold theorem to the period-doubling bifurcation of the limit cycle \cite{Kuznetsov2023Elements}.

\subsection{Critical behavior of entropy-production fluctuations} \label{ep-fluctuation}
Here, we derive the critical behavior of entropy-production fluctuation $D$ and 
determine its critical exponents $(\alpha_{+}, \alpha_{-})$. 
We have already derived the formula for the entropy-production fluctuations, Eq.~\eqref{fluctuation-formula-ep}, 
in Sec.~\ref{fluctuation-formula}.
Using the Floquet-mode decomposition of the fundamental matrix, we write
\begin{equation}
    \begin{aligned}
        \nu^{\mathrm{P1 / P2}}_{\rho}(\bm{x}_{\mathrm{P1 / P2}}(s)) & = \eta_{\rho} + \sum_{i} \nu^{\mathrm{P1 / P2}, i}_{\rho}(\bm{x}_{\mathrm{P1 / P2}}(s)) \\
        \nu^{\mathrm{P1 / P2}, i}_{\rho}(\bm{x}_{\mathrm{P1 / P2}}(s)) & = \sum_{\ell, \ell'} \int_{s}^{\tau} \d u \pdv{}!{x_\ell} j_{\mathrm{d}}(\bm{x}_{\mathrm{P1 / P2}}(u)) \ab[\Phi^{\mathrm{P1 / P2}}_{\theta, i}(u, s)]_{\ell, \ell'} \nabla_{\rho}^{\ell'}.     
    \end{aligned}
\end{equation}
where the superscripts $\mathrm{P1}$ and $\mathrm{P2}$ refer to the
limit cycles $\bm{x}_{\mathrm{P1}}(t,\theta)$ and
$\bm{x}_{\mathrm{P2}}(t,\theta)$, respectively.
The decomposition contains three types of modes: (i) the critical mode $i = 1$, (ii) the noncritical modes $i = 2, \dots, n-1$, and (iii) the orbit mode $i = n$. 
The entropy-production fluctuation $D$ contains
the individual mode contributions, the contribution from $\eta_\rho$,
and their cross terms.
If any mode contribution diverges near the critical point, the terms involving
the bounded contribution $\eta_\rho$ and the cross terms with noncritical
modes are subleading compared with the squared contribution of the diverging
mode.
Therefore, to determine the leading critical divergence, we focus on
\begin{equation}
    D_{i} := \lim_{\tau \to \infty} \frac{1}{\tau} \mathbb{E}_{\mu} \ab[\int_{0}^{\tau} \d s \sum_{\rho} \ab|\nu^{\mathrm{P1 / P2}, i}_{\rho}(\bm{x}_{\mathrm{P1 / P2}}(s))|^2 \mathcal{A}_{\rho}(\bm{x}_{\mathrm{P1 / P2}}(s))]. 
\end{equation}

First, we consider the orbit mode $i=n$.
The contribution of the orbit mode is given by
\begin{equation}
    \begin{aligned}
       D_{n} & = \lim_{\tau \to \infty} \frac{1}{\tau} \mathbb{E}_{\mu} \ab[\int_{0}^{\tau} \d s \sum_{\rho} \mathcal{A}_{\rho}(\bm{x}_{\mathrm{P1}}(s)) \ab|\sum_{\ell, \ell'} \int_{s}^{\tau} \d u \pdv{}!{x_\ell} j_{\mathrm{d}}(\bm{x}_{\mathrm{P1}}(u)) \ab[\Phi^{\mathrm{P1}}_{\theta, n}(u, s)]_{\ell, \ell'} \nabla_{\rho}^{\ell'}|^2] \\
    \end{aligned}
\end{equation}
for the period-one limit cycle $\bm{x}_{\mathrm{P1}}(t, \theta)$. 
Since the orbit mode satisfies $\Phi^{\mathrm{P1}}_{\theta, n}(u, s) = \bm{v}_{n, \theta}(u) \bm{w}^\dagger_{n, \theta}(s) \propto \dot{\bm{x}}_{\mathrm{P1}}(u, \theta)\bm{w}^\dagger_{n, \theta}(s)$, 
we obtain
\begin{equation}
    \begin{aligned}
        D_{n}
        & \propto \lim_{\tau \to \infty} \frac{1}{\tau} \mathbb{E}_{\mu} \ab[ \int_{0}^{\tau} \d s 
        \ab[j_{\mathrm{d}}(\bm{x}_{\mathrm{P1}}(\tau)) - j_{\mathrm{d}}(\bm{x}_{\mathrm{P1}}(s))]^2
        \sum_{\rho} \sum_{\ell, \ell'}  
        \ab[\bm{w}^\dagger_{n, \theta}(s) ]_{\ell} \nabla_{\rho}^{\ell} 
        \mathcal{A}_{\rho}(\bm{x}_{\mathrm{P1}}(s)) 
        \nabla_{\rho}^{\ell'} \ab[\bm{w}^\dagger_{n, \theta}(s) ]^*_{\ell'} ]. 
    \end{aligned}
\end{equation}
which remains finite near the critical point and therefore does not contribute to the critical divergence.
The same argument applies to the period-two limit cycle $\bm{x}_{\mathrm{P2}}(t,\theta)$.

Next, we consider the noncritical modes $i=2,\dots,n-1$.
The fundamental-matrix components associated with the noncritical modes take the form
\begin{equation}
    \begin{aligned}
        \ab[\Phi_{i, \theta}^{\mathrm{P1/ P2}}(t, s)]_{\ell, \ell'} = e^{\lambda(t - s)} R_{\ell, \ell'}(t, s)
    \end{aligned}
\end{equation}
where $R_{\ell,\ell'}(t,s)$ is bounded and $\lambda<0$ denotes the real part of the corresponding Floquet exponent.
Assuming that the concentrations along the limit cycles remain finite and
strictly positive near the bifurcation point,
$\partial_{x_\ell}j_{\mathrm{d}}(\bm{x}_{\mathrm{P1/P2}}(u))$ and
$\mathcal{A}_\rho(\bm{x}_{\mathrm{P1/P2}}(s))$ remain bounded.
Using these properties, we obtain an upper bound on the contribution of each noncritical mode:
\begin{equation}
    \begin{aligned}
        D_{i}
        & = \lim_{\tau \to \infty} \frac{1}{\tau} \mathbb{E}_{\mu} \ab[ \int_{0}^{\tau} \d s \sum_{\rho} \mathcal{A}_{\rho}(\bm{x}_{\mathrm{P1 / P2}}(s)) 
        \ab|\sum_{\ell, \ell'} \int_{s}^{\tau} \d u \pdv{}!{x_\ell} j_{\mathrm{d}}(\bm{x}_{\mathrm{P1 / P2}}(u)) 
        e^{\lambda(u - s)} R_{\ell, \ell'}(u, s) \nabla_{\rho}^{\ell'}|^2 ] \\
        & \leq \lim_{\tau \to \infty}  \frac{M}{\tau} \int_{0}^{\tau} \d s 
        \ab( \int_{s}^{\tau} \d u e^{\lambda(u - s)} )^2 \\
        & \leq \frac{M}{\lambda^2}, 
    \end{aligned}
\end{equation}
where $M$ is a finite constant independent of $\theta$ sufficiently close
to $\theta_c$.
Since the noncritical exponents remain bounded away from zero at
$\theta=\theta_c$, this upper bound remains finite near the bifurcation
point. Therefore, the noncritical modes do not contribute to the critical
divergence of $D$.

Finally, we turn to the critical mode, which is the only mode that can
produce the critical divergence.
\begin{equation}
    D_{1} = \lim_{\tau \to \infty} \frac{1}{\tau} \mathbb{E}_{\mu} \ab[\int_{0}^{\tau} \d s \sum_{\rho} \ab|\nu^{\mathrm{P1 / P2}, 1}_{\rho}(\bm{x}_\mathrm{P1 / P2}(s))|^2 \mathcal{A}_{\rho}(\bm{x}_\mathrm{P1 / P2}(s))]. 
\end{equation}
We now evaluate this contribution separately on the two sides of the bifurcation point.

\subsubsection{Period-one regime }
We first investigate the period-one regime $\theta<\theta_c$.
In this regime, the period-one limit cycle $\bm{x}_{\mathrm{P1}}(t,\theta)$ is stable, and
the contribution of the critical mode to the entropy-production fluctuations is given by
\begin{equation}
    \begin{aligned}
        D_{1}
        & =\lim_{\tau \to \infty} \frac{1}{\tau} 
        \mathbb{E}_{\mu} \ab[
            \int_{0}^{\tau} \d s \sum_{\rho} \ab|\nu^{\mathrm{P1}, 1}_{\rho}(\bm{x}_\mathrm{P1}(s))|^2 \mathcal{A}_{\rho}(\bm{x}_\mathrm{P1}(s))
        ], \\
        \nu^{\mathrm{P1}, 1}_{\rho}(\bm{x}_{\mathrm{P1}}(s)) 
        & = \sum_{\ell, \ell'} \int_{s}^{\tau} \d u \pdv{}!{x_\ell} j_{\mathrm{d}}(\bm{x}_{\mathrm{P1}}(u))
        e^{\lambda_{\mathrm{P1}}(u - s)} 
        \ab[\tilde{\bm{v}}_\theta(u)]_{\ell}
        \ab[\tilde{\bm{w}}^\dagger_\theta(s)]_{\ell'} 
        \nabla_{\rho}^{\ell'}.   
    \end{aligned}
    \label{ep-fluctuation-critical-mode}
\end{equation}
Since
$\sum_{\ell} \pdv{}!{x_\ell} j_{\mathrm{d}}(\bm{x}_{\mathrm{P1}}(u))
[\tilde{\bm{v}}_\theta(u)]_{\ell}$
is $T$-antiperiodic, it admits the Fourier expansion
\begin{equation}
    \sum_{\ell} \pdv{}!{x_\ell} j_{\mathrm{d}}(\bm{x}_{\mathrm{P1}}(u)) \ab[\tilde{\bm{v}}_\theta(u)]_{\ell} 
    = \sum_{m = -\infty}^{\infty} \mathcal{F}_{2m+1} e^{i \omega (m + \frac{1}{2}) u },
    \quad \omega := \frac{2 \pi}{T}, 
    \label{Fourier-expansion-F}
\end{equation}
where $\mathcal{F}_{2m + 1}$ denotes the Fourier coefficient given by 
\begin{equation}
    \mathcal{F}_{2m + 1} := \frac{1}{2T} \int_{0}^{2T} \d u  \sum_{\ell} \pdv{}!{x_\ell} 
    j_{\mathrm{d}}(\bm{x}_{\mathrm{P1}}(u)) \ab[\tilde{\bm{v}}_\theta(u)]_{\ell}  e^{- i\omega (m + \frac{1}{2}) u }, 
\end{equation}
while the even Fourier coefficients satisfy $\mathcal{F}_{2m}=0$ because of the $T$-antiperiodicity.
Substituting the Fourier expansion \eqref{Fourier-expansion-F} into $\nu^{\mathrm{P1},1}_{\rho}(\bm{x}_{\mathrm{P1}}(s))$, 
we obtain
\begin{equation}
    \nu^{\mathrm{P1}, 1}_{\rho}(\bm{x}_{\mathrm{P1}}(s))
        = 
        \sum_{m = -\infty}^{\infty} 
        \frac{\mathcal{F}_{2m+1}}{\lambda_{\mathrm{P1}} + i \omega (m + \frac{1}{2})}
        \ab[e^{i \omega (m + \frac{1}{2})\tau}
        e^{\lambda_{\mathrm{P1}} (\tau - s)} 
        - e^{i \omega (m + \frac{1}{2})s}]
        \sum_{\ell'} 
        \ab[\tilde{\bm{w}}^\dagger_\theta(s)]_{\ell'} 
        \nabla_{\rho}^{\ell'}.
        \label{integrated-v}
\end{equation}
We next consider the factor
$
    \sum_{\rho} \sum_{\ell, \ell'} 
    \ab[\tilde{\bm{w}}^\dagger_\theta(s)]_{\ell} 
    \nabla_{\rho}^{\ell}
    \mathcal{A}_{\rho}(\bm{x}_{\mathrm{P1}}(s))
    \nabla_{\rho}^{\ell'}
    \ab[\tilde{\bm{w}}^\dagger_\theta(s)]^*_{\ell'} 
$
appearing in Eq.~\eqref{ep-fluctuation-critical-mode}.
This factor is $T$-periodic, and we introduce its Fourier expansion
\begin{equation}
    \sum_{\rho} \sum_{\ell, \ell'} 
    \ab[\tilde{\bm{w}}^\dagger_\theta(s)]_{\ell} 
    \nabla_{\rho}^{\ell}
    \mathcal{A}_{\rho}(\bm{x}_{\mathrm{P1}}(s))
    \nabla_{\rho}^{\ell'}
    \ab[\tilde{\bm{w}}^\dagger_\theta(s)]^*_{\ell'} 
    = \sum_{n = -\infty}^{\infty} \mathcal{D}_{n} e^{i n \omega s}
    \label{Fourier-expansion-D}
\end{equation}
where the Fourier coefficients $\mathcal{D}_n$ are given by
\begin{equation}
    \mathcal{D}_{n} := \frac{1}{T} \int_{0}^{T} \d s \sum_{\rho} \sum_{\ell, \ell'} 
    \ab[\tilde{\bm{w}}^\dagger_\theta(s)]_{\ell} 
    \nabla_{\rho}^{\ell}
    \mathcal{A}_{\rho}(\bm{x}_{\mathrm{P1}}(s))
    \nabla_{\rho}^{\ell'}
    \ab[\tilde{\bm{w}}^\dagger_\theta(s)]^*_{\ell'} e^{- i n \omega s}. 
\end{equation}

Substituting Eqs.~\eqref{integrated-v} and \eqref{Fourier-expansion-D} into \eqref{ep-fluctuation-critical-mode}, 
we obtain
\begin{equation}
    \begin{aligned}
        D_{1}
        = &  
        \sum_{m, m' = -\infty}^{\infty} 
        \sum_{n = -\infty}^{\infty}
        \frac{
            \mathbb{E}_{\mu}[\mathcal{F}_{2m+1} \mathcal{D}_{n} \mathcal{F}_{2m'+1}^*]
        }{
            \ab(\lambda_{\mathrm{P1}} + i \omega (m + \frac{1}{2}))
            \ab(\lambda_{\mathrm{P1}} - i \omega (m' + \frac{1}{2}))
        } \\
        & \times
        \lim_{\tau \to \infty}
        \frac{1}{\tau} \int_{0}^{\tau} \d s 
        \ab[e^{i \omega (m + \frac{1}{2})\tau}
        e^{\lambda_{\mathrm{P1}} (\tau - s)} 
        - e^{i \omega (m + \frac{1}{2})s}] 
        e^{i n \omega s}
        \ab[e^{- i \omega (m' + \frac{1}{2})\tau}
        e^{\lambda_{\mathrm{P1}} (\tau - s)} 
        - e^{- i \omega (m' + \frac{1}{2})s}]
    \end{aligned}
\end{equation}
In the long-time limit $\tau\to\infty$, the integral evaluates to
\begin{equation}
    \lim_{\tau \to \infty} \frac{1}{\tau} \int_{0}^{\tau} \d s 
        \ab[e^{i \omega (m + \frac{1}{2})\tau}
        e^{\lambda_{\mathrm{P1}} (\tau - s)} 
        - e^{i \omega (m + \frac{1}{2})s}] 
        e^{i n \omega s}
        \ab[e^{- i \omega (m' + \frac{1}{2})\tau}
        e^{\lambda_{\mathrm{P1}} (\tau - s)} 
        - e^{- i \omega (m' + \frac{1}{2})s}]
    = \delta_{m - m' + n, 0}. 
\end{equation}
Therefore, the critical-mode contribution to the entropy-production fluctuations is given by
\begin{equation}
    D_{1}
    = 
    \sum_{m, m' = -\infty}^{\infty} 
    \frac{
        \mathbb{E}_{\mu}[\mathcal{F}_{2m+1} \mathcal{D}_{m' - m } \mathcal{F}_{2m'+1}^*]
    }{
        \ab(\lambda_{\mathrm{P1}} + i \omega (m + \frac{1}{2}))
        \ab(\lambda_{\mathrm{P1}} - i \omega (m' + \frac{1}{2}))
    }.
    \label{ep-fluctuation-period-one}
\end{equation}
Since $m$ and $m'$ are integers,
$m+\frac{1}{2}$ and $m'+\frac{1}{2}$ never vanish.
Therefore, the denominators in Eq.~\eqref{ep-fluctuation-period-one}
remain finite even in the limit $\lambda_{\mathrm{P1}}\to0$.
Hence, the critical-mode contribution does not diverge as
$\theta\to\theta_c^{-}$. 
Therefore, the entropy-production fluctuations do not exhibit a critical divergence in the period-one regime.
Thus, the critical exponent $\alpha_{-}$ defined in Eq.~\eqref{ep-fluctuation-critical-exponent} is
\begin{equation}
    \alpha_{-} = 0_{-}. 
\end{equation}

\subsubsection{Period-two regime}
We next investigate the period-two regime $\theta>\theta_c$.
In this regime, the stable limit cycle is the period-two limit cycle
$\bm{x}_{\mathrm{P2}}(t,\theta)$.
The contribution of the critical mode to the entropy-production fluctuations is given by
\begin{equation}
    \begin{aligned}
        D_{1} 
        & = \lim_{\tau \to \infty}\frac{1}{\tau} \mathbb{E}_{\mu}\ab[\int_{0}^{\tau} \d s \sum_{\rho} 
        \ab|\nu^{\mathrm{P2}, 1}_{\rho}(\bm{x}_\mathrm{P2}(s))|^2 \mathcal{A}_{\rho}(\bm{x}_\mathrm{P2}(s))], \\
        \nu^{\mathrm{P2}, 1}_{\rho}(\bm{x}_\mathrm{P2}(s))
        & = \sum_{\ell, \ell'} \int_{s}^{\tau} \d u 
        \pdv{}!{x_\ell} j_{\mathrm{d}}(\bm{x}_{\mathrm{P2}}(u))
        e^{\lambda_{\mathrm{P2}}(u - s)} 
        \ab[\tilde{\bm{u}}_\theta(u)]_{\ell}
        \ab[\tilde{\bm{s}}^\dagger_\theta(s)]_{\ell'} 
        \nabla_{\rho}^{\ell'}.   
    \end{aligned}
\end{equation}
As in the period-one regime, we introduce the Fourier expansions
\begin{equation}
    \begin{aligned}
        & \sum_{\ell} \pdv{}!{x_\ell} j_{\mathrm{d}}(\bm{x}_{\mathrm{P2}}(u)) \ab[\tilde{\bm{u}}_\theta(u)]_{\ell} = \sum_{k = -\infty}^{\infty} \mathcal{G}_{k} e^{i k \omega u / 2}, \\
        & \mathcal{G}_{k} := \frac{1}{2T} \int_{0}^{2T} \d u \sum_{\ell} \pdv{}!{x_\ell} j_{\mathrm{d}}(\bm{x}_{\mathrm{P2}}(u)) \ab[\tilde{\bm{u}}_\theta(u)]_{\ell} e^{- i k \omega u / 2}. 
    \end{aligned}
\end{equation}
and 
\begin{equation}
    \begin{aligned}
        & \sum_{\rho} \sum_{\ell, \ell'} \ab[\tilde{\bm{s}}^\dagger_\theta(s)]_{\ell} \nabla_{\rho}^{\ell} \mathcal{A}_{\rho}(\bm{x}_{\mathrm{P2}}(s)) \nabla_{\rho}^{\ell'} \ab[\tilde{\bm{s}}^\dagger_\theta(s)]^*_{\ell'} = \sum_{n = -\infty}^{\infty} \mathcal{H}_{n} e^{i n \omega s / 2 }, 
        \\
        & \mathcal{H}_{n} := \frac{1}{2T} \int_{0}^{2T} \d s \sum_{\rho} \sum_{\ell, \ell'} \ab[\tilde{\bm{s}}^\dagger_\theta(s)]_{\ell} \nabla_{\rho}^{\ell} \mathcal{A}_{\rho}(\bm{x}_{\mathrm{P2}}(s)) \nabla_{\rho}^{\ell'} \ab[\tilde{\bm{s}}^\dagger_\theta(s)]^*_{\ell'} e^{- i n \omega s / 2}. 
    \end{aligned}
\end{equation}
Using these Fourier expansions, we obtain
\begin{equation}
    \begin{aligned}
        D_{1}
        & = 
        \sum_{k, k' = -\infty}^{\infty}
        \frac{
            \mathbb{E}_{\mu}\ab[\mathcal{G}_{k} \mathcal{H}_{k' - k} \mathcal{G}_{k'}^*]
        }{(\lambda_{\mathrm{P2}} + i \omega k / 2 ) (\lambda_{\mathrm{P2}} - i \omega k' / 2 )}
    \end{aligned}
\end{equation}
The term with $k=k'=0$ can give rise to a divergence. We denote this potentially dominant contribution by
\begin{equation}
    \begin{aligned}
        D_{\mathrm{dom}} 
        & := 
        \frac{\mathbb{E}_{\mu}\ab[\mathcal{G}_{0} \mathcal{H}_{0} \mathcal{G}_{0}^*]}{\lambda_{\mathrm{P2}}^2}. 
    \end{aligned}
    \label{dominant-term}
\end{equation}

We now expand the numerator
$\mathcal{G}_{0}\mathcal{H}_{0}\mathcal{G}_{0}^*$
of Eq.~\eqref{dominant-term} in powers of $\sqrt{\theta-\theta_c}$.
Using the asymptotic expansion of the period-two solution obtained in
Eq.~\eqref{double-period-solution}, we expand the fundamental matrix as
\begin{equation}
    \Phi(t, s) = \Phi^{(0)}(t, s) + \sqrt{\theta - \theta_c} \Phi^{(1)}(t, s) + O(\theta - \theta_c)
    \label{fundamental-matrix-expansion}
\end{equation}
where the terms are given by 
\begin{align}
    \Phi^{(0)}(t, s) & := \mathcal{T} \exp \ab(\int_{s}^{t} \d u \mathbb{S}(\bm{x}_{c}(u) )), \\
    \Phi^{(1)}(t, s) & := \int_{s}^{t} \d u  \Phi^{(0)}(t, u) H(\bm{x}_{c}(u), \tilde{\bm{v}}_{\theta_c}(u)) \Phi^{(0)}(u, s). 
\end{align}
Here, we define the matrix-valued function
\begin{equation}
    \ab[H(\bm{x}_{c}(u), \tilde{\bm{v}}_{\theta_c}(u))]_{ij} =\sum_{k} \left.\pdv{\ab[ \mathbb{S}_{\theta_c}(\bm{x})]_{ij}}{x_{k}} \right|_{\bm{x} = \bm{x}_{c}(u)} \ab[\tilde{\bm{v}}_{\theta_c}(u)]_{k}. 
\end{equation}
Using the expansion of the fundamental matrix in Eq.~\eqref{fundamental-matrix-expansion}, we obtain
\begin{equation}
    \begin{aligned}
        \tilde{\bm{u}}_\theta(t)
        &=
        \tilde{\bm{u}}_{\theta_c}(t)
        +
        \sqrt{\theta-\theta_c}\,
        \tilde{\bm{u}}^{(1)}_{\theta_c}(t)
        +
        O(\theta-\theta_c),
        \\
        \tilde{\bm{s}}^\dagger_\theta(t)
        &=
        \tilde{\bm{s}}^\dagger_{\theta_c}(t)
        +
        \sqrt{\theta-\theta_c}\,
        \tilde{\bm{s}}^{(1),\dagger}_{\theta_c}(t)
        +
        O(\theta-\theta_c).
    \end{aligned}
\end{equation}
We write the corresponding expansions of $\mathcal{G}_{0}$ and $\mathcal{H}_{0}$ as
\begin{equation}
    \begin{aligned}
        \mathcal{G}_{0} & = \mathcal{G}^{(0)}_{0} + \sqrt{\theta - \theta_c} \mathcal{G}^{(1)}_{0} + O(\theta - \theta_c),  \\   
        \mathcal{H}_{0} & = \mathcal{H}^{(0)}_{0} + \sqrt{\theta - \theta_c} \mathcal{H}^{(1)}_{0} + O(\theta - \theta_c).  \\    
    \end{aligned}
\end{equation}
The lowest-order term of $\mathcal{G}_{0}$ vanishes because of the
$T$-periodicity of $\bm{x}_{c}(t)$ and the $T$-antiperiodicity of
$\tilde{\bm{u}}_{\theta_c}(t)$:
\begin{equation}
    \begin{aligned}
        \mathcal{G}^{(0)}_{0} 
        & = \frac{1}{2T} \int_{0}^{2T} \d u \sum_{\ell} \pdv{}!{x_\ell} j_{\mathrm{d}}(\bm{x}_{c}(u)) \ab[\tilde{\bm{u}}_{\theta_c}(u)]_{\ell} \\
        & = \frac{1}{2T} 
        \int_{0}^{T} \d u \sum_{\ell} \pdv{}!{x_\ell} j_{\mathrm{d}}(\bm{x}_{c}(u)) \ab[\tilde{\bm{u}}_{\theta_c}(u)]_{\ell}      
        + \frac{1}{2T}  
        \int_{0}^{T} \d u' \sum_{\ell} \pdv{}!{x_\ell} j_{\mathrm{d}}(\bm{x}_{c}(u' + T)) \ab[\tilde{\bm{u}}_{\theta_c}(u' + T)]_{\ell} \\
        & = 0
    \end{aligned}
\end{equation}
In general, this cancellation due to antiperiodicity does not persist at
higher orders. We therefore consider the generic case
\begin{equation}
    \mathcal{G}^{(1)}_{0}\neq0.
\end{equation}
On the other hand, the lowest-order term of $\mathcal{H}_{0}$ is nonzero:
\begin{equation}
    \mathcal{H}^{(0)}_{0} 
    = \frac{1}{2T} \int_{0}^{2T} \d s \sum_{\rho} \sum_{\ell, \ell'} \ab[\tilde{\bm{s}}^\dagger_{\theta_c}(s)]_{\ell} \nabla_{\rho}^{\ell} \mathcal{A}_{\rho}(\bm{x}_{c}(s)) \nabla_{\rho}^{\ell'} \ab[\tilde{\bm{s}}^\dagger_{\theta_c}(s)]^*_{\ell'} 
    \neq 0 
\end{equation}
Therefore, the numerator of the dominant contribution scales as
\begin{equation}
    \mathcal{G}_{0}\mathcal{H}_{0}\mathcal{G}_{0}^*
    =
    O(\theta-\theta_c).
\end{equation}
Since $\lambda_{\mathrm{P2}}$ scales as
$\lambda_{\mathrm{P2}}\approx-2(\theta-\theta_c)/T$,
the dominant contribution to the entropy-production fluctuations diverges as
\begin{equation}
    D_{\mathrm{dom}}
    \propto
    \frac{1}{|\theta-\theta_c|}.
\end{equation}
Therefore, the entropy-production fluctuations diverge as
$|\theta-\theta_c|^{-1}$ in the period-two regime, and the corresponding
critical exponent is
\begin{equation}
    \alpha_{+}=1.
\end{equation}

\subsection{Critical behavior of parametric response of entropy production rate} \label{ep-response}
We derive the critical behavior of the parametric response of the entropy-production rate,
\begin{equation}
    \begin{aligned}
        \pdv{}!{\theta} \sigma
        & := \pdv{}{\theta} \lim_{\tau \to \infty} \frac{1}{\tau} \lim_{\Omega \to \infty} \frac{1}{\Omega} \ab<\Sigma(\Gamma)> \\
        & = \pdv{}{\theta} \lim_{\tau \to \infty} \frac{1}{\tau} \int_{0}^{\tau} \d t \sum_{\rho} \eta_{\rho} \ab(J_{+\rho}(\bm{x}(t)) - J_{-\rho}(\bm{x}(t)))    
    \end{aligned}    
\end{equation}
and determine the corresponding critical exponents $(\beta_{-},\beta_{+})$.
In both regimes, the corresponding limit-cycle solutions are $2T$-periodic, and hence we obtain
\begin{equation}
    \pdv{}!{\theta} \sigma  = \pdv{}{\theta} \frac{1}{2T} \d t \int_{0}^{2T}\sum_{\rho} \eta_{\rho} \ab(J_{+\rho}(\bm{x}_{\mathrm{P1 / P2}}(t, \theta)) - J_{-\rho}(\bm{x}_{\mathrm{P1 / P2}}(t, \theta))). 
\end{equation}

In the period-one regime, the limit cycle $\bm{x}_{\mathrm{P1}}(t, \theta)$ is given by Eq.~\eqref{asymptotic-period-one-solution} near the critical point. 
Under mass-action kinetics, the reaction currents $J_{+\rho}(\bm{x})$ and $J_{-\rho}(\bm{x})$ are monomials in the concentrations $\bm{x}$.
Since the period-one limit cycle depends smoothly on $\theta$ and the
mass-action currents are polynomial functions of the concentrations,
the $2T$-averaged entropy-production rate admits a regular expansion in
integer powers of $\theta-\theta_c$.
Therefore, $\partial_\theta\sigma$ does not diverge as $\theta\to\theta_c$, and the corresponding critical exponent is
\begin{equation}
    \beta_{-} = 0_{-}. 
\end{equation}

In the period-two regime, the limit cycle $\bm{x}_{\mathrm{P2}}(t, \theta)$ 
is given by Eq.~\eqref{double-period-solution}. 
Its component-wise representation is given by 
\begin{equation}
    \ab[\bm{x}_{\mathrm{P2}}(t, \theta)]_{\ell} = \ab[\bm{x}_c(t)]_{\ell} + \sqrt{\theta - \theta_c} \ab[\tilde{\bm{v}}_{\theta_c}(t)]_{\ell} + O(\theta - \theta_c). 
\end{equation}
Let $\phi(t)$ be an arbitrary $T$-periodic function satisfying $\phi(t+T)=\phi(t)$.
We consider the general $2T$-integral of the product of $\phi(t)$ and the components $\ab[\tilde{\bm{v}}_{\theta_c}(t)]_{\ell}$ given by
\begin{equation}
    \int_{0}^{2T} \d t \phi(t) \prod_{\ell \in K} \ab[\tilde{\bm{v}}_{\theta_c}(t)]_{\ell}
\end{equation}
where $K$ is a finite sequence of indices in $\mathcal{S}$, with repetitions allowed.
Since each component $\ab[\tilde{\bm{v}}_{\theta_c}(t)]_{\ell}$ is $T$-antiperiodic, the integral reduces to 
\begin{equation}
    \int_{0}^{2T} \d t \phi(t) \prod_{\ell \in K} \ab[\tilde{\bm{v}}_{\theta_c}(t)]_{\ell}
    = \begin{cases}
        2\int_{0}^{T} \d t \phi(t) \prod_{\ell \in K} \ab[\tilde{\bm{v}}_{\theta_c}(t)]_{\ell}, & |K|: \text{ even} \\
        0, & |K|: \text{ odd}
    \end{cases}
\end{equation}
where $|K|$ denotes the length of the sequence. 
Since each factor of $\sqrt{\theta-\theta_c}$ is accompanied by one factor of $\tilde{\bm{v}}_{\theta_c}(t)$,
all terms containing odd powers of $\sqrt{\theta-\theta_c}$ vanish upon averaging over $2T$.
Therefore, the $2T$-averaged entropy-production rate contains only integer
powers of $\theta-\theta_c$ near the bifurcation point.
Consequently, $\partial_\theta\sigma$ remains finite and the corresponding critical exponent is
\begin{equation}
    \beta_{+} = 0_{-}. 
\end{equation}

\section{Numerical models and observation protocols}
In this section, we describe the numerical procedures used to evaluate entropy-production fluctuations and the reaction models employed in our simulations.

To evaluate the entropy production, we numerically integrate the chemical Langevin equations (CLEs) for the integrated reaction currents,
\begin{equation}
    \begin{aligned}
        \dot{Q}^{\mathrm{CLE}}_{\rho} &= J_{\rho}(\bm{x}^{\mathrm{CLE}}(t)) + \Omega^{-1/2} \sqrt{\mathcal{A}_{\rho}(\bm{x}^{\mathrm{CLE}}(t))} \cdot \xi_{\rho}(t), \\
        x_{\ell}^{\mathrm{CLE}}(t) &= x_{\ell}^{\mathrm{CLE}}(0) + \sum_{\rho} \nabla_{\rho}^{\ell} Q_{\rho}^{\mathrm{CLE}}(t).
    \end{aligned}
    \label{current-CLE}
\end{equation}
Here, $Q_{\rho}^{\mathrm{CLE}}(t)$ denotes the integrated current associated with reaction channel $\rho$, and $x_{\ell}^{\mathrm{CLE}}(t)$ denotes the concentration of species $\ell$. The noises $\xi_{\rho}(t)$ are independent Gaussian white noises with unit intensity,
and the symbol $\cdot$ denotes the It\^o product.

For a given initial concentration $\bm{x}_0$, we numerically
integrate Eq.~\eqref{current-CLE} using the Euler--Maruyama method~\cite{higham2001algorithmic, rackauckas2017differentialequations},
with a fixed time step $\Delta t$ over an observation time $\tau$,
starting from $\bm{x}^{\mathrm{CLE}}(0)=\bm{x}_0$
and $Q_{\rho}^{\mathrm{CLE}}(0)=0$.
The trajectory-wise entropy production is evaluated as
\begin{equation}
    \Sigma^{\mathrm{CLE}}(\tau)
    =
    \Omega
    \sum_{\rho}
    \eta_{\rho}
    Q_{\rho}^{\mathrm{CLE}}(\tau).
\end{equation}
For each fixed initial concentration $\bm{x}_0$, we generate $N_{\mathrm e}$ independent noise realizations and obtain an ensemble of entropy-production values $\{\Sigma_i^{\mathrm{CLE}}(\tau)\}_{i=1}^{N_{\mathrm e}}$. The variance over this ensemble provides an estimate of the conditional variance $V_{\bm{x}_0}(\Omega,\tau)$.

We employ this procedure for the three sets of simulations corresponding to Figs.~2, 3(a), and 3(b). 
For Figs.~2 and 3(b), the initial concentration $\bm{x}_0$ at each parameter value is obtained by integrating the deterministic rate equations for $1000$ time units from a prescribed initial condition. 
All CLE trajectories are then initialized from this same state $\bm{x}_0$ for each parameter value to evaluate the conditional variance.
For Fig.~3(a), we additionally average the conditional variance over initial concentrations distributed according to the SRB measure $\mu_{\mathrm{SRB}}$. 
To approximate this measure, we integrate a single deterministic trajectory for $\tau_{\bm{x}_0} + 500$ time units, discard the first $500$ time units as a transient, 
and sample $N_{\bm{x}_0}$ initial concentrations at equally spaced time intervals over the remaining $\tau_{\bm{x}_0}$ time units. For each sampled initial concentration, we estimate the conditional variance using $N_{\mathrm e}$ independent CLE trajectories. We then average these conditional variances over the $N_{\bm{x}_0}$ sampled initial concentrations. 

Below, we specify the Willamowski--R\"ossler model and the simulation parameters, followed by additional results for the Gaspard model.
\subsection{Willamowski-R\"ossler model}

The Willamowski--R\"ossler model \cite{WillamowskiRoessler1980} exhibits chaotic behavior through a period-doubling cascade. Its reaction scheme is given by 
\begin{equation}
	\begin{aligned}
		A_{1} + X & \xrightleftharpoons[k_{-1}]{k_{+1}} 2 X, & 
		X + Y &\xrightleftharpoons[k_{-2}]{k_{+2}} 2Y, & 
		A_5 + Y &\xrightleftharpoons[k_{-3}]{k_{+3}} A_{2}, \\
		X + Z &\xrightleftharpoons[k_{-4}]{k_{+4}} A_{3}, & 
		A_{4} + Z &\xrightleftharpoons[k_{-5}]{k_{+5}} 2Z
	\end{aligned}
    \label{Willamowski-Rossler model}
\end{equation}
where the concentrations of $X$, $Y$, and $Z$ vary in time, whereas $A_i$, $i=1,\dots,5$, are chemostatted species whose concentrations are set to unity: $a_{i} = 1$ for $i = 1, 2, \dots 5$.
The model parameters used in the simulations are given by $k_{+1} = 30.0, k_{-1} = 0.25,  k_{-2} = 10^{-4}, k_{+3} = 10.0, k_{-3} = 10^{-3}, k_{+4} = 1.0, k_{-4} = 10^{-3}, k_{+5} = 16.5, k_{-5} = 0.5$ 
and we use $k_{+2}$ as the control parameter, $\theta=k_{+2}$.
Successive local maxima of $z(t)$ resolve the period-doubling branches in Fig.~2.
In the period-$2^n$ regime, $2^n$ distinct local maxima appear at a fixed value of $\theta$, corresponding to a period-$2^n$ limit cycle. 
Successive splitting of these branches illustrates the period-doubling cascade toward chaos.

Figs.~2(b), 2(c), and 3(b) in the main text were obtained
with simulation parameters
$N_{\mathrm e}=512$, $\Delta t=0.002$, and $\tau=400000$,
whereas Fig.~3(a) was obtained with
$N_{\mathrm e}=100$, $\Delta t=0.002$,
$N_{\bm{x}_0}=100$, $0 \leq \tau \leq 30$,
and $\tau_{\bm{x}_0}=500$. For Fig.~2(c), we perform a finite-size scaling analysis
assuming the scaling form
$V_{\bm{n}_0}(\theta)/(\Omega\tau)
= \Omega^y f(\Omega^x(\theta-\theta_c))$.
The scaling plot is constructed using
$\theta_c=0.92445$, $x=0.62$, and $y=0.59$.

\begin{figure}[ht]
    \centering
    \includegraphics[width=0.4\linewidth]{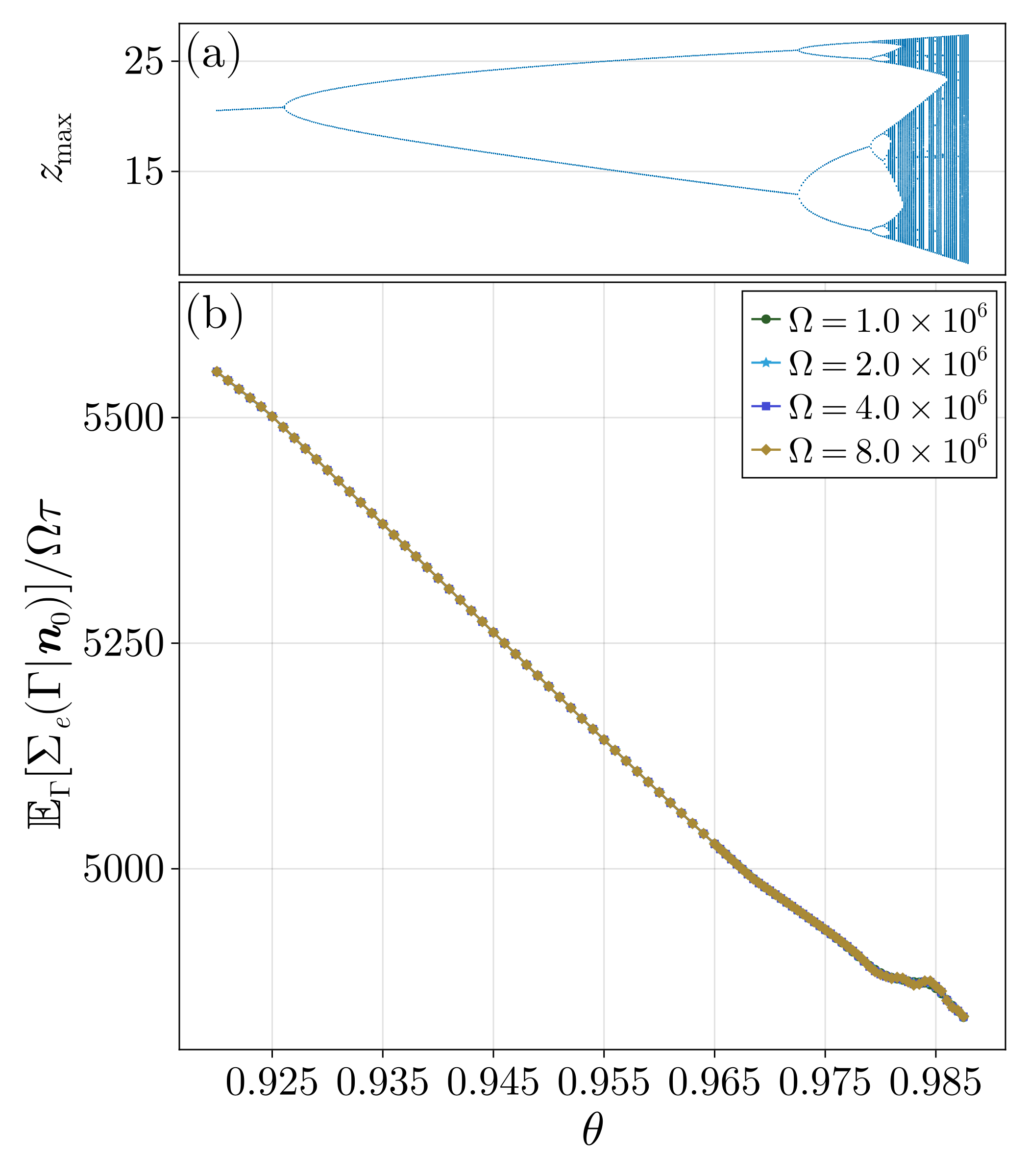}
    \caption{
        Mean entropy-production rate
        as a function of the control parameter
        $\theta=k_{+2}$ for the Willamowski-R\"ossler model.
    }
    \label{mean entropy production WR}
\end{figure}

We also plot the mean entropy-production rate
as a function of the control parameter $\theta=k_{+2}$
in Fig.~\ref{mean entropy production WR},
using the same simulation parameters as those
for the variance calculations in Fig 2 of the main text.

\subsection{Gaspard model}
The Gaspard model was introduced in Ref.~\cite{Gaspard10.1063/5.0025350ChaosEntropyProduction} and its reaction scheme is given by
\begin{equation}
    \begin{aligned}
        A_1 + X &\xrightleftharpoons[k_{-1}]{k_{+1}} 2X, & 
        A_2 + 2X &\xrightleftharpoons[k_{-2}]{k_{+2}} 3X, & 
        X + Y &\xrightleftharpoons[k_{-3}]{k_{+3}} A_3 + Y, \\
        X + Z &\xrightleftharpoons[k_{-4}]{k_{+4}} A_4 + Z, & 
        A_5 + X + Y &\xrightleftharpoons[k_{-5}]{k_{+5}} X + 2Y, & 
        A_6 + Y + Z &\xrightleftharpoons[k_{-6}]{k_{+6}} 2Y + Z, \\
        Y &\xrightleftharpoons[k_{-7}]{k_{+7}} A_7, & 
        A_8 + X &\xrightleftharpoons[k_{-8}]{k_{+8}} X + Z, & 
        A_9 + 2Z &\xrightleftharpoons[k_{-9}]{k_{+9}} 3Z, \\
        Z &\xrightleftharpoons[k_{-10}]{k_{+10}} A_{10}, 
    \end{aligned}
    \label{Gaspard model}
\end{equation}
where $A_i$, $i=1,\dots,10$, are chemostatted species whose concentrations are taken to be unity: $ a_i = 1$ for $ i = 1,\ldots,10$.
We use the following model parameters: 
$
    k_{+1} = 0.6,
    k_{+3} = 0.5,
    k_{+4} = 1,
    k_{+5} = 1,
    k_{+6} = 0.3, 
    k_{+7} = 1.3,
    k_{+8} = 100,
    k_{+9} = 300,
    k_{+10} = 500, 
    k_{-9} = 50, 
    k_{-\rho} = 0.01,
    $ for $
    \rho = 1,\ldots,8,10,
$
and we use $k_{+2}$ as the control parameter, $\theta = k_{+2}$.   
The Gaspard model exhibits the following dynamical regimes \cite{Gaspard10.1063/5.0025350ChaosEntropyProduction}:
\begin{equation}
    \begin{aligned}
        \theta &\lesssim 0.363
        && \text{fixed-point regime}, \\
        0.363 \lesssim \theta &\lesssim 0.494
        && \text{limit-cycle regime}, \\
        0.494 \lesssim \theta
        &&& \text{chaotic regime}.
    \end{aligned}
\end{equation}

\begin{figure}[ht]
    \centering
    \includegraphics[width = 0.45\linewidth]{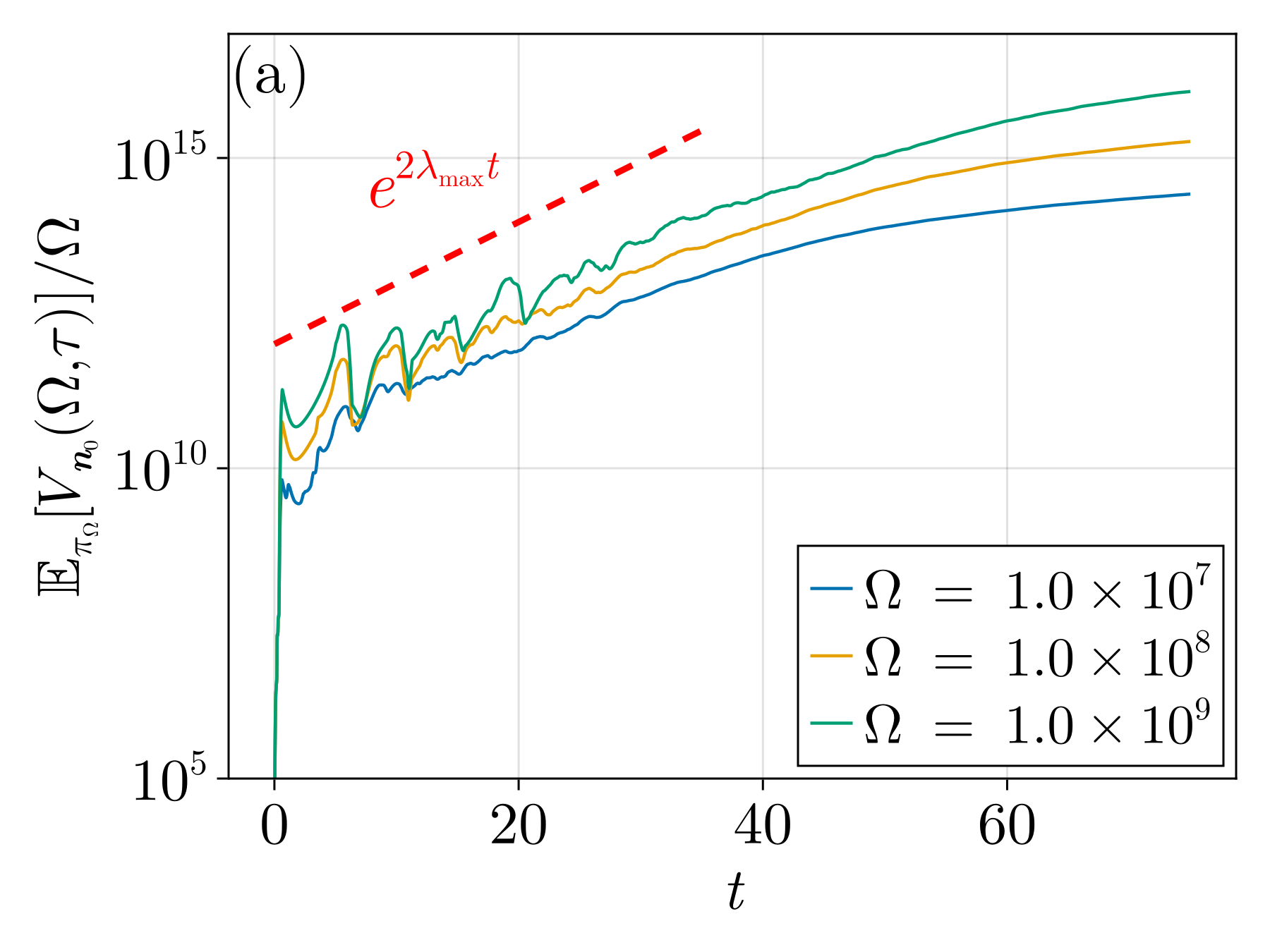}
    \includegraphics[width = 0.45\linewidth]{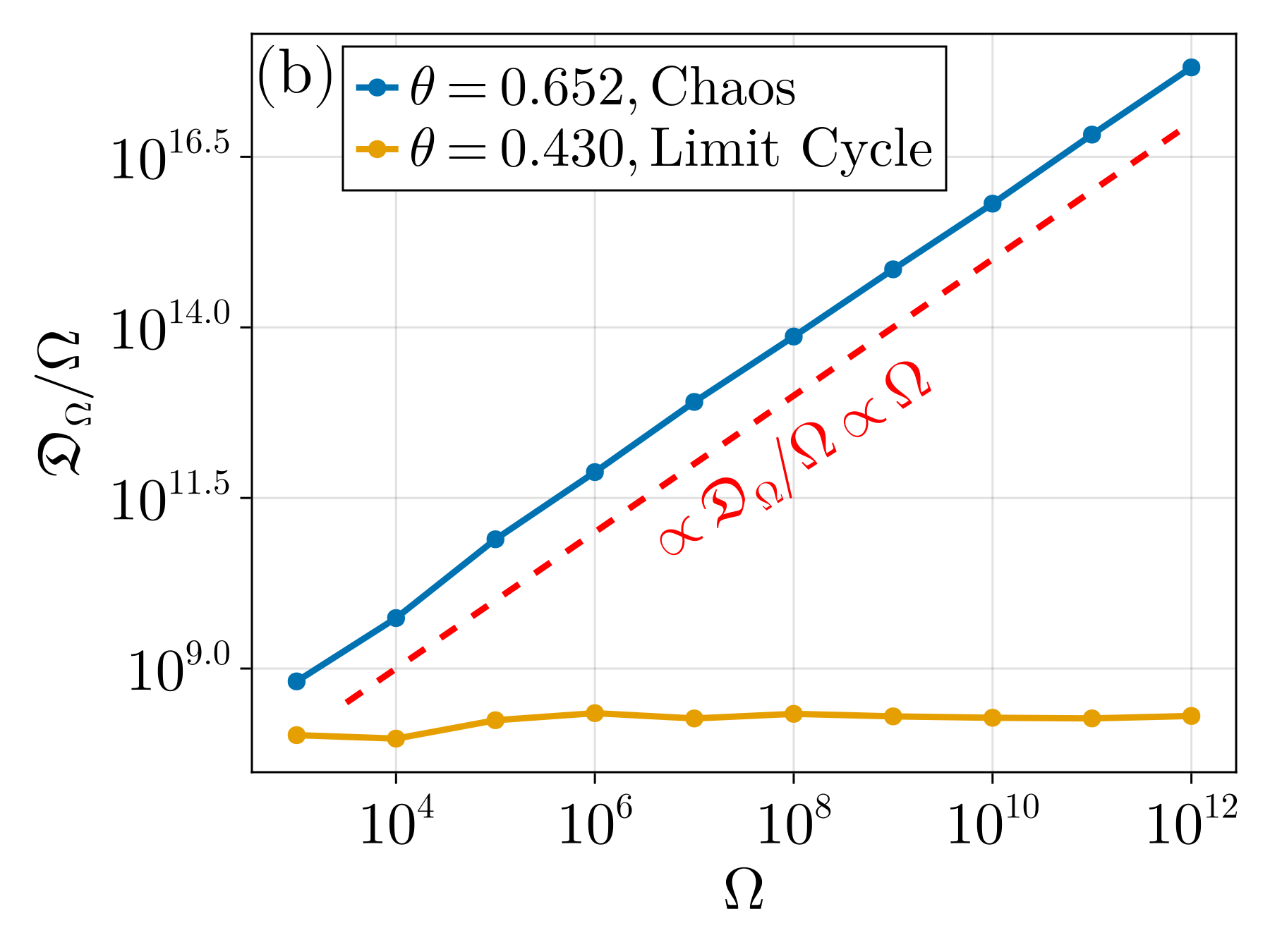}
    \caption{
        The behavior of entropy-production fluctuations for the Gaspard model in the chaotic regime.
        (a): $\mathbb{E}_{\pi_{\Omega}}[{V}_{{\bm n}_0}(\Omega, \tau)]/\Omega$ at chaotic regime $\theta = 0.652$. 
        (b): $\mathfrak{D}_{\Omega}/\Omega$ as a function of volume $\Omega$ for the chaotic case and the limit-cycle case.
    }
    \label{gaspard}
\end{figure}

\begin{figure}[ht]
    \centering
    \includegraphics[width=0.45\linewidth]{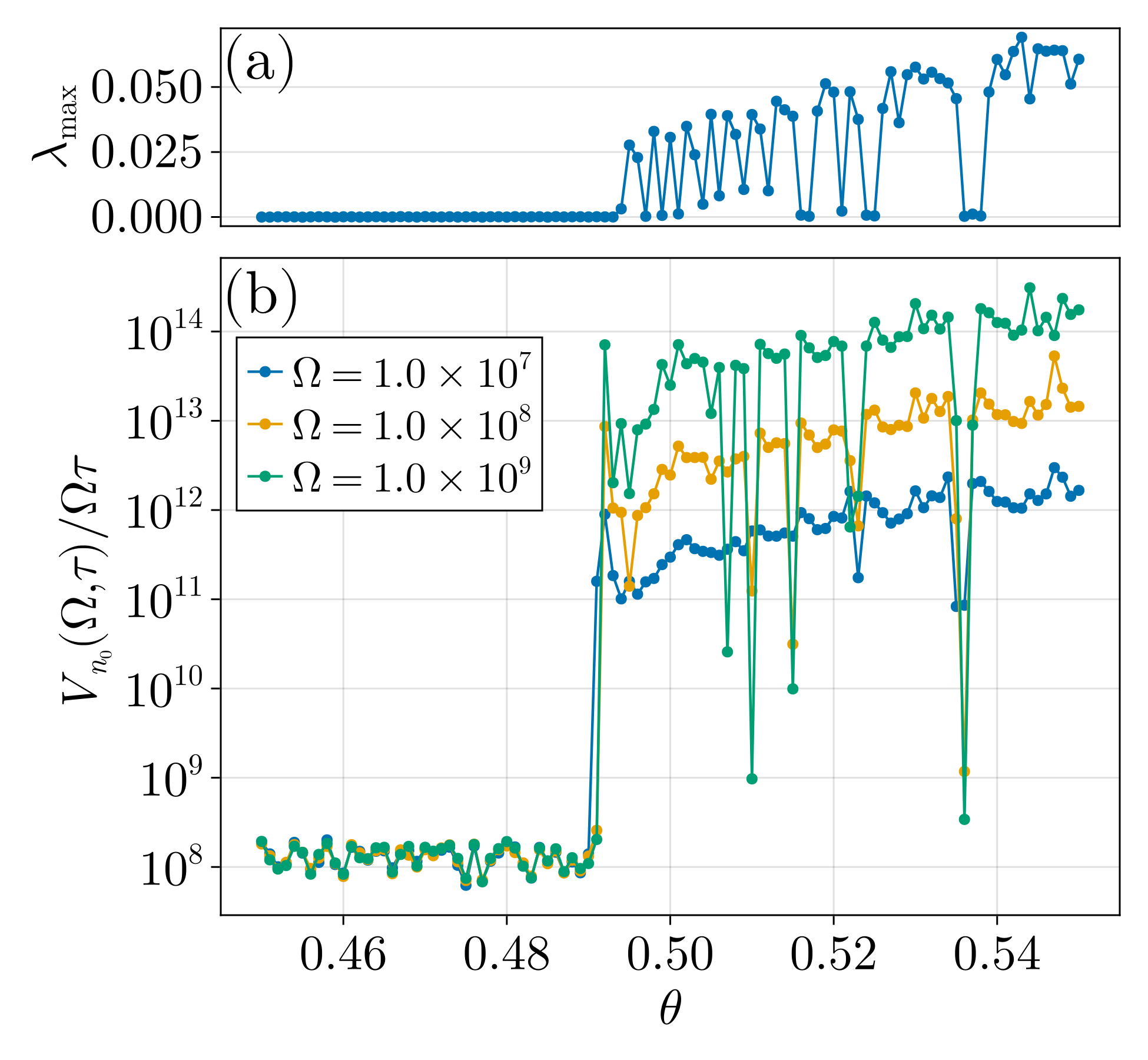}
    \includegraphics[width=0.45\linewidth]{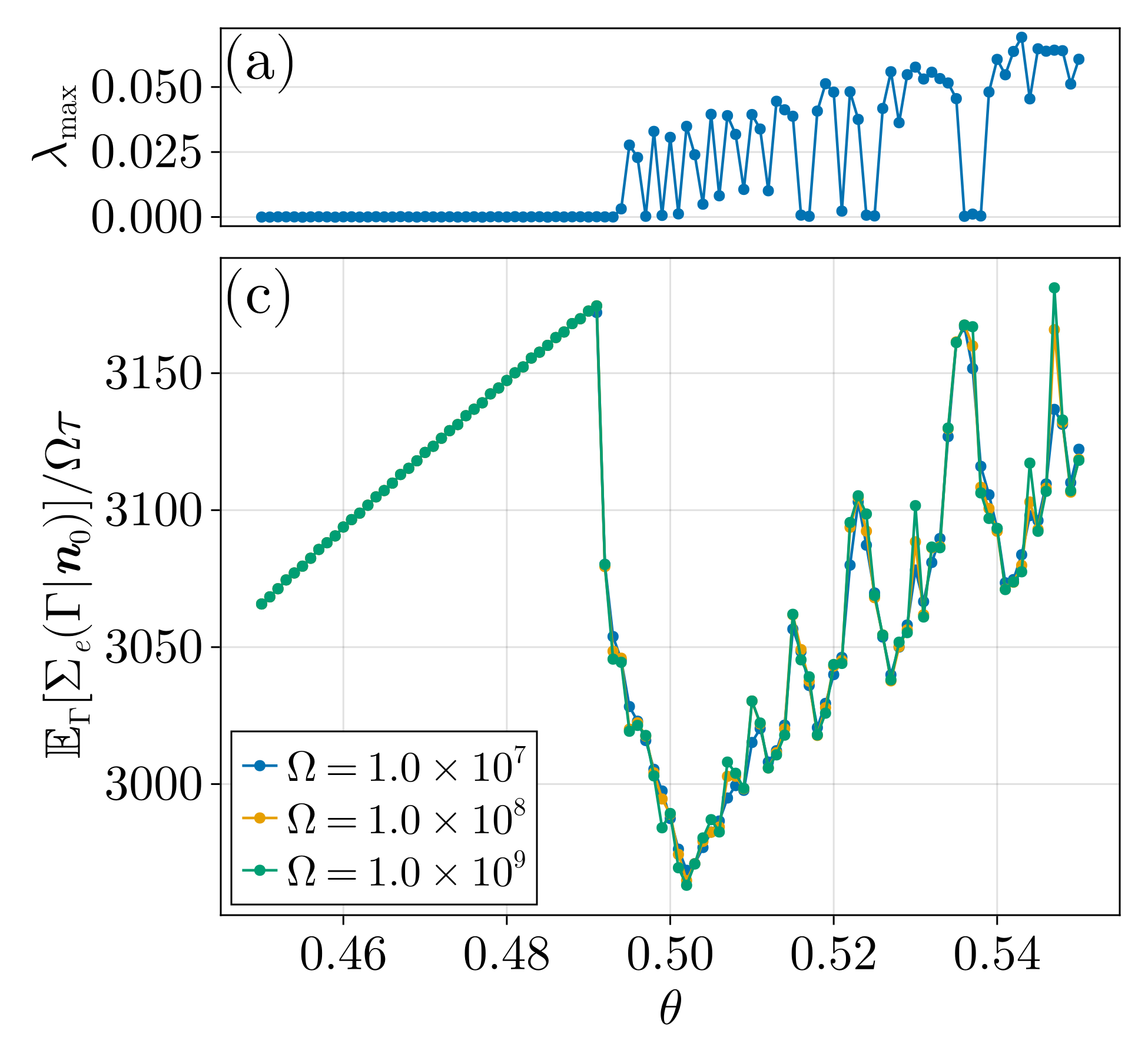}
    \caption{
        Entropy produciton fluctuations and its mean in the Gaspard model as a function of the control parameter $\theta = k_{+2}$. 
        (a) Maximum Lyapunov exponents. The same data are shown in both the left and right panels for clarity.
        (b) Entropy produciton fluctuations
        (c) Mean entropy-production rate. 
    }
    \label{fluc mean entropy production Gaspard}
\end{figure}

We investigate entropy-production fluctuations in the chaotic regime of the Gaspard model, following the same procedure as for the Willamowski--R\"ossler model.
Fig.~\ref{gaspard}(a) shows the time dependence of $\mathbb{E}_{\pi_{\Omega}}[V_{\bm{n}_0}(\Omega,\tau)]/\Omega$ for several system volumes. The quantity exhibits exponential growth over the time interval shown.
Fig.~\ref{gaspard}(b) shows the volume dependence of entropy-production fluctuations in both the chaotic and limit-cycle regimes. In the chaotic regime, the fluctuations exhibit quadratic scaling, $\mathfrak{D}_{\Omega}=O(\Omega^2)$, whereas in the limit-cycle regime, they scale linearly, $\mathfrak{D}_{\Omega}=O(\Omega)$.
These results illustrate that finite-time amplification and quadratic volume scaling are not restricted to chaos arising from a period-doubling cascade.
We also plot entropy-production fluctuations and its mean 
as a function of the control parameter $\theta=k_{+2}$ in Fig.~\ref{fluc mean entropy production Gaspard}. 
Fig.~\ref{fluc mean entropy production Gaspard}(b) shows
the entropy-production fluctuations in both the limit-cycle and chaotic regimes,
clearly demonstrating their system-size dependence throughout the chaotic regime.
For completeness, Fig.~\ref{fluc mean entropy production Gaspard}(c)
also shows the mean entropy-production rate,
for which corresponding results have already been reported
in Ref.~\cite{Gaspard10.1063/5.0025350ChaosEntropyProduction}.

Figs.~\ref{gaspard}(b) and \ref{fluc mean entropy production Gaspard} were obtained with $N_{\mathrm e}=618$, $\Delta t=0.002$, and $\tau=10000$. 
Fig.~\ref{gaspard}(a) was obtained with the following simulation parameters: $N_{\mathrm e}=500$, $\Delta t=0.002$, $N_{\bm{x}_0} = 400$, $0 \leq \tau \leq 75$ and $\tau_{\bm{x}_0} = 10000$.

\section{Deterministic diffusion constant and its stochastic counterpart}

We numerically investigate the finiteness of the deterministic diffusion constant $\mathfrak{D}_{\mathrm{d}}$ in both the Willamowski-R\"ossler model \eqref{Willamowski-Rossler model} and the Gaspard model \eqref{Gaspard model}.
We numerically integrate the deterministic dynamics governed by 
\begin{equation}
    \dot{Q}_{\rho} = J_{\rho}(\bm{x}(t)), \quad x_{\ell}(t) = x_{\ell}(0) + \sum_{\rho} \nabla_{\rho}^{\ell} Q_{\rho}(t).
\end{equation}
We numerically integrate these equations using the fifth-order explicit Runge–Kutta method (Tsit5) with a fixed time step $\Delta t = 0.002$.
Along each trajectory, we evaluate the instantaneous entropy-production rate $j_{\mathrm{d}}(t) = \sum_{\rho} \eta_{\rho} J_{\rho}(\bm{x}(t))$. 
We estimate the autocorrelation function $C_{\mathrm{d}}(t)$
of the entropy-production rate by averaging over deterministic trajectories with initial conditions sampled from the SRB measure $\mu_{\mathrm{SRB}}$. 
To approximate sampling from $\mu_{\mathrm{SRB}}$, we generate a single deterministic trajectory for $\tau_{\bm{x}(0)} + 500$ time units, discard the first $500$ time units as a transient, 
and sample $N_{\bm{x}(0)}$ initial conditions at equally spaced time intervals over the remaining $\tau_{\bm{x}_0}$ time units. 
From the estimated autocorrelation function, we evaluate the finite-time deterministic diffusion constant
\begin{equation}
    \mathfrak{D}_{d}(t) = 2 \int_{0}^{t} \ab(1 - \frac{s}{t}) C_{\mathrm{d}}(s) \d s
\end{equation}
to examine the finiteness of $\mathfrak{D}_{d} = \lim_{\tau \to \infty}\mathfrak{D}_{d}(\tau)$. 

\begin{figure}[ht]
    \centering
    \includegraphics[width=0.49\linewidth]{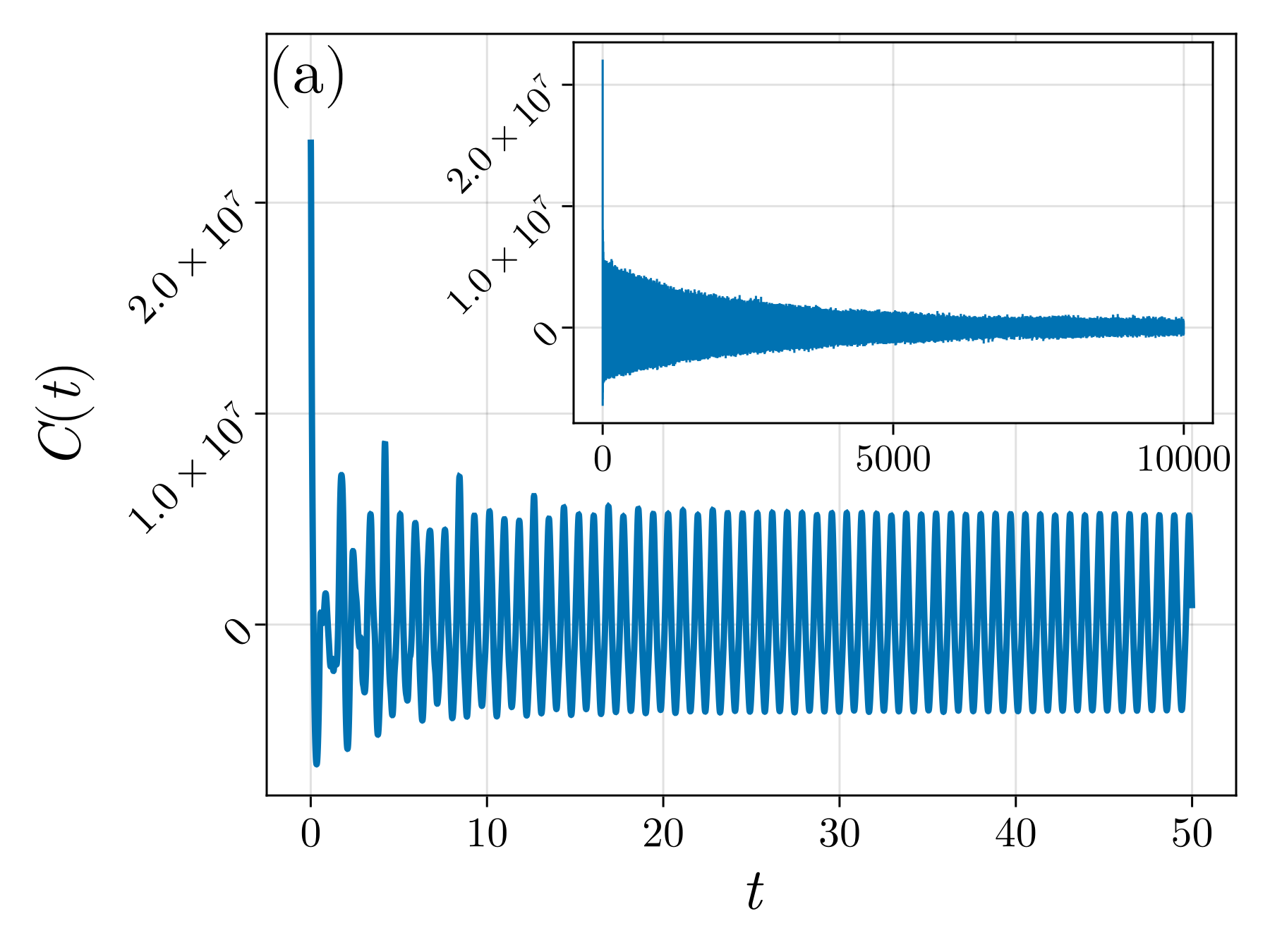}
    \includegraphics[width=0.49\linewidth]{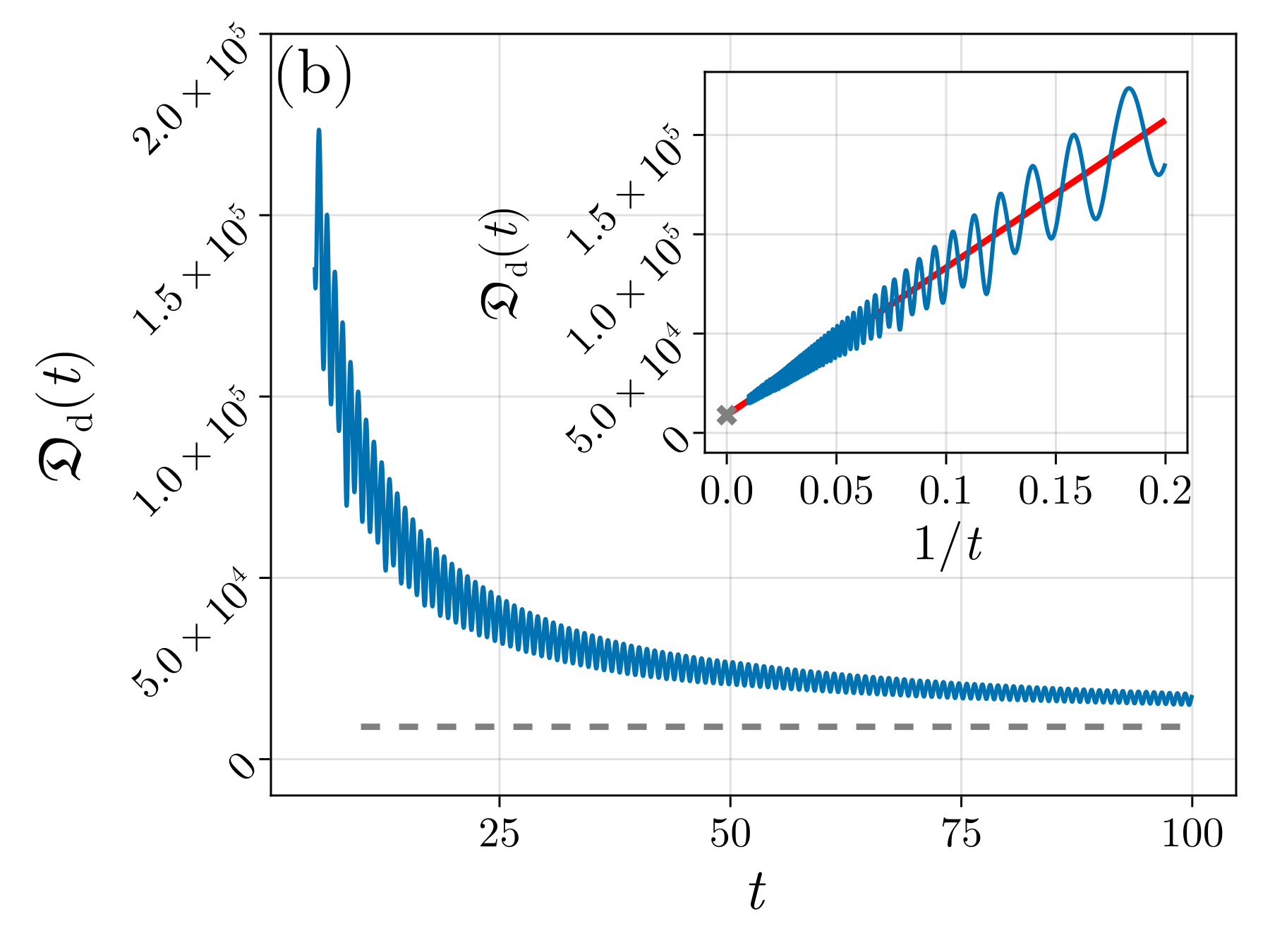}
    \caption{
        Numerical results for the Willamowski-R\"ossler model.
        (a) Autocorrelation function $C_{\mathrm{d}}(t)$
        of the entropy-production rate.
        The inset shows the longer-time behavior
        estimated using fewer initial conditions and lower resolution.
        (b) Finite-time deterministic diffusion constant
        $\mathfrak{D}_{\mathrm{d}}(t)$.
        The inset shows $\mathfrak{D}_{\mathrm{d}}(t)$
        as a function of $1/t$.
        The red solid line in the inset is a linear fit
        to the form $\mathfrak{D}_{\mathrm{d}}(t)=a+b/t$,
        shown as a guide to the eye.
        The gray dot marks the extrapolated intercept $a$
        at $1/t=0$, and the gray dashed line in the main panel
        indicates the corresponding value.
    }
    \label{autocorr Willamowski Rossler}
\end{figure}

We plot the estimated autocorrelation function $C_{\mathrm{d}}(t)$
and finite-time diffusion constant $\mathfrak{D}_{\mathrm{d}}(t)$
for the Willamowski-R\"ossler model in
Figs.~\ref{autocorr Willamowski Rossler}(a) and (b), respectively.
For these calculations, we set the control parameter to $k_{+2}=0.99$.
For the numerical simulations, we set
$\tau_{\bm{x}(0)} = 1000000$ and
$N_{\bm{x}(0)} = 1000000$,
and evaluate $C_{\mathrm{d}}(t)$ and
$\mathfrak{D}_{\mathrm{d}}(t)$.
Although $C_{\mathrm{d}}(t)$ appears not to decay
within the time window shown in
Fig.~\ref{autocorr Willamowski Rossler}(a),
its decay becomes evident over the longer time window
shown in the inset.
For this longer-time calculation,
we reduce the number of sampled initial conditions to
$N_{\bm{x}(0)}=100000, \tau_{\bm{x}(0)} = 100000$
and use a coarser temporal resolution
$\Delta t_{\mathrm{save}}=0.01$,
while keeping the integration time step fixed at
$\Delta t=0.002$.
To examine the convergence of $\mathfrak{D}_{\mathrm{d}}(t)$,
we plot it against $1/t$ in the inset of
Fig.~\ref{autocorr Willamowski Rossler}(b).
The numerical data suggest a finite-time correction
proportional to $1/t$, which can be distinguished
from the asymptotic constant contribution in this representation.
The data extrapolate to a finite intercept at $1/t=0$, 
supporting the finiteness of $\mathfrak{D}_{\mathrm{d}}$.

\begin{figure}[ht]
    \centering
    \includegraphics[width=0.49\linewidth]{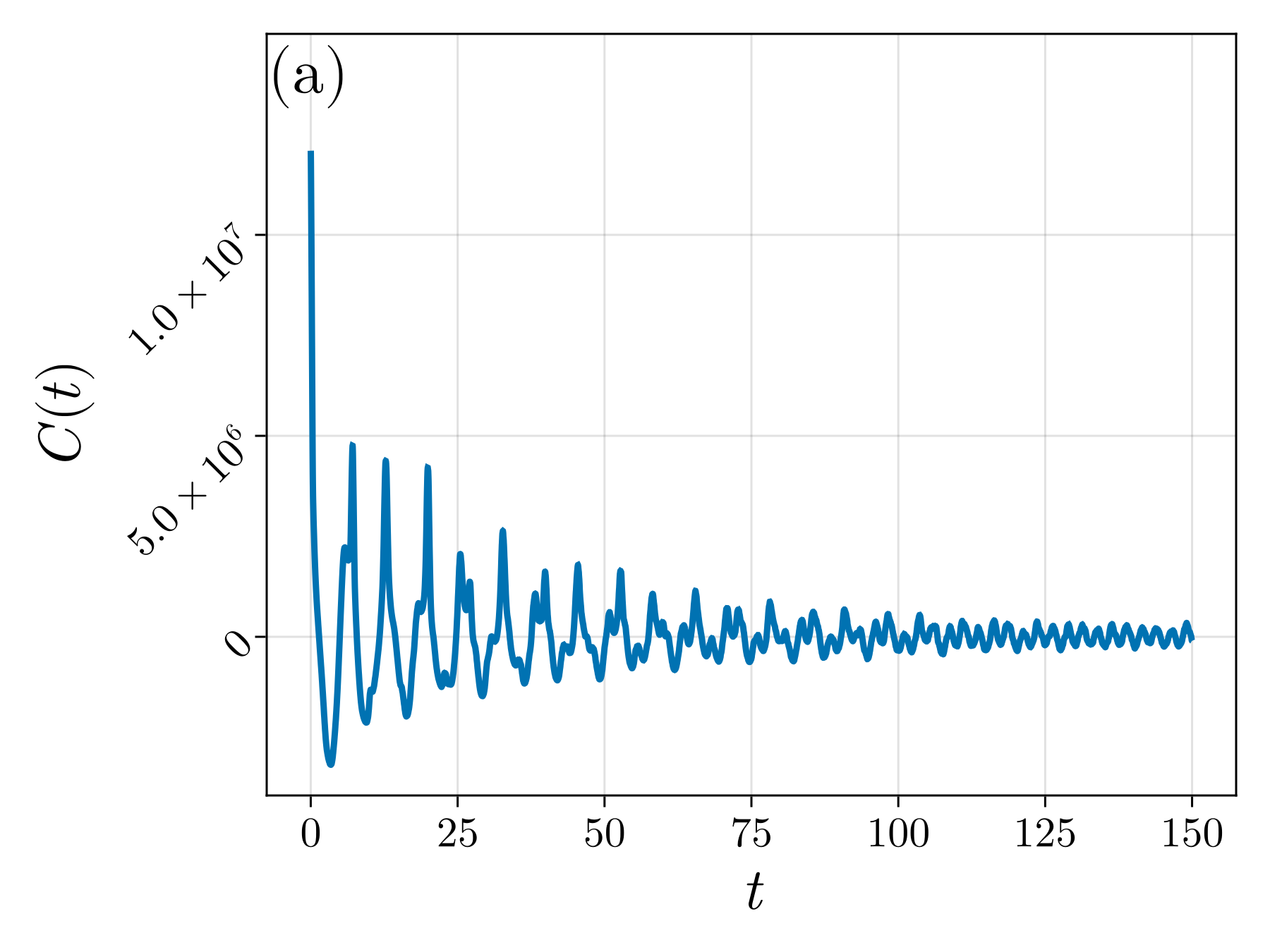}
    \includegraphics[width=0.49\linewidth]{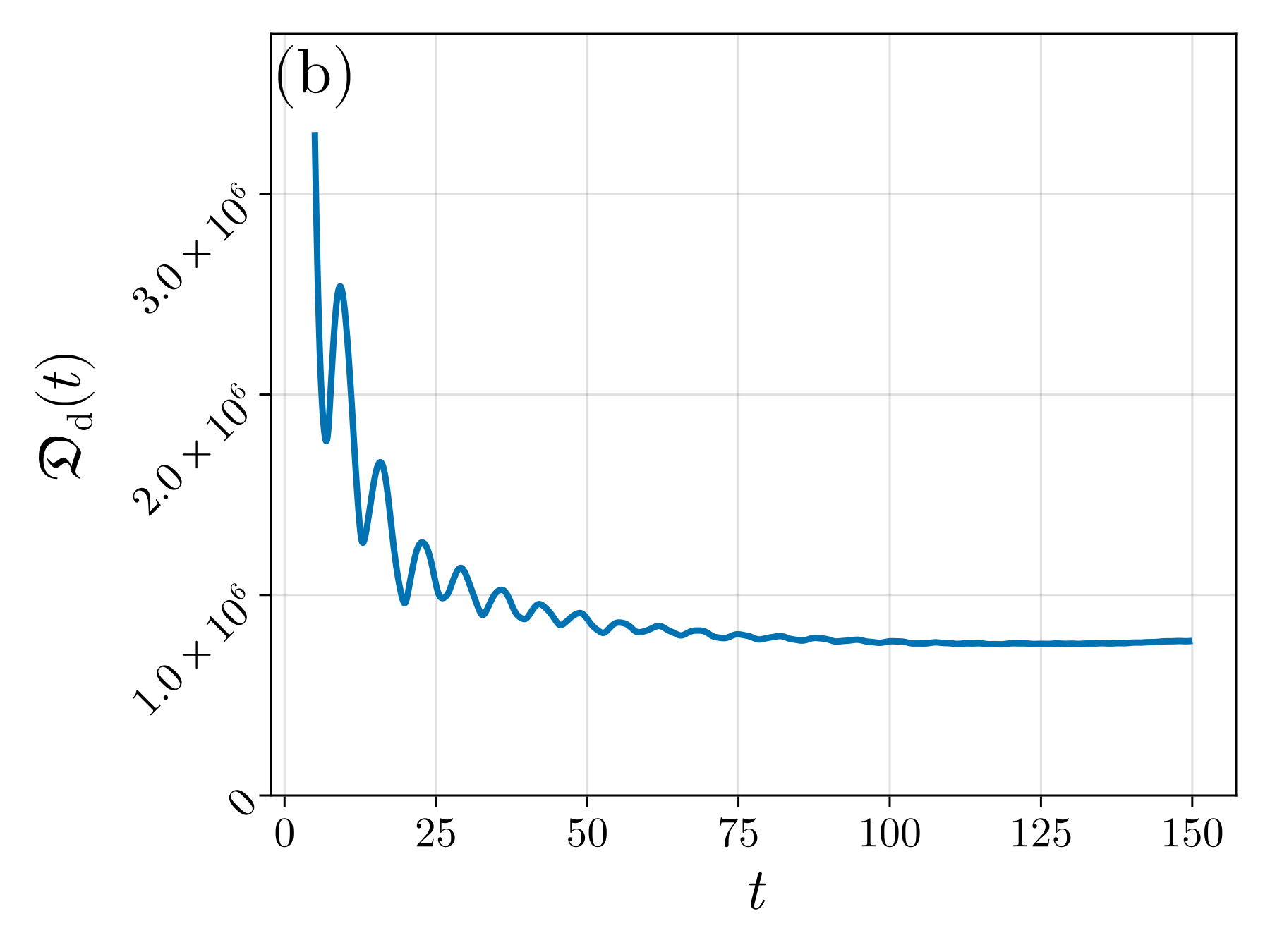}
    \caption{
        Numerical results for the Gaspard model.
        (a) Autocorrelation function $C_{\mathrm{d}}(t)$
        of the entropy-production rate.
        (b) Finite-time deterministic diffusion constant
        $\mathfrak{D}_{\mathrm{d}}(t)$.
    }
    \label{autocorr Gaspard}
\end{figure}

We perform the same analysis for the Gaspard model.
The estimated autocorrelation function $C_{\mathrm{d}}(t)$
and finite-time diffusion constant $\mathfrak{D}_{\mathrm{d}}(t)$
are shown in Figs.~\ref{autocorr Gaspard}(a) and (b), respectively.
For these calculations, we set the control parameter to $k_{+2}=0.652$.
For the numerical simulations, we set
$\tau_{\bm{x}(0)} = 100000$ and
$N_{\bm{x}(0)} = 5000$,
and evaluate $C_{\mathrm{d}}(t)$ and
$\mathfrak{D}_{\mathrm{d}}(t)$.
As shown in Fig.~\ref{autocorr Gaspard}(b),
$\mathfrak{D}_{\mathrm{d}}(t)$ approaches a constant value
at long times, providing numerical evidence
for the finiteness of $\mathfrak{D}_{\mathrm{d}}$.

We also compare the deterministic diffusion constant $\mathfrak{D}_{\mathrm{d}}$ with the rescaled stochastic diffusion constant $\mathfrak{D}_{\Omega}/\Omega^2$.

\begin{figure}[ht]
    \centering
    \includegraphics[width = 0.49\linewidth]{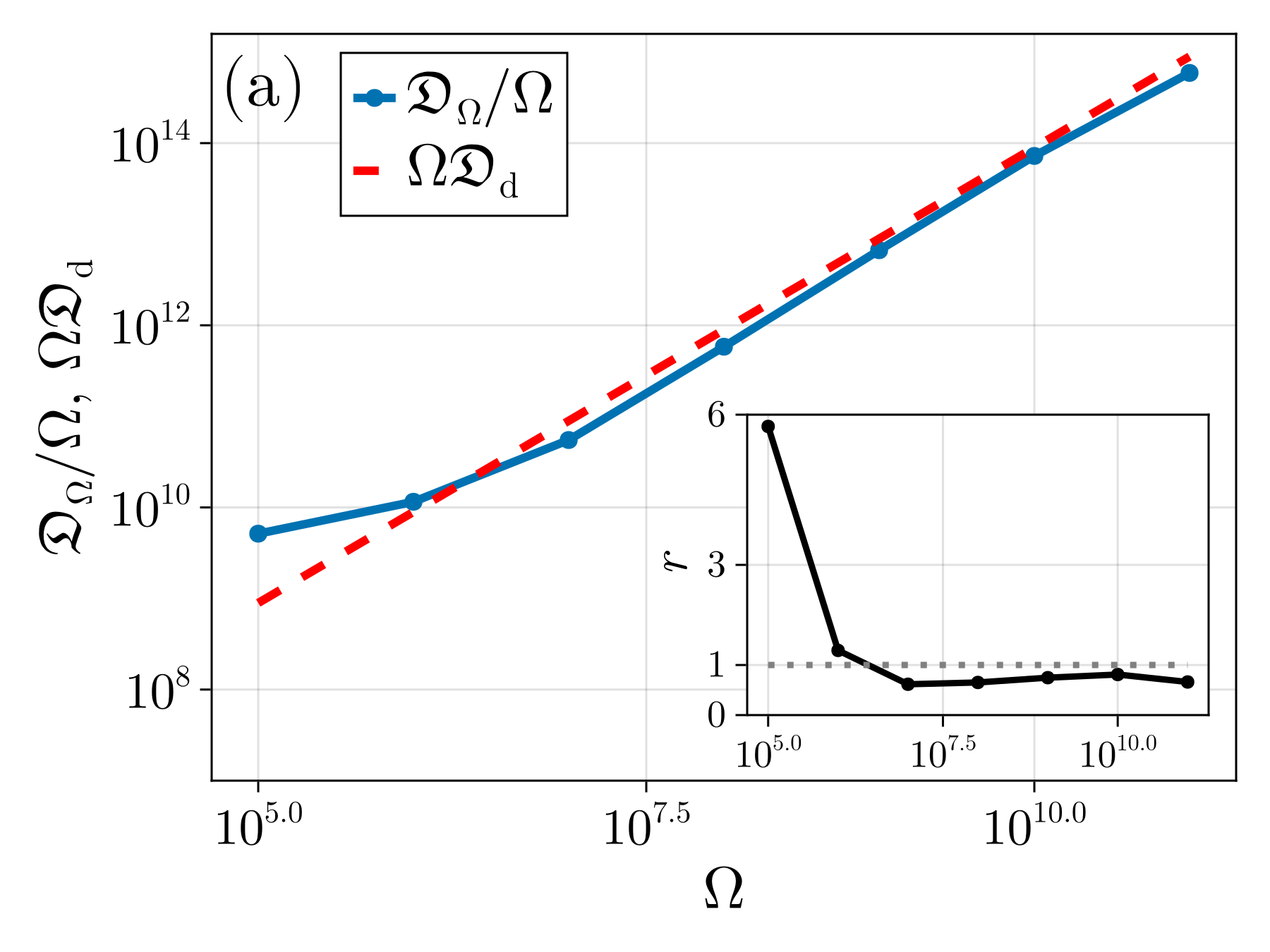}
    \includegraphics[width = 0.49\linewidth]{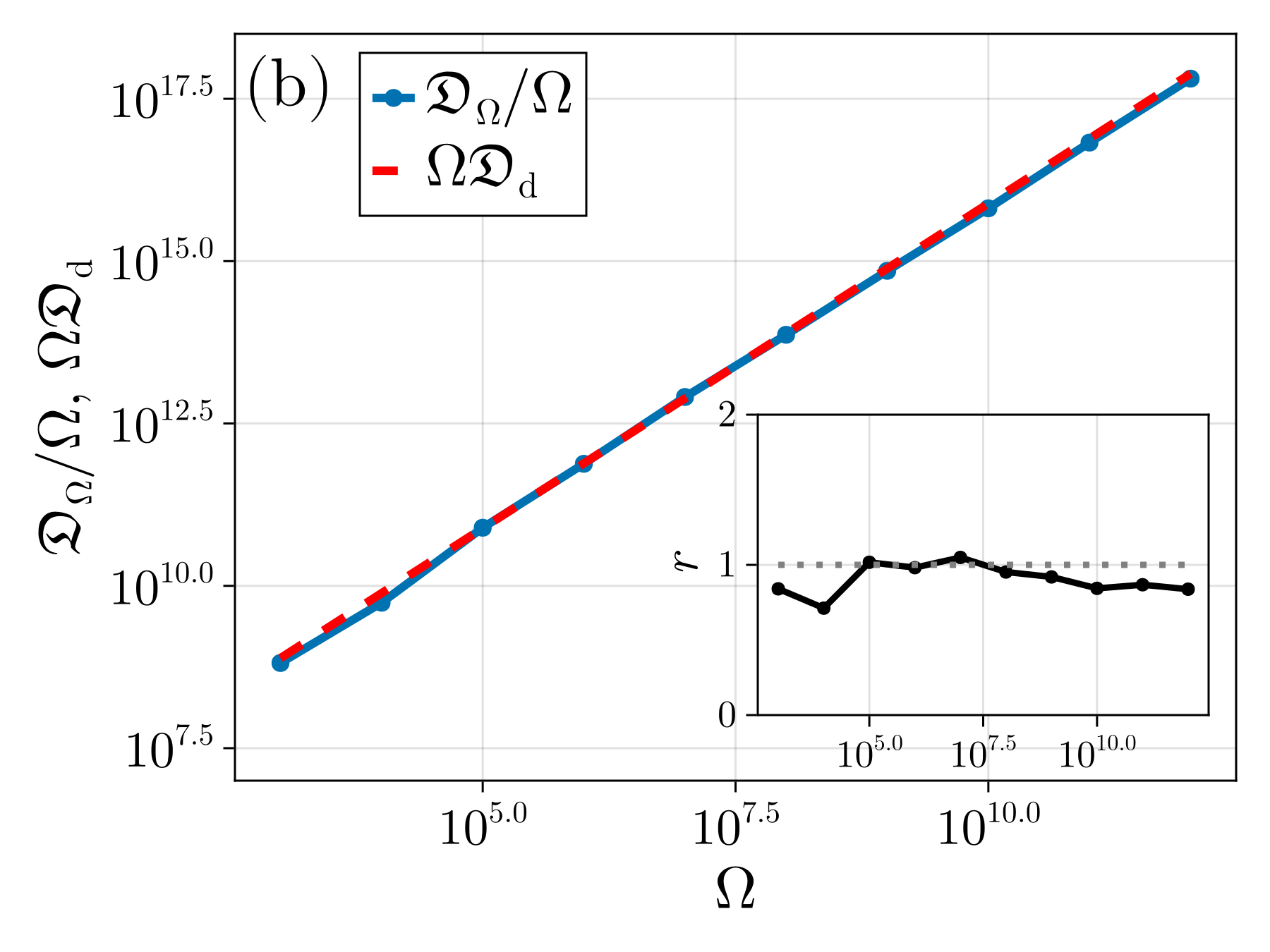}
    \caption{
    Comparison of the deterministic and stochastic diffusion constants.
    (a) Willamowski--R\"ossler model.
    (b) Gaspard model.
    In both panels, $\Omega\mathfrak{D}_{\mathrm{d}}$
    and $\mathfrak{D}_{\Omega}/\Omega$
    are plotted as functions of the system volume $\Omega$.
    The insets show the ratio
    $r = \mathfrak{D}_{\Omega}/(\mathfrak{D}_{\mathrm{d}}\Omega^2)$.
    For the Willamowski--R\"ossler model,
    $\mathfrak{D}_{\mathrm{d}}$ is estimated by extrapolation
    to the infinite-time limit,
    whereas for the Gaspard model,
    its value at the final observation time is used.
    }
    \label{compare}
\end{figure}

Fig.~\ref{compare}(a) compares $\Omega\mathfrak{D}_{\mathrm{d}}$ and $\mathfrak{D}_{\Omega}/\Omega$ for the Willamowski--R\"ossler model. 
Here, we use the extrapolated value of $\mathfrak{D}_{\mathrm{d}}$ obtained from the intercept at $1/t=0$ in the inset of Fig.~\ref{autocorr Willamowski Rossler}(b). 
The inset of Fig.~\ref{compare}(a) shows the ratio $r := \mathfrak{D}_{\Omega}/(\mathfrak{D}_{\mathrm{d}}\Omega^2)$. 
The two quantities agree in order of magnitude. More quantitatively, the ratios are $0.65$, $0.75$, $0.81$, and $0.66$ for $\Omega=10^{8},10^{9},10^{10}$, and $10^{11}$, respectively. 
The remaining quantitative discrepancy may be partly attributed to the finite observation time used to estimate $\mathfrak{D}_{\Omega}$, 
whereas $\mathfrak{D}_{\mathrm{d}}$ is estimated by extrapolation to the infinite-time limit.

We perform the same comparison for the Gaspard model. Since $\mathfrak{D}_{\mathrm{d}}(t)$ converges rapidly, we use its value at the final observation time in Fig.~\ref{autocorr Gaspard}(b), without extrapolation. 
Fig.~\ref{compare}(b) compares $\Omega\mathfrak{D}_{\mathrm{d}}$ and $\mathfrak{D}_{\Omega}/\Omega$, with the ratio $\mathfrak{D}_{\Omega}/(\mathfrak{D}_{\mathrm{d}}\Omega^2)$ shown in the inset. 
For larger system volumes, the ratios are $0.92$, $0.84$, $0.87$, and $0.84$ for $\Omega=10^9,10^{10},10^{11}$, and $10^{12}$, respectively, remaining close to unity.
These values remain close to unity over a wide range of system volumes, indicating quantitative agreement between $\mathfrak{D}_{\mathrm{d}}$ and $\mathfrak{D}_{\Omega}/\Omega^2$.

\section{Detailed proof of Theorem~\ref{theo1}}
\label{sec:sm_theorem1_proof}

We give the details of the proof outlined in the End Matter (EM).
Throughout this section, $\mu=\mu_{\rm SRB}$ is the ergodic invariant measure in Theorem~\ref{theo1}, and $\bm x(t)=\phi_t\bm x_0$ follows the mass-action dynamics~\eqref{deterministic eq}.
The limit $\Omega\to\infty$ has already been taken in $\upsilon_{\bm x_0}(\tau)$.
Thus $\upsilon_{\bm x_0}(\tau)$ is a deterministic function of the initial concentration $\bm x_0$; convergence in probability below refers to sampling $\bm x_0$ from $\mu$, not to sampling reaction noise.
We use the Euclidean norm in the concentration coordinates and its induced operator and row-vector norms.
Compactness makes $\bm g=\partial_{\bm x}j_d$, the reaction activities, and the Jacobian of the drift bounded along the invariant set.
The multiplicative ergodic theorem therefore gives forward and backward Lyapunov regularity on an invariant set of full $\mu$-measure, on which we work below.

\subsection{Entropy response along an unstable trajectory}
\label{sec:sm_theo1_response}

Let $E(t)=E(\bm x(t))$ denote the full Oseledets subspace corresponding to $\lambda_{\rm max}>0$, including its multiplicity.
It is a space of concentration perturbations, with
$\bm\Phi(t,s)E(s)=E(t)$.
The response to a perturbation introduced at time $s$ is
\begin{equation}
R_\tau(s):=\int_s^\tau dt\,\bm g(\bm x(t))^T\bm\Phi(t,s),
\qquad
\upsilon_{\bm x_0}(\tau)
=\int_0^\tau ds\sum_\rho A_\rho(\bm x(s))
\bigl[\eta_\rho+R_\tau(s)\bm\nabla_\rho\bigr]^2.
\label{eq:sm_theo1_variance_representation}
\end{equation}
The second identity is Eq.~\eqref{lcformula} of the main text.
The term $R_\tau(s)\bm\nabla_\rho$ accounts for the later change in entropy transfer caused by the concentration change in reaction $\rho$; $\eta_\rho$ is its direct contribution.
The nonnegative integrand will allow us to obtain a lower bound using only noise generated during a fixed initial interval.

Define the row vector acting on $E(t)$ by
\begin{equation}
\bm q(t):=\int_{-\infty}^{t}du\,
\bm g(\bm x(u))^T\left.\bm\Phi(u,t)\right|_{E(t)}.
\label{eq:sm_theo1_q_definition}
\end{equation}
For each fixed $t$ and $0<\delta<\lambda_{\rm max}$, backward regularity gives
$\|\bm\Phi(t-u,t)|_{E(t)}\|\le C_{\delta,t}e^{-(\lambda_{\rm max}-\delta)u}$ for $u\ge0$, after enlarging a finite constant to include short times.
Since $\bm g$ is bounded, the integral converges.
Splitting it at $s$ and using the composition law gives
\begin{align}
\bm q(\tau)\left.\bm\Phi(\tau,s)\right|_{E(s)}
&=\int_{-\infty}^{s}du\,\bm g(\bm x(u))^T
  \left.\bm\Phi(u,s)\right|_{E(s)}
  +\int_s^\tau du\,\bm g(\bm x(u))^T
  \left.\bm\Phi(u,s)\right|_{E(s)}\nonumber\\
&=\bm q(s)+\left.R_\tau(s)\right|_{E(s)}.
\label{eq:sm_theo1_response_identity}
\end{align}
This is an identity between row vectors on $E(s)$.
It separates the response amplified from $s$ to $\tau$ from a contribution fixed by the past at $s$.
Unstable stretching alone is insufficient if its coefficient $\bm q(\tau)$ cancels; the next step controls this possibility.

\subsection{Zero invariant measure of the zero-response set}
\label{sec:sm_theo1_zero_set}

Write
$\mathcal C:=\{\bm x:\bm g(\bm x)^T|_{E(\bm x)}\ne0\}$.
The coupling hypothesis is $\mu(\mathcal C)>0$.
To analyze zeros of $\bm q(t)$ without comparing vectors in different spaces, transport it to $E(0)$:
\begin{equation}
\bm H(t):=\bm q(t)\left.\bm\Phi(t,0)\right|_{E(0)}
=\bm q(0)+\int_0^t du\,\bm g(\bm x(u))^T
\left.\bm\Phi(u,0)\right|_{E(0)},
\qquad
\dot{\bm H}(t)=\bm g(\bm x(t))^T
\left.\bm\Phi(t,0)\right|_{E(0)}.
\label{eq:sm_theo1_transported_response}
\end{equation}
For mass action, both $\bm F$ and $j_d$ are polynomials in the concentrations.
Consequently $\bm x(t)$, $\bm\Phi(t,0)$, and the components of $\bm H(t)$ in a fixed basis of $E(0)$ are analytic in time.
This argument uses the finite integral in Eq.~\eqref{eq:sm_theo1_transported_response}; it does not require differentiating an infinite integral or assuming that $E(\bm x)$ is analytic as a function of $\bm x$.

Ergodicity and $\mu(\mathcal C)>0$ imply that almost every trajectory visits $\mathcal C$.
At such a time, $\dot{\bm H}(t)\ne0$, because $\bm\Phi(t,0)$ maps $E(0)$ bijectively onto $E(t)$.
Thus $\bm H$ is not identically zero along a typical trajectory.
At least one component is an analytic function that is not identically zero, so its zeros are isolated; the common zeros of all components are a subset of these zeros.
Moreover, $\bm q(t)=0$ if and only if $\bm H(t)=0$.
It follows that a typical trajectory spends zero time at $\bm q=0$, although it may cross zero infinitely often as $t\to\infty$.

To convert this statement about time into a statement about initial states, set
$\bm Q(\bm x):=\bm q(0;\bm x)$ and
$Z:=\{\bm x:\bm Q(\bm x)=0\}$.
The autonomous invertible flow determines the entire past from its current state, so
$\bm q(t;\bm x_0)=\bm Q(\phi_t\bm x_0)$.
For any fixed $T>0$, invariance and interchange of the two nonnegative integrals give
\begin{equation}
T\mu(Z)
=\int_0^T dt\,\mu(\phi_t^{-1}Z)
=\int d\mu(\bm x_0)\int_0^T dt\,
\mathbf 1_Z(\phi_t\bm x_0)
=0.
\label{eq:sm_theo1_zero_measure}
\end{equation}
The last equality follows from the isolated-zero property on almost every trajectory.
In particular, $0<\|\bm Q\|<\infty$ almost surely.
No density of $\mu$ in concentration space is required.

The distribution of $\|\bm q(\tau)\|$ is stationary.
Hence, for every $\delta>0$,
\begin{align}
\mu\left\{\frac{\bigl|\ln\|\bm q(\tau)\|\bigr|}{\tau}>\delta\right\}
&=\mu\{\|\bm Q\|<e^{-\delta\tau}\}
 +\mu\{\|\bm Q\|>e^{\delta\tau}\}
\longrightarrow0.
\label{eq:sm_theo1_q_log_probability}
\end{align}
The first term tends to $\mu(Z)=0$ as its threshold shrinks to zero, and the second tends to zero because $\bm Q$ is finite almost surely.
The first term alone is the estimate in Eq.~\eqref{eq:theo1_q_probability} of the EM.
This is a probability statement at each observation time: an exponentially small terminal response occurs for a vanishing fraction of stationary initial states.
It does not exclude isolated cancellations along an individual trajectory, and requires no moment assumption on $\ln\|\bm Q\|$.

\subsection{A lower bound from noise generated near the initial time}
\label{sec:sm_theo1_lower}

Fix $T>0$ independently of $\tau$, and consider $\tau\ge T$.
Define the least stretching within the leading subspace and the response at the initial time by
\begin{equation}
m_\tau:=\min_{\substack{\bm v\in E(0)\\\|\bm v\|=1}}
\|\bm\Phi(\tau,0)\bm v\|,
\qquad
r_\tau:=\|R_\tau(0)|_{E(0)}\|.
\label{eq:sm_theo1_m_and_r}
\end{equation}
If $k=\dim E(0)$, the restricted propagator is represented by a $k\times k$ matrix in orthonormal bases of $E(0)$ and $E(\tau)$, and $m_\tau$ is its smallest singular value.
The same singular-value bound applies to row vectors:
$\|\bm q(\tau)\bm\Phi(\tau,0)|_{E(0)}\|\ge m_\tau\|\bm q(\tau)\|$.
Applying $\|a-b\|^2\ge\|a\|^2/2-\|b\|^2$ to Eq.~\eqref{eq:sm_theo1_response_identity} at $s=0$ gives
\begin{equation}
r_\tau^2\ge\frac12m_\tau^2\|\bm q(\tau)\|^2-\|\bm q(0)\|^2.
\label{eq:sm_theo1_initial_response_bound}
\end{equation}
We use the quadratic inequality directly, since a linear lower bound that may be negative cannot simply be squared.

A response at the single time $s=0$ contributes nothing by itself to the time integral for the variance.
The composition law supplies a bound over an interval of nonzero length.
Set
\begin{equation}
K_T:=\sup_{0\le s\le T}\|\bm\Phi(s,0)|_{E(0)}\|,
\qquad
L_T:=\sup_{0\le s\le T}\|R_s(0)|_{E(0)}\|.
\label{eq:sm_theo1_window_constants}
\end{equation}
Both are finite, $K_T\ge1$, and neither depends on $\tau$.
Restricting $R_\tau(s)\bm\Phi(s,0)=R_\tau(0)-R_s(0)$ to $E(0)$, and using the operator-norm inequality, yields for every $0\le s\le T$,
\begin{align}
K_T^2\|R_\tau(s)|_{E(s)}\|^2
&\ge\|(R_\tau(0)-R_s(0))|_{E(0)}\|^2
\ge\frac12r_\tau^2-L_T^2.
\label{eq:sm_theo1_uniform_response_bound}
\end{align}
This bound remains valid when its right-hand side is negative.

The noise covariance in concentration space is
\begin{equation}
\bm B(s):=\sum_\rho A_\rho(\bm x(s))\bm\nabla_\rho\bm\nabla_\rho^T.
\label{eq:sm_theo1_noise_matrix}
\end{equation}
For the positive concentrations and reversible channels considered here, the reaction activities are positive.
The absence of conservation laws means that the stoichiometric vectors span the concentration tangent space.
Thus, for every nonzero $\bm w$,
$\bm w^T\bm B(s)\bm w=\sum_\rho A_\rho(\bm x(s))(\bm w^T\bm\nabla_\rho)^2>0$.
Continuity on $[0,T]$ gives $\bm B(s)\ge bI$ with $b=b(\bm x_0,T)>0$.
Only this finite interval is used; a common positive lower bound over the whole attractor is unnecessary.

Retaining this interval in Eq.~\eqref{eq:sm_theo1_variance_representation}, and applying $(a+b)^2\ge a^2/2-b^2$ to the indirect and direct entropy contributions, gives
\begin{align}
\upsilon_{\bm x_0}(\tau)
&\ge\frac12\int_0^T ds\,R_\tau(s)\bm B(s)R_\tau(s)^T-C_\eta
\nonumber\\
&\ge\frac b2\int_0^T ds\,\|R_\tau(s)|_{E(s)}\|^2-C_\eta
\nonumber\\
&\ge\frac{bT}{4K_T^2}r_\tau^2-\frac{bTL_T^2}{2K_T^2}-C_\eta
\nonumber\\
&\ge c\,m_\tau^2\|\bm q(\tau)\|^2-C,
\label{eq:sm_theo1_variance_lower}
\end{align}
where
\begin{equation}
C_\eta:=\int_0^T ds\sum_\rho A_\rho(\bm x(s))\eta_\rho^2,
\qquad
c:=\frac{bT}{8K_T^2}>0,
\qquad
C:=C_\eta+\frac{bT}{4K_T^2}
\bigl(\|\bm q(0)\|^2+2L_T^2\bigr).
\label{eq:sm_theo1_lower_constants}
\end{equation}
In the second line we used $\|R_\tau(s)\|\ge\|R_\tau(s)|_{E(s)}\|$.
The last two lines use Eqs.~\eqref{eq:sm_theo1_uniform_response_bound} and \eqref{eq:sm_theo1_initial_response_bound}, respectively.
The constants $c$ and $C$ are finite for almost every initial state and are independent of $\tau$, although they need not be uniform in $\bm x_0$.
Physically, the noise injected during the initial interval already excites perturbations whose later amplification provides the required variance.

\subsection{Lyapunov growth estimates and the upper bound}
\label{sec:sm_theo1_growth}

We justify both the growth of $m_\tau$ and the propagator bound used in the EM.
Choose a regular basis $\bm v_1,\ldots,\bm v_d$ at $t=0$, with Lyapunov exponents $\lambda_i$ counted with multiplicity.
Arrange it so that the first $k$ vectors span $E(0)$ and have exponent $\lambda_{\rm max}$.
Define
\begin{equation}
\bm V=(\bm v_1,\ldots,\bm v_d),
\qquad
\bm\Lambda=\operatorname{diag}(\lambda_1,\ldots,\lambda_d),
\qquad
\bm C(t)=\bm\Phi(t,0)\bm V e^{-\bm\Lambda t}.
\label{eq:sm_theo1_regular_frame}
\end{equation}
Each column of $\bm C(t)$ has subexponential norm.
Regularity also controls volume growth:
$\ln|\det\bm\Phi(t,0)|=t\sum_i\lambda_i+o(t)$.
It follows that $\ln|\det\bm C(t)|=o(t)$.
The cofactor formula for the inverse then shows that both $\|\bm C(t)\|$ and $\|\bm C(t)^{-1}\|$ grow at most subexponentially.
This inverse bound is important: growth of each fixed vector alone would not control a minimizing direction that changes with time.

For a unit vector $\bm v\in E(0)$, write $\bm v=\bm V\bm a$, where only the first $k$ components of $\bm a$ are nonzero.
Then $\bm\Phi(t,0)\bm v=e^{\lambda_{\rm max}t}\bm C(t)\bm a$ and
$1/\|\bm V\|\le\|\bm a\|\le\|\bm V^{-1}\|$.
These inequalities hold uniformly over all such $\bm v$, so taking the minimum gives
\begin{equation}
\frac{e^{\lambda_{\rm max}t}}{\|\bm V\|\,\|\bm C(t)^{-1}\|}
\le m_t\le
e^{\lambda_{\rm max}t}\|\bm C(t)\|\,\|\bm V^{-1}\|,
\qquad
\frac{\ln m_t}{t}\longrightarrow\lambda_{\rm max}.
\label{eq:sm_theo1_minimum_stretching}
\end{equation}
Thus $m_t=e^{\lambda_{\rm max}t+o(t)}$, including when the maximal exponent has multiplicity greater than one.
No constant finite-time stretching rate is assumed.

For any $\epsilon>0$, a finite initial-state-dependent constant $L_\epsilon$ can be chosen so that
$\|\bm C(t)\|,\|\bm C(t)^{-1}\|\le L_\epsilon e^{\epsilon t/2}$ for all $t\ge0$.
The composition law now gives
\begin{align}
\bm\Phi(t,s)&=\bm C(t)e^{\bm\Lambda(t-s)}\bm C(s)^{-1},\nonumber\\
\|\bm\Phi(t,s)\|&\le L_\epsilon^2 e^{\epsilon(t+s)/2}e^{\lambda_{\rm max}(t-s)}
=K_\epsilon e^{\epsilon s}e^{(\lambda_{\rm max}+\epsilon/2)(t-s)},
\qquad t\ge s\ge0,
\label{eq:sm_theo1_propagator_bound}
\end{align}
where $K_\epsilon=L_\epsilon^2$.
The factor $e^{\epsilon s}$ retains the possible nonuniformity with respect to the starting time.
With $G=\sup\|\bm g\|<\infty$ and $0<\epsilon<\lambda_{\rm max}$, integration gives
\begin{align}
\|R_\tau(s)\|
&\le GK_\epsilon e^{\epsilon s}
\int_s^\tau dt\,e^{(\lambda_{\rm max}+\epsilon/2)(t-s)}\nonumber\\
&\le\frac{GK_\epsilon}{\lambda_{\rm max}+\epsilon/2}
e^{(\lambda_{\rm max}+\epsilon/2)\tau}
e^{-(\lambda_{\rm max}-\epsilon/2)s}.
\label{eq:sm_theo1_response_upper}
\end{align}
Using $(a+b)^2\le2a^2+2b^2$ in Eq.~\eqref{eq:sm_theo1_variance_representation}, bounded activities, and
$\int_0^\tau ds\,e^{-(2\lambda_{\rm max}-\epsilon)s}\le(2\lambda_{\rm max}-\epsilon)^{-1}$, we obtain
\begin{equation}
\upsilon_{\bm x_0}(\tau)
\le C_0\tau+C_\epsilon e^{(2\lambda_{\rm max}+\epsilon)\tau},
\qquad
\limsup_{\tau\to\infty}\frac1\tau\ln\upsilon_{\bm x_0}(\tau)
\le2\lambda_{\rm max}
\quad\text{for $\mu$-almost every $\bm x_0$}.
\label{eq:sm_theo1_variance_upper}
\end{equation}
The constants are independent of $\tau$; $C_\epsilon$ may depend on $\bm x_0$.

\subsection{Convergence in probability and the divergence of $D$}
\label{sec:sm_theo1_probability}

It remains to handle the initial-state-dependent constants and the subtraction in the lower bound.
Define $Y_\tau:=c\,m_\tau^2\|\bm q(\tau)\|^2$.
Equations~\eqref{eq:sm_theo1_q_log_probability} and \eqref{eq:sm_theo1_minimum_stretching}, together with $0<c<\infty$ almost surely, give
\begin{equation}
\frac{\ln Y_\tau}{\tau}
=\frac{\ln c}{\tau}+2\frac{\ln m_\tau}{\tau}
 +2\frac{\ln\|\bm q(\tau)\|}{\tau}
\xrightarrow[\tau\to\infty]{\mu\text{-prob.}}2\lambda_{\rm max}.
\label{eq:sm_theo1_amplified_response_probability}
\end{equation}
The terms need not be independent, and no integrability of $c$, $C$, or their logarithms is used.
For $0<\delta<2\lambda_{\rm max}$, put $a_\tau=e^{(2\lambda_{\rm max}-\delta)\tau}$.
Since $\upsilon_{\bm x_0}(\tau)\ge Y_\tau-C$, the event inclusion
\begin{equation}
\{\upsilon_{\bm x_0}(\tau)<a_\tau\}
\subset
\{Y_\tau<2a_\tau\}\,\cup\,\{C>a_\tau\}
\label{eq:sm_theo1_subtraction_control}
\end{equation}
shows that its probability tends to zero.
The first event becomes improbable by Eq.~\eqref{eq:sm_theo1_amplified_response_probability}; the second does because $C$ is finite almost surely and $a_\tau\to\infty$.
The almost-sure upper logarithmic bound in Eq.~\eqref{eq:sm_theo1_variance_upper} likewise makes the probability of exceeding $2\lambda_{\rm max}+\delta$ tend to zero.
Combining the two sides proves
\begin{equation}
\lim_{\tau\to\infty}
\mu\left\{\bm x_0:
\left|\frac1\tau\ln\upsilon_{\bm x_0}(\tau)-2\lambda_{\rm max}\right|>\delta
\right\}=0
\qquad\text{for every $\delta>0$}.
\label{eq:sm_theo1_probability_conclusion}
\end{equation}

Finally, nonnegativity of the variance implies
\begin{equation}
\E_\mu[\upsilon_{\bm x_0}(\tau)]
\ge a_\tau\,\mu\{\upsilon_{\bm x_0}(\tau)\ge a_\tau\}
=[1-o(1)]e^{(2\lambda_{\rm max}-\delta)\tau}.
\label{eq:sm_theo1_mean_lower}
\end{equation}
Using the stationary-limit identification stated in the main text, this gives
\begin{equation}
D=\lim_{\tau\to\infty}\frac{\E_\mu[\upsilon_{\bm x_0}(\tau)]}{\tau}
=\infty,
\qquad
\liminf_{\tau\to\infty}\frac1\tau
\ln\E_\mu[\upsilon_{\bm x_0}(\tau)]\ge2\lambda_{\rm max}.
\label{eq:sm_theo1_diffusion_divergence}
\end{equation}
The last inequality follows by letting $\delta\downarrow0$.
The mean lower bound suffices; no interchange of an expectation and a long-time limit is needed.
The theorem determines the typical logarithmic growth rate and a lower bound on the rate of the arithmetic mean; equality of these two rates is not required for $D=\infty$.

\end{document}